\documentclass[fleqn,usenatbib]{mnras}

\usepackage[T1]{fontenc}
\usepackage[utf8]{inputenc}
\usepackage{lmodern}
\usepackage{newtxtext,newtxmath}
\usepackage{graphicx}
\usepackage{natbib}
\bibpunct{(}{)}{;}{a}{}{,} 
\setcitestyle{round}
\usepackage{ulem}
\usepackage{textcomp}
\usepackage{gensymb}
\usepackage{longtable}
\usepackage{threeparttable}
\usepackage{lipsum}
\usepackage{float}
\usepackage{todonotes}
\usepackage{amsmath}
\usepackage{etoolbox}
\usepackage{url}
\usepackage{mathtools}
\usepackage{colortbl}
\usepackage{booktabs}
\usepackage{subcaption}
\usepackage[labelfont=bf]{caption}

\DeclareRobustCommand{\VAN}[3]{#2}
\let\VANthebibliography\thebibliography
\def\thebibliography{\DeclareRobustCommand{\VAN}[3]{##3}\VANthebibliography}

\usepackage[most]{tcolorbox}
\newcommand{\spice}{\textsc{SPICE}}

\newcommand{\CII}{\ensuremath{\mbox{[\ion{C}{ii}]}}}

\newcommand{\OIII}{\ensuremath{\mbox{[\ion{O}{iii}]}}}
\newcommand{\LOIII}{$L_{\textup{\OIII}}$}

\newcommand{\HI}{\ion{H}{i}}
\newcommand{\HII}{\ion{H}{ii}}

\newcolumntype{Y}{>{\centering\arraybackslash}X}
\newcolumntype{W}[1]{>{\centering\arraybackslash}p{#1}}

\title[\OIII\ in high$-z$ galaxies with \spice]{New constraints on stellar feedback through \OIII\ emission: interpreting ALMA and JWST observations with \texttt{SPICE} simulations.}

\author[B. Casavecchia et al.]{
Benedetta Casavecchia,$^{1}$\thanks{E-mail: benecasa@mpa-garching.mpg.de}
Aniket Bhagwat,$^{1}$
Benedetta Ciardi,$^{1}$
C\'eline P\'eroux$^{2,3}$
and Tiago Costa$^{4}$
\\
$^{1}$Max-Planck-Institut f\"ur Astrophysik, Karl-Schwarzschild-Str. 1, 85748 Garching b. M\"unchen, Germany\\
$^{2}$European Southern Observatory, Karl-Schwarzschild-Str. 2, 85748 Garching b. M\"unchen, Germany\\
$^{3}$Aix Marseille Université, CNRS, Laboratoire d’Astrophysique de Marseille (LAM) UMR 7326, 13388 Marseille, France\\
$^{4}$Newcastle University, School of Mathematics, Statistics and Physics, Herschel Building Newcastle upon Tyne NE1 7RU, UK
}

\date{Accepted XXX. Received YYY; in original form ZZZ}

\pubyear{\the\year{}}

\begin{document}
\label{firstpage}
\pagerange{\pageref{firstpage}--\pageref{lastpage}}
\maketitle

\begin{abstract}
ALMA and JWST detected emission lines from the interstellar medium of star-forming galaxies during the Epoch of Reionization, reaching redshifts up to $z=14$. Among these, \OIII\ lines offer a powerful diagnostic of metal enrichment, gas ionization, and the impact of stellar feedback in galaxies at $z>6$. Modeling such emission in cosmological simulations is challenging due to the wide range of spatial scales and physical processes involved. To address this, we develop a post-processing pipeline that implements a sub-grid model for \OIII\ line emission within the \texttt{SPICE} radiation-hydrodynamical simulations. The simulation explores three supernova feedback prescriptions—\texttt{bursty-sn}, \texttt{smooth-sn}, and the hypernova-based \texttt{hyper-sn}. We examine how these feedback modes affect metal enrichment, neutral gas fraction, size and morphology of the ionized halos traced by \OIII\ emission in the optical and FIR rest frame. We find that \OIII\ emission predominantly arises from gas that is both shock-heated and radiatively ionized. We further characterize the mass–metallicity relation and the correlation between neutral gas fraction and \OIII\ luminosity. \texttt{Bursty-sn} efficiently ionizes gas but enriches galaxies less effectively by $z = 5$. This leads to fewer bright \OIII\ emitters than in the \texttt{smooth-sn} model, with both \texttt{bursty-sn} and \texttt{hyper-sn} showing suppressed luminosity functions. Spatially resolved \OIII\ emission shows that \texttt{smooth-sn} generally produces more compact galaxies and slightly higher $V /\sigma$ values, though with overlap among models. Our results demonstrate that \OIII\ emission is a sensitive tracer of stellar feedback at high redshift and highlight the need for observations reaching fainter luminosities where feedback effects are strongest.
\end{abstract}




\section{Introduction}

Directly observing the first galaxies is a crucial step toward understanding the formation and evolution of cosmic structures. With the advent of telescopes such as ALMA and JWST, it is now possible to detect galaxies at the Epoch of Reionization (EoR), formed less than a billion years after the Big Bang \citep{LeFevre20, Bouwens22, Bouwens23a, Bouwens23b, Naidu22, Harikane23, Robertson23}. These observations have revealed a massive population of UV-bright galaxies, with $M_{UV} > -18$ at $z > 9$, raising fundamental questions about the physical processes regulating star formation and feedback during the EoR  \citep{Finkelstein22, Donnan23a, Donnan23b, Carniani24, Castellano24, Napolitano24, Whitler25}.

Understanding the origin of such high UV luminosities has become a central challenge in the study of early galaxy formation, and several scenarios have been put forward to explain these observations. AGN activity in massive early halos has been suggested as a possible explanation for the high UV luminosities. \citep{Trinca24, Napolitano25}. Alternatively, models with reduced dust attenuation—possibly due to efficient gas and dust expulsion by stellar or AGN feedback—have been proposed to explain the observed UV brightness \citep{Ferrara23, Ferrara24}. Other scenarios suggest stochasticity in the stellar feedback, an increased star formation efficiency \citep[SFE,][]{Dekel23, Harikane23, Ceverino24, Gelli24, Gelli25, Li24, Basu25, Feldmann25}, or, alternatively, a top-heavy initial mass function \citep[IMF,][]{Inayoshi22, Trinca24}. However, the UV luminosity function alone is a degenerate diagnostic, as multiple scenarios can reproduce the observed distributions, making it difficult to identify the dominant physical processes at play.

Because the physical assumptions underlying each scenario affect the gas temperature, density, ionization state, and metallicity, they lead to distinct predictions for metal emission lines in the interstellar medium (ISM) and circumgalactic medium \citep[CGM, ][]{Peroux24}. Metal lines, therefore, provide complementary constraints that are essential to break the degeneracy inherent in UV luminosity functions. In addition, many of the brightest nebular lines are emitted at
rest-frame optical and infrared wavelengths, where dust attenuation
is reduced compared to the UV continuum. As a result,
luminosity functions constructed from metal emission lines provide a powerful alternative probe of early galaxy populations \citep{Matthee23, Meyer24, Wold25}. In this context, characterizing the ISM of galaxies during the EoR remains one of the new frontiers in observational studies of galaxy formation. Deep ALMA observations of forbidden infrared (IR) lines have enabled the detection of ISM emission lines at redshifts $z > 12$ \citep{Carniani24, Schouws24, Zavala24}. These observations have revolutionized the field by revealing early metal enrichment at $z > 5$ through \CII\ and \OIII\ detections \citep{Faisst20, Bouwens22, Witstok22, Witstok25, Casavecchia24, Herrera-Camus25}, and by uncovering high \OIII/\CII\ line ratios indicative of a high ionization parameter, likely driven by young stellar populations with top-heavy IMFs or leakage of ionizing photons \citep{Katz22, Fujimoto24, Nyhagen24, Nakazato25, Schouws25}. In addition, ALMA has revealed the presence of both turbulent and cold rotating disks \citep{Lelli21, Parlanti23, Rowland24}, molecular gas rich galaxies \citep{Dessauges-Zavadsky20, Aravena24, Casavecchia25, Salvestrini25}, as well as rapid dust production already in place at these redshifts \citep{Fudamoto20, Inami22, Li24}.

Complementary to these far-IR diagnostics, the advent of JWST NIRCam and NIRSpec instruments has enabled rest-frame optical spectroscopy up to $z \sim 14$, revealing ubiquitous strong nebular emission lines such as \OIII\ at $\lambda \sim 4690,5008$ \AA, [O\,\textsc{ii}] at $3727, 3729$ \AA\ and hydrogen (H) Balmer lines \citep{Sun22a, Rigby23, Naidu25}. These lines originate from the hot, ionized gas around young, massive stars and are sensitive to the physical conditions in \ion{H}{ii} regions, and more in general to the properties of the ISM, such as density, temperature, and metallicity, as well as to the intensity of the local ionizing radiation. Current observations with JWST reveal young stellar populations, highly ionized ISMs and low metallicities at $z = 5-7$ \citep{Schaerer22, Tacchella22, Taylor22, Brinchmann23, Carnall23, Curti23, Katz23, Rhoads23, Trump23}. Additionally, some studies find flat gas-phase metallicity gradients, consistent with turbulent ISM from stellar feedback and galaxy interactions \citep{Vallini24, Venturi24}. These dynamical processes are also reflected in the observed morphologies of EoR galaxies, which often show signs of clumpy star formation, disturbed kinematics, and ongoing mergers \citep{DeGraaff24, Danhaive25}.

Among the emission lines observable at high redshift, \OIII\ is particularly valuable due to its brightness and dependency on ISM conditions such as electron temperature, ionization state, and metallicity. Detected both in the far-IR with ALMA (88 $\mu$m) and in the optical with JWST (4690, 5008 \AA), it offers an insight of the ionized ISM, and traces regions of intense star formation. Missions like the Spectro-Photometer for the History of the Universe, Epoch of Reionization and Ices Explorer \citep[SPHEREx,][]{Dore14} and Atacama Large Aperture Submillimetre Telescope \citep[AtLAST,][]{Klaassen20} will further expand this effort by enabling large-scale line-intensity mapping surveys of emission lines across cosmic time. However, the interpretation of \OIII\ emission is non-trivial, as it depends on radiative transfer, gas dynamics, and feedback. Hydrodynamic simulations with emission line modeling are thus essential to link observed \OIII\ luminosities to galaxy properties.

High-resolution simulations of the ISM in galactic patches have proven crucial for modeling the relevant physical processes, particularly for studying the impact of stellar feedback \citep{Forbes16, Hu17, Emerick18, Lahen19, Agertz20, Tress20, Andersson24, Fotopoulou24} and non-ionizing far-ultraviolet (FUV) radiation \citep{Rathjen24} on cold gas. These simulations minimize assumptions about ionization processes, and instead aim for a more accurate and direct modeling of ionization and emission processes, down to the resolution limit. Following this approach, zoom-in simulations of individual galaxies provide a detailed look at how these physical processes play out within the context of galaxy formation and evolution. Such simulations offer the opportunity to explore how stellar feedback and ionization mechanisms evolve and interact on galactic scales \citep{Ma18, Ma19, Pallottini19, Vallini21, Nyhagen24, Schimek24b, Schimek24a, Kannan25, Rey25}. However, the number of galaxies included in such zoom-in studies is often small, limiting the ability to conduct comprehensive statistical analyses that relate line emission to galactic properties across a wide range of environments. Cosmological simulations that model line emission from galaxies provide a broader parameter space to explore correlations between the physical properties of galaxies, such as star formation rate (SFR), stellar mass, metallicity, and their emission properties. Examples of radiation hydrodynamics (RHD) simulations specifically modeling emission lines at the epoch of reionization include CoDa \citep{Ocvirk16, Ocvirk20, Lewis22, Lee25}, AURORA \citep{Pawlik17}, COLDSIM \citep{Maio22, Maio23}, THESAN \citep{Kannan22, Kannan22b, Garaldi24}, SPHINX \citep{Rosdahl18, Rosdahl22} and MEGATRON \citep{Katz24, Katz25}. However, their high computational cost often limits the range of initial conditions and physical models explored. This is especially critical when studying \OIII\ emission, where observable properties are often degenerate, since strongly affected by the adopted feedback models, subgrid prescriptions, radiative transfer assumptions, and emission modeling techniques. Having a set of cosmological simulations that produce statistically significant galaxy samples while systematically exploring different parameter configurations, and keeping the other initial conditions and modeling choices fixed, is fundamental to assess the impact of these parameters on observable properties.

In this paper, we present the results from modeling \OIII\ in the SPICE simulations \citep[][from now on refered to as BH24]{Bhagwat24a}. By varying the timing and energy of supernova explosions, these cosmological RHD simulations aim at understanding how different stellar feedback assumptions impact the evolution of galaxies through the epoch of reionization. The strength of SPICE lies in its variety of feedback models with self-consistent radiative transfer and the ability to conduct comparative studies while keeping all other factors constant. In this study we investigate how different stellar feedback modes affect \OIII\ emission, and how this is related to other galactic properties. This approach enables a more robust interpretation of current observations with ALMA and JWST, and helps to refine predictions for future data from SPHEREx and AtLAST.

In Section \ref{sect:Method} we summarize the simulations employed in this study and show the modeling for \OIII\ emission. In Section \ref{sect:Results} we present our results and compare them with observations. We discuss the impact of our findings in Section \ref{sect:Discussion}, and in Section \ref{sect:Conclusions} we outline the main conclusions of this work.

\section{Method} 
\label{sect:Method}

In this section, we summarize the key features of the SPICE simulations used in this work, and describe the method employed to calculate the galactic \OIII\ content and the associated luminosity.

\subsection{\spice\ simulations}
\label{sect: Spice simulations}

For our analysis, we employ the SPICE simulation suite, as described in BH24, which includes a tailored implementation designed to model the formation and evolution of the first galaxies in the Universe. A key strength of these simulations lies in its three distinct runs, each exploring the effects of varying the parameters of the same feedback mechanism on galaxies while keeping all other parameters fixed.

The simulations are implemented using RAMSES-RTZ \citep{Rosdahl13, Rosdahl15}, a radiation-hydrodynamic version of the adaptive mesh refinement code RAMSES \citep{Teyssier02}. The simulated volume is \( V_\mathrm{box} = (10 \, \mathrm{cMpc} \, h^{-1})^3 \) with \( 512^3 \) dark matter particles, each with an average mass of \( 6.38 \times 10^5 \, \rm M_\odot \). The initial conditions are set using the MUSIC-MonofonIC generator, starting from \( z = 30 \). 
The simulations adopt a \(\Lambda\)CDM cosmological framework with density parameters \(\Omega_\Lambda = 0.6901\), \(\Omega_m = 0.3099\), and \(\Omega_b = 0.0489\). The present-day Hubble parameter is assumed to be \( H_0 = 67.74 \, \mathrm{km} \, \mathrm{s}^{-1} \, \mathrm{Mpc}^{-1} \), while the power spectrum has a spectral index of \( n_s = 0.9682 \) and a normalization over an \( 8 \, \mathrm{Mpc} \, h^{-1} \) sphere of \(\sigma_8 = 0.8159\) \citep{Planck16}.

We treat cooling from hydrogen (H), helium (He), and metals separately. The ionization states of H and He in each cell are computed consistently with the local ionizing radiation, and the associated cooling and heating rates. These include contributions from Bremsstrahlung, photoionization, dielectronic recombination, Compton cooling by the cosmic microwave background, and collisional ionization and excitation. For cells with \( T > 10^4 \, \mathrm{K} \), cooling from metals is calculated by interpolating CLOUDY tables \citep{Ferland17}, assuming photoionization equilibrium with the ultraviolet background \citep{HaardtMadau96}. For \( T < 10^4 \, \mathrm{K} \), fine-structure cooling rates from \cite{Rosen95} are applied, with a temperature floor set at approximately \( 300 \, \mathrm{K} \). A single metallicity value is assumed throughout the simulation, and individual elemental abundances are not tracked. Furthermore, in \texttt{SPICE} there is an initial gas metallicity floor of $3.2 \times 10^{-4}$ Z$_{\odot}$, which is used to mimic metal enrichment from Pop-III stars and compensate for the missing molecular hydrogen cooling \citep{Wise12}.

Following \cite{Nickerson18}, dust is implemented by assuming a local dust number density given by \( n_d \equiv \left( \frac{Z}{\rm Z_\odot} \right) f_d n_{\mathrm{H}} \), where \( Z \) is the local metallicity, \( f_d = 1 - x_{\mathrm{H\,II}} \) is the fraction of neutral hydrogen, and $n_{\mathrm{H}}$ is the hydrogen number density. This ensures that the presence of dust depends not only on the metal enrichment of the ISM, but also on the local ionization state of the gas, effectively preventing dust formation in highly ionized regions.

\begin{figure*}
\begin{center}
  \begin{tabular}{c}
    \includegraphics[width = \textwidth]{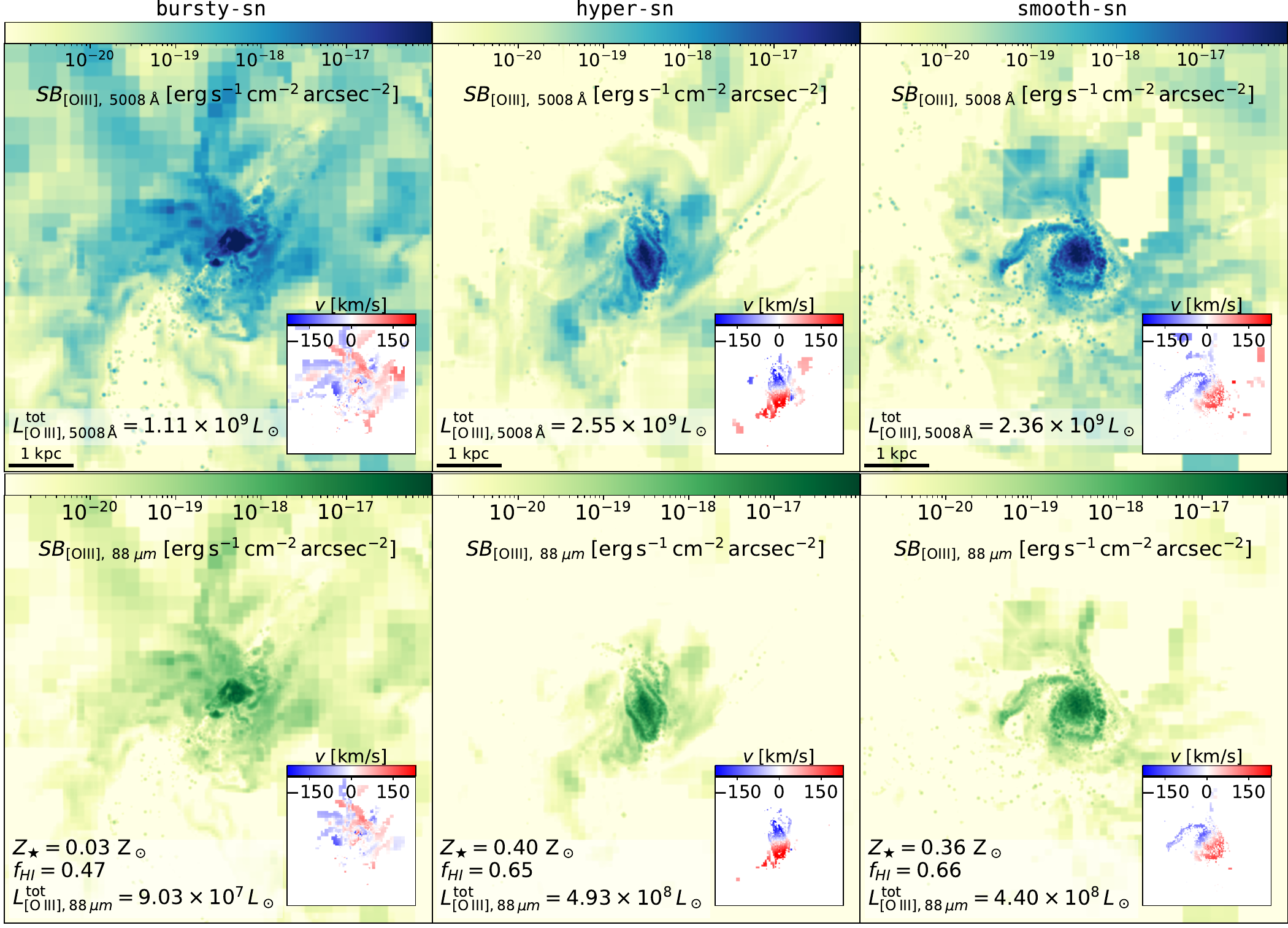}
  \end{tabular}
  \caption{Maps of surface brightness in \OIII\ integrated along the line of sight at 5008 \AA\ (top row) and 88 $\mu$m (bottom row) for the most massive galaxy at $z=7$. The first column shows the map for the \texttt{bursty-sn} model, the second for \texttt{hyper-sn}, and the third for \texttt{smooth-sn}. In each panel, the insets in the bottom-right corner display the velocity field maps derived from the luminosities at 5008 \AA\ (top) and 88 $\mu$m (bottom). Additional labels indicate the total \OIII\ luminosities at 5008 \AA\ (top) and 88 $\mu$m (bottom), the stellar metallicity, and the neutral–gas fraction.}
  \label{fig:OIII_grid}
\end{center}
\end{figure*}

Star formation in the simulation is implemented using a sub-grid multi-freefall model, which calculates a non-local star formation efficiency based on the turbulent state of the gas (see \cite{Kretschmer20}, BH24 for more details). A Chabrier initial mass function (IMF) is assumed, resulting in approximately 30\% of the initial stellar mass being recycled into the interstellar medium ISM, with 5\% of the ejected mass in the form of metals. Once a stellar particle forms, mechanical supernova feedback is implemented by injecting radial momentum into the surrounding gas cells, corresponding to the terminal momentum of the snowplow phase of the supernova remnant. Stellar particles reach a mass resolution of \( 970 \, \rm M_\odot \).

The \texttt{SPICE} simulations adopt three distinct stellar feedback models, while all other initial conditions and parameters remain unchanged during the full simulation, enabling a robust comparative study. The main characteristics of the three feedback models are as follows:  

\begin{itemize}  
    \item \texttt{bursty-sn}: All supernovae (SN) associated with a stellar particle explode simultaneously 10 Myr after the particle's formation, with an energy of \( 2 \times 10^{51} \, \mathrm{ergs} \).  

    \item \texttt{smooth-sn}: The timing of SN explosions depends on the progenitor star’s mass. In this model, SN events occur over a time range of 3–10 Myr after the stellar particle's formation, each releasing a constant energy of \( 2 \times 10^{51} \, \mathrm{ergs} \).  

    \item \texttt{hyper-sn}: The timing of SN explosions is the same as in the \texttt{smooth-sn} model, but the explosion energy varies between \( 10^{50} \) and \( 2 \times 10^{51} \, \mathrm{ergs} \). Additionally, a fraction of these explosions occurs as hypernovae (HN), releasing \( 10^{52} \, \mathrm{ergs} \). In the simulation the fraction of HN is:
    \begin{equation}
         f_{\rm HN} = {\rm max} \left [ 0.5 \; {\rm exp} \left ( \frac{-Z_{\star}}{0.001} \right ), 0.01 \right ]
    \end{equation}
    Where $Z_{\star}$ is the metallicity of the stellar particle and $f_{\rm HN}$ the fraction of HN explosions.
    This type of supernova is thought to occur under low-metallicity conditions, making the fraction of hypernovae generally negligible at lower redshift.
\end{itemize}  

Of particular importance for modeling \OIII\ emission, is the ability of RAMSES-RT to simultaneously solve hydrodynamic equations and radiative transport. For our study, the radiation produced by a stellar population is divided into five photon groups: three in the UV, one in the optical, and one in the infrared. The division into three UV groups is determined by the ionization potentials of \(\mathrm{H\,I}\), \(\mathrm{He\,I}\), and \(\mathrm{He\,II}\). For each stellar particle, luminosities dependent on mass, age, and metallicity are extracted self-consistently from spectral energy distribution (SED) models provided by BPASSv2.2.1 \citep{Eldridge17, Stanway18}. Photons from each group are injected into the host cell of the stellar particle. All photons contribute radiation pressure from photoionization and dust on the gas cell, while only UV photons can additionally photoionize and photoheat the gas. Each radiation group is assigned specific dust absorption and scattering opacities \citep[see][]{Rosdahl13}.


\subsection{Emission of \OIII\ modeling}

In this section we describe the method adopted to compute the \OIII\ emission in the \texttt{SPICE} galaxies, where each galaxy is defined following the same procedure as in BH24, using the Adapta\texttt{HOP} halo finder \citep{Aubert04}. To determine the total luminosity of a galaxy, we sum the contributions from all individual gas cells, $L_{[\rm OIII]}^{tot} = \sum_{n_{\rm cell}=0}^{N_{\rm tot}}L_{[\rm OIII]}^{cell}$. For the sake of simplicity, we omit the apex when we refer to the single cells. The luminosity of each cell depends on its physical properties, namely: electron number density ($n_e$), temperature and number density of \ion{O}{iii}. While the first two quantities are directly provided as outputs of the simulation, the \ion{O}{iii} number density is derived through an ionization balance calculation. 
For this, the first step is to derive the oxygen abundance for each cell, which we assume to be proportional to the total metallicity:
\begin{equation}
n_{\mathrm{O}} = 3.31 \times 10^{-4} Z n_{{\rm gas}},
\label{eq: n_O}
\end{equation}
where $n_{\mathrm{O}}$ and $n_{{\rm gas}}$ are the number densities of oxygen and total gas, and the coefficient $3.5 \times 10^{-4}$ accounts for the relative abundance of oxygen with respect to the total metallicity \citep{Grevesse98}. In \texttt{SPICE} the metallicity of each galaxy, $Z_{\rm gas}$, is calculated as the sum of the mass of the elements heavier than He for each gas particle divided by the sum of the gas particles mass. The mass of metals in gas particles is a fraction of the total mass returned from stellar particles to the gas according to the IMF reported in \cite{Chabrier03}. In the same way, we calculate [O/H] by adding up the contribution of each gas particle contained in the galaxies. However, in \texttt{SPICE} a single metallicity value is saved for each particle instead of the contribution from each atomic species. For this reason, we calculate the mass of oxygen by adopting a fixed value for the relative abundance of oxygen with respect to all the metals, reported in eq. \ref{eq: n_O}, and subsequently normalising by the solar abundances. Since observations of \OIII\ emission lines are easier to conduct, and often total metallicity estimates require a number of assumptions for the relative abundances, this second method for estimating gas metallicities is closer to the estimates obtained with ALMA and JWST. We emphasise that while a mass-weighted total metallicity for each galaxy is a value commonly predicted by simulations, this quantity is not accessible by observations.
We further assume that oxygen exists only in the forms $\mathrm{\ion{O}{i}}, \mathrm{\ion{O}{ii}}, \mathrm{\ion{O}{iii}}$, and $\mathrm{\ion{O}{iv}}$, i.e.
$n_{\mathrm{O}} = n_{\mathrm{\ion{O}{i}}} + n_{\mathrm{\ion{O}{ii}}} + n_{\mathrm{\ion{O}{iii}}} + n_{\mathrm{\ion{O}{iv}}}$.

The number density of each oxygen ion $X$ is determined by solving the ionization equilibrium equation, which balances ionization and recombination processes:
\begin{equation}
n_X (\Gamma_X + \beta_X n_e) = \alpha_{X+1} n_{X+1} n_e,
\label{eq:ion bal}
\end{equation}
where $\Gamma_X$ is the photoionization rate, $\beta_X$ is the collisional ionization rate, and $\alpha_{X+1}$ is the recombination coefficient, which includes both radiative and dielectronic recombination\footnote{We follow the same approach of \cite{Oppenheimer13} and \cite{Katz22RTZ}, and use the collisional recombination rates from \cite{Badnell06}, the collisional ionization rates from \cite{Voronov97} and the cross-sections for the photoionization rates from \cite{Verner96}.}. In our post-processing, the production of O$^{++}$ is dominated by photoionization of gas at $T \gtrsim 10^4$ K, driven by stellar radiation from young stars.

To model \OIII\ emission, we follow the 5-level population approach described in \cite{Draine11}. This modeling approach has been adopted in previous numerical studies \citep[e.g.,][]{Katz22RTZ, Yang24, Yang24b} and has been validated against CLOUDY simulations, showing good agreement.
We compute the level populations of O$^{++}$ by solving the statistical equilibrium equations for its five lowest energy levels. In this framework, the luminosity of a transition between levels $i \rightarrow j$ is given by:

\begin{equation}
L_{ij} = \Delta E_{ij} A_{ij} n_i f_{\ion{H}{ii}} V_{\rm cell},
\end{equation}

where $\Delta E_{ij}$ is the energy of the transition, $A_{ij}$ is the Einstein coefficient for spontaneous decay, $n_i$ is the number density of O$^{++}$ ions in the upper level $i$, $V_{\rm cell}$ is the cell volume, and $f_{\text{\ion{H}{ii}}}$ is the fraction of $V_{\rm cell}$ occupied by ionized gas. The \OIII\ emission primarily originates from compact \ion{H}{ii} regions surrounding young stellar populations. When these regions are not spatially resolved in the simulation, we estimate the fraction of the cell volume occupied by ionized gas as:

\begin{equation}
f_{\text{\ion{H}{ii}}} = \min \left( 1, \frac{V_{\text{\ion{H}{ii}}}}{V_{\text{cell}}} \right),
\end{equation}

where the volume of the ionized region is given by $V_{\text{\ion{H}{ii}}} = \dot{N}_{\mathrm{\ion{H}{i}}}/(\alpha_{B, \mathrm{\ion{H}{ii}}} n_e n_{\mathrm{H}})$, with $\dot{N}_{\mathrm{\ion{H}{i}}}$ the rate of ionizing photons emitted by stellar particles, and $\alpha_{B, \mathrm{\ion{H}{ii}}}$ the case B recombination coefficient of hydrogen \citep{Verner96}. When $V_{\text{\ion{H}{ii}}} \geq V_{\text{cell}}$, we set $f_{\text{\ion{H}{ii}}} = 1$, corresponding to a fully resolved Strömgren region.

Assuming steady-state conditions ($\mathrm{d}n_i/\mathrm{d}t = 0$), the level populations are obtained by solving the system of equations:
\begin{equation}
\frac{dn_i}{dt} = \sum_{j \neq i} n_j R_{ji} - n_i \sum_{j \neq i} R_{ij} = 0,
\label{eq:system}
\end{equation}

for $i = 1,2,3, 4$, where $R_{ij}$ represents the total transition rate from level $i$ to level $j$. These rates include both radiative and collisional processes, and are defined as:
\begin{equation}
R_{ij} =
\begin{cases}
A_{ij} + n_e k_{ij}(T), & \text{if } i > j \\
n_e k_{ij}(T), & \text{if } i < j,
\end{cases}
\end{equation}

where $k_{ij}(T)$ are the collisional (de-)excitation coefficients, which depend on the gas temperature, and $n_e$ is the electron number density. Together with the closure condition $\sum_i n_i = n_{\ion{O}{iii}}$, the system of equation \ref{eq:system} yields the relative level populations $n_i/n_{\ion{O}{iii}}$, from which the emissivity of each transition is computed.
To compare our results with ALMA and JWST observations, we focus on three key \OIII\ emission lines: the 88 $\mu$m transition $^3P_1 \to ^3P_0$, the 4960 \AA\ transition $^1D_2 \to ^3P_1$, and the 5008 \AA\ transition $^1D_2 \to ^3P_2$.
We denote the corresponding line luminosities as $L_{10} \equiv L_{\textup{\OIII},\,88\,\mu\mathrm{m}}$, $L_{31} \equiv L_{\textup{\OIII},\,4960\,\textup{\AA}}$, and $L_{32} \equiv L_{\textup{\OIII},\,5008\,\textup{\AA}}$.

 To account for dust attenuation, we apply a correction to the intrinsic \OIII\ luminosities following a Monte Carlo radiative transfer approach similar to that implemented in \textsc{RASCAS} \citep{MichelDansac20}. From each stellar particle, we cast 100 rays isotropically out to the virial radius of the halo and compute the dust optical depth along each line of sight using the dust model described in BH24. The final \OIII\ luminosity of each galaxy is then obtained by averaging over all viewing angles, yielding a solid-angle–weighted attenuation factor that captures the orientation dependence of the dust distribution. From this point onward, we refer to "dust-corrected" or "attenuated" \OIII\ emission when dust effects are included. In Appendix~\ref{appendixA}, we present the $L_{\rm [OIII]}$–SFR relation as a consistency check, showing that our \OIII\ modeling produces results consistent with both observational constraints and previous numerical studies.

\begin{figure}
\begin{center}
  \begin{tabular}{c}
     \includegraphics[width = 1.\columnwidth]{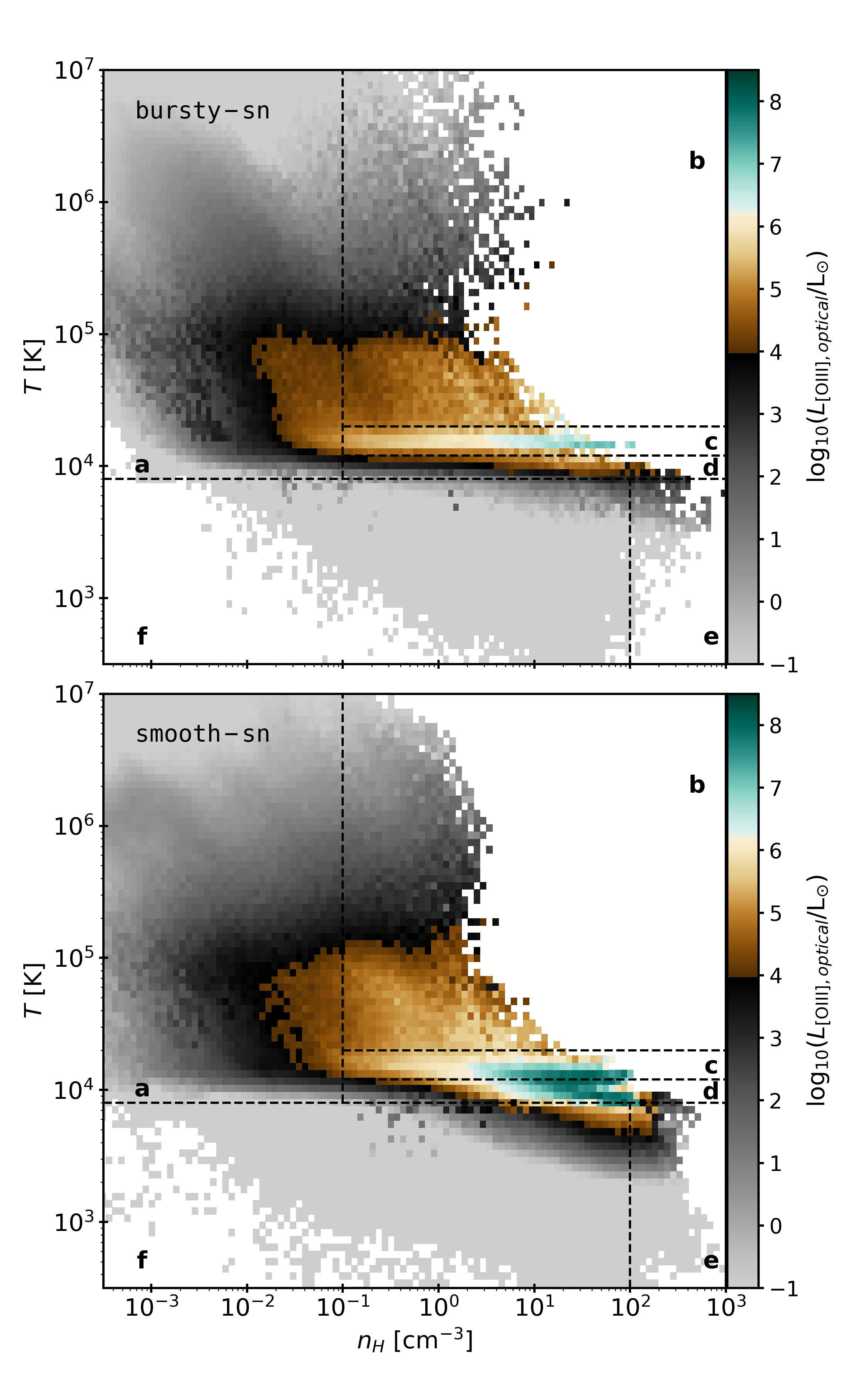}
  \end{tabular}
  \caption{Temperature–density diagram of gas cells in the most massive galaxy at $z = 5$, for the \texttt{bursty-sn} (top panel) and \texttt{smooth-sn} (bottom panel) models. The 2D distribution is color-coded by the sum of intrinsic optical \OIII\ luminosities in each density–temperature bin. The plot reveals a multiphase interstellar medium, structured as follows: diffuse, hot gas produced by stellar feedback (\textbf{a}); shocked gas (\textbf{b}); warm gas ionized by radiative feedback (\textbf{c}); warm gas in thermal equilibrium (\textbf{d}); cold, dense gas near the star formation threshold (\textbf{e}); thermally unstable gas transitioning between warm and cold phases (\textbf{f}). In the \texttt{bursty-sn} model, the bulk of the optical \OIII\ emission is confined to region \textbf{c}, while in the \texttt{smooth-sn} case, \OIII-emitting gas spans a broader temperature range, extending toward cooler phases.
}
  \label{fig:OIII_phasediagram}
\end{center}
\end{figure}

As an illustration, in Figure~\ref{fig:OIII_grid}, we present the intrinsic \OIII\ surface brightness maps for the most massive halo at $z = 7$ across the three simulations. In all panels the \OIII\ emission is inhomogeneous, tracing a complex multiphase structure within each galaxy, far from the idealised picture of a uniform medium. For the same dark matter halo, the three feedback models lead to visually different \OIII\ intensities and spatial distributions. This primarily reflects diversity in galaxy morphology, neutral gas fraction, and stellar metallicity. Moreover, the kinematics traced by \OIII\ emission also differ among the models: while the \texttt{hyper-sn} and \texttt{smooth-sn} runs produce a rotating disk of ionized gas, in the \texttt{bursty-sn} case for several halos turbulent motions dominate the velocity field. All these aspects will be analyzed for the full simulation volume at multiple redshifts and presented in Section~\ref{sect:Results}.

\section{Results}
\label{sect:Results}

In this section we report the results concerning the mass-metallicity relation and \OIII\ luminosity functions for the three stellar feedback models in SPICE. In addition, we show the dependence of \LOIII\ on other galactic properties such as SFR, metallicity and neutral gas fraction.

\subsection{Origin of \OIII\ emission}
\label{sect:OIII origin}

\begin{figure}
\begin{center}
  \begin{tabular}{c}
     \includegraphics[width = \columnwidth]{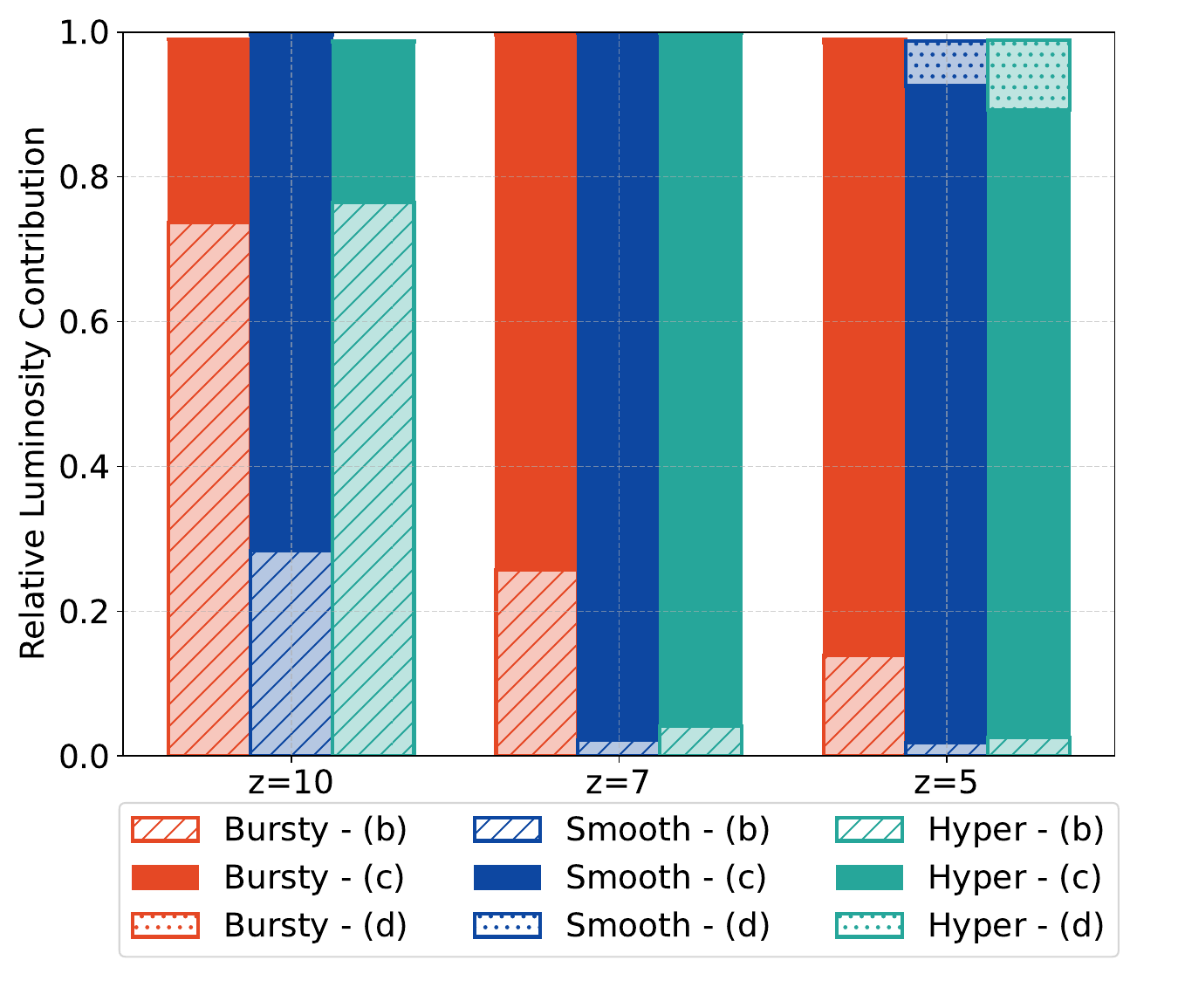}
  \end{tabular}
  \caption{
Contribution from each gas phase to the total intrinsic optical \OIII\ luminosity, at redshifts $z = 10$, 7, and 5. Red, green, and blue bars represent results from the \texttt{bursty-sn}, \texttt{hyper-sn}, and \texttt{smooth-sn} feedback models, respectively. Following the notation introduced in Figure \ref{fig:OIII_phasediagram}, the histogram segments with diagonal stripe patterns correspond to the percentage of $L_{[\mathrm{O\textsc{iii}], optical}}$ originating from phase \textbf{b}, solid-pattern regions represent the contribution from phase \textbf{c}, while dotted-pattern regions indicate gas in thermal equilibrium in zone \textbf{d}. In general, gas impacted by radiative feedback dominates the \OIII\ emission at $z = 7$ and 5, while at $z = 10$ this holds only for the \texttt{smooth-sn} model. In contrast, for the \texttt{bursty-sn} and \texttt{hyper-sn} models, shock-heated gas dominates the emission at that redshift.
}
  \label{fig:OIII_contribution}
\end{center}
\end{figure}

An important aspect of our analysis is understanding how different phases of the gas contribute to the total intrinsic \OIII\ emission. As an illustrative example, we analyze the most massive halo at $z = 5$ in the \texttt{bursty-sn} and \texttt{smooth-sn} models, and present in 
Figure~\ref{fig:OIII_phasediagram} the associated phase diagram, with each pixel color-coded by the cumulative \LOIII\ at 5008~\AA\ and 4960~\AA\ from gas cells in the corresponding bin with a cut in the color scale at $\log_{10} L_{[\mathrm{O\textsc{iii}], optical}} = 0$.
This example shows that the optical \OIII\ emission originates from a multiphase medium, and that the distribution of the emitting gas in temperature--density space depends on the underlying stellar feedback model. 

We define gas phases in the temperature--density diagram by using an approach similar to the one reported in \cite{Marinacci19}, and assess their relative contribution to the total $L_{[\mathrm{O\textsc{iii}], optical}}$.
Region \textbf{a} corresponds to diffuse and hot gas ($n_{\mathrm{H}} < 0.1\,\mathrm{cm}^{-3}$ and $T > 8000$ K), typically associated with supernova-driven outflows \citep[see review by][]{McQuinn16}. Region \textbf{b} includes hot and denser gas with $n_{\mathrm{H}} \geq 0.1\,\mathrm{cm}^{-3}$ and $T > 20000$ K, shocked by strong stellar feedback. Region \textbf{c} is characterized by warm and dense gas ($20000\,\mathrm{K} \geq T > 12000\,\mathrm{K}$, $n_{\mathrm{H}} \geq 0.1\,\mathrm{cm}^{-3}$), which is populated by gas directly impacted by stellar radiative feedback. Usually, the bulk of the ISM mass resides in region \textbf{d}, a warm, ionized phase near thermal equilibrium at $T \sim 10^4\,\mathrm{K}$ and high densities ($n_{\mathrm{H}} \geq 0.1\,\mathrm{cm}^{-3}$). Regions \textbf{e} and \textbf{f} contain cold gas ($T \leq 8000\,\mathrm{K}$). However, in our simulation only gas with $n_{\mathrm{H}} \geq 10\,\mathrm{cm}^{-3}$ can form stars.

The optical \OIII\ emission primarily originates from warm ($T \sim 8000$–$20000$ K), ionized gas at high densities ($n_{\mathrm{H}} \geq 1\, \mathrm{cm}^{-3}$), where collisional excitation of O$^{++}$ ions is efficient \citep{Tadhunter89, Binette12}. While the overall structure of the phase diagram is broadly similar between the two simulations, with most of the \OIII-bright gas confined within this temperature and density range, 
the thermodynamical and chemical properties of this gas are strongly shaped by stellar feedback, which regulates its temperature, density, and metallicity. As a result, different feedback models distribute the gas differently across the phase diagram, and alter which regions dominate the total \LOIII\ emission. Among the regions of the phase diagram that contribute significantly to the \OIII\ emission, regions \textbf{c} and \textbf{d} are the ones where the impact of feedback produces the most evident differences between the two simulations. Gas at $T \geq 10^4$ K is primarily ionized by stellar radiative feedback \citep{Rey25} within \HII\ regions, and is a key contributor to the \OIII\ emission budget.

To assess whether the differences observed in the most massive halo are also present in a more statistically significant sample, we extend our analysis to the full \texttt{SPICE} galaxy population. For every galaxy we compute the contribution of each gas phase (as defined by regions \textbf{a}--\textbf{f}) to the total \LOIII\ in the optical. These contributions are then summed across all galaxies to obtain a global view of how different ISM and CGM components contribute to $L_{[\mathrm{O\textsc{iii}], optical}}$ at various cosmic times. The resulting fractional contributions for the optical emission are shown in Figure~\ref{fig:OIII_contribution} at $z=10$, 7, and 5 for phases \textbf{b}, \textbf{c}, and \textbf{d}, which contribute more than 96 $\%$ of the total \LOIII\ across all feedback models and redshifts considered. Where bars do not reach a normalized value of 1, the missing fraction corresponds to the combined contribution of the remaining three gas phases (a, e and f).

\begin{figure*}
\begin{center}
  \begin{tabular}{c}
    \includegraphics[width = \textwidth]{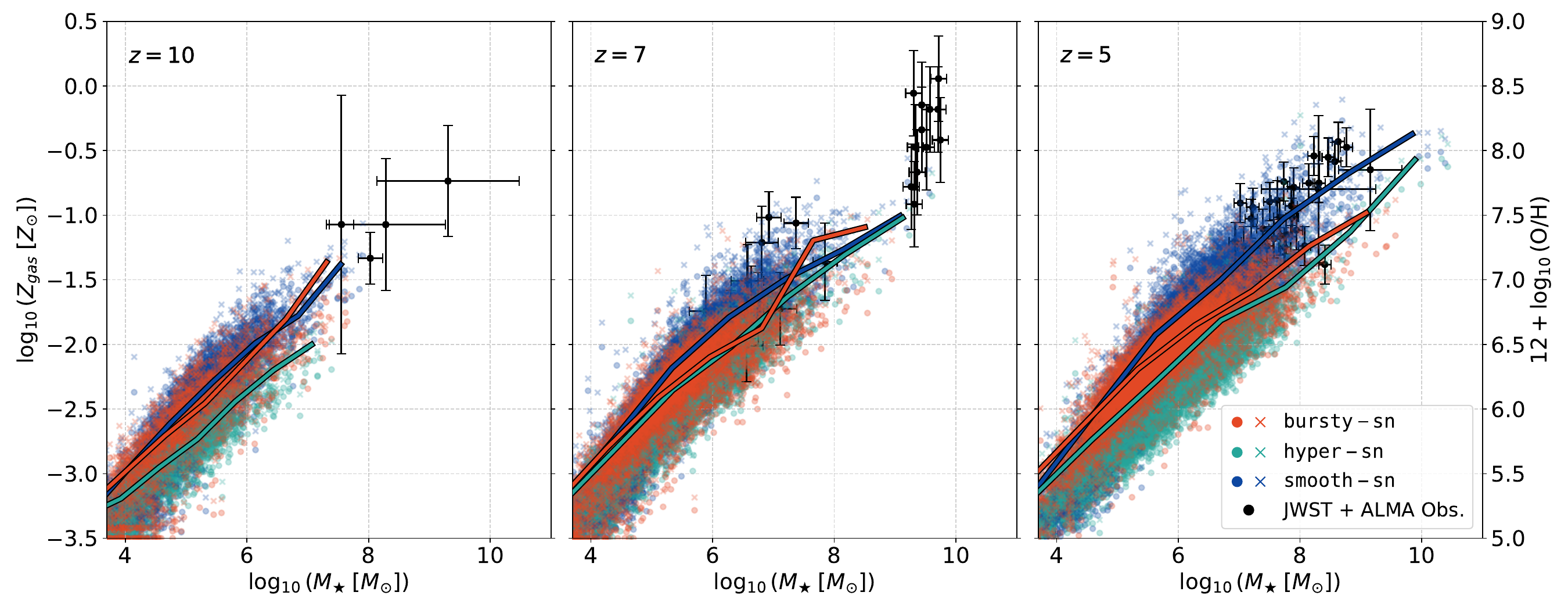}
  \end{tabular}
  \caption{The left y-axis shows the gas metallicities (dots), while the right y-axis indicates metallicities expressed in terms of [O/H] (crosses). Solid lines show the optical \OIII\ luminosity-weighted mean relations in bins of stellar mass. Red, green, and blue points correspond to galaxies in the \texttt{bursty-sn}, \texttt{hyper-sn}, and \texttt{smooth-sn} feedback models, respectively. Observational estimates from ALMA and JWST are shown in black \citep{Curti23, Nakajima23, Chemerynska24, Rowland26}. Across all redshifts and feedback models, \texttt{SPICE} galaxies follow linear mass–metallicity relation, in good agreement with extrapolations of observational trends.}
  \label{fig:MZ relation}
\end{center}
\end{figure*}

We find that the contribution of each gas phase to the total $L_{[\mathrm{O\textsc{iii}], optical}}$ varies significantly with redshift and depends on the adopted feedback model. At $z=10$, the \OIII\ emission in the \texttt{bursty-sn} and \texttt{hyper-sn} simulations is dominated (up to 90$\%$) by phase \textbf{b}, corresponding to shocked gas primarily heated by supernova feedback. This indicates that stellar feedback is particularly efficient at injecting energy into the ISM at early times, leading to widespread shock-heating and strong ionization of the ISM. In contrast, in the \texttt{smooth-sn} model the dominant contribution at the same redshift originates from region \textbf{c}, which traces warm and dense gas photoionized by radiative feedback in \HII\ regions. At lower redshifts, the dominant emitting phase becomes region \textbf{c} across all feedback models. This transition reflects the growth of galaxies into more massive and stable systems that do not get destroyed by stellar feedback, allowing the warm, photoionized gas phase to build up and dominate the \OIII\ emission. By $z = 5$, the relative contribution of region \textbf{c} begins to decline in the \texttt{hyper-sn} and \texttt{smooth-sn} runs, while it remains roughly constant in \texttt{bursty-sn}. In contrast, the contribution from gas in thermal equilibrium at $T \sim 10^4,\mathrm{K}$ —i.e., phase \textbf{d}—becomes increasingly important. This trend is particularly evident in the \texttt{hyper-sn} and \texttt{smooth-sn} models, where the milder stellar feedback allows a larger fraction of the ISM to settle into this warm, equilibrium phase. Although this gas is intrinsically less efficient at producing \OIII\ emission compared to phase \textbf{c}, its growing mass fraction leads it to contribute significantly to the total luminosity. The emergence of this phase-dominated emission suggests that galaxies at this epoch have evolved sufficiently to maintain more stable ISM conditions and deeper potential wells, which help mitigate the disruptive impact of stellar feedback. All the results shown in this section apply also to the 88 $\mu$m \OIII\ transition, as the emission in both cases is primarily regulated by the abundance of O$^{++}$, which resides in warm photoionised gas.

In the following sections, we show how these differences in the origin of \OIII\ emission among the three feedback models impact a range of observable properties, including the luminosity function, line kinematics, and key scaling relations such as the \LOIII-metallicity correlation.

\subsection{Mass metallicity relation}


In Figure \ref{fig:MZ relation} we show the mass–metallicity relation of galaxies at redshifts 10, 7, and 5. We quantify metallicity using two different measures: the gas metallicity, $Z_{\rm gas}$, defined as the total mass of metals in the gas divided by the total gas mass of the galaxy, and [O/H], obtained by summing the oxygen mass of all gas particles in the galaxy and normalized by the mass of hydrogen.

Our results indicate that for all three models there is a strong correlation between stellar mass and metallicity, which is expected from the assembly of stars in galaxies and the associated metal enrichment. This relation extends to increasingly higher values of metallicity and stellar mass towards lower redshifts, as a consequence of galaxy growth through mergers and gas inflows. 
All stellar feedback models show similar trends, with a scatter that spans up to 0.5 dex for the median values. Interestingly, at all redshifts, galaxies in \texttt{smooth-sn} reach higher metallicities, as also discussed in BH24, with $M_{\star}$ ($Z_{\rm gas}$) up to 1 dex (0.5 dex) higher than in \texttt{bursty-sn} for the same dark matter halo mass. The \texttt{hyper-sn} model shows the lowest amplitude at $z = 10$, while at lower redshifts ($z = 7$ and $5$) it follows trends similar to the other models. By $z = 5$, it resembles \texttt{bursty-sn} at $M_{\star} \lesssim 10^9 \mathrm{M}_{\odot}$, while also extending towards the high-mass, high-metallicity regime of \texttt{smooth-sn}. Indeed, the high HN rate combined with the shallower potential wells lead to more effective quenching for less massive halos, while for more massive ones the effect of HN becomes negligible, leading to a faster growth of the stellar mass.

\begin{figure*}
\begin{center}
  \begin{tabular}{c}
     \includegraphics[width = \textwidth]{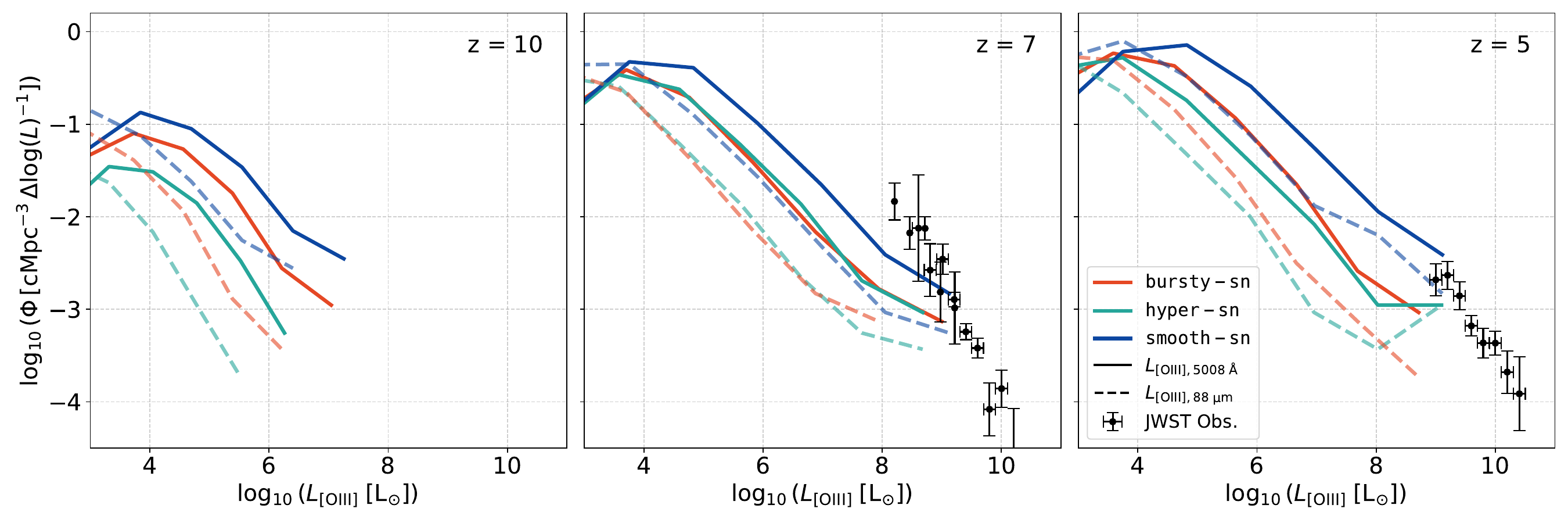}
  \end{tabular}
  \caption{Dust-corrected luminosity functions of \OIII\ at $z = 10$, 7, and 5 for the \texttt{bursty-sn} (red), \texttt{hyper-sn} (green), and \texttt{smooth-sn} (blue) feedback models. Solid and dashed lines show median values for the emission at 5008 \AA, and 88 $\mu$m, respectively. Black points represent JWST observational constraints for the 5008 \AA\ LFs \citep{Matthee23, Meyer24, Wold25}. For both transitions, the \texttt{smooth-sn} model consistently produces a larger number of \OIII-bright galaxies.}
  \label{fig:LF_OIII}
\end{center}
\end{figure*}

The mass-metallicity relation in the three feedback models seems to reproduce well the trend and the scatter observed with JWST and ALMA \citep{Curti23, Nakajima23, Chemerynska24, Rowland26}. However, we emphasise that observations of the relation alone are not sufficient to distinguish which type of feedback dominates in high redshift galaxies, as shown also for MEGATRON galaxies \citep{Choustikov25}. Moreover, observed oxygen-based metallicity estimates are sensitive to the brightest regions of the galaxy, i.e. \HII\ regions, and can overestimate the total metallicity by up to 0.5 dex \citep{Katz23}. Since the effects of stellar feedback are degenerate in the mass-metallicity relation —- both in simulations and observations -— in the next section we shift our focus to the \OIII\ luminosity functions.

\subsection{Luminosity functions}
\label{sect: LF}


In Figure \ref{fig:LF_OIII} we show the redshift evolution of the dust attenuated luminosity functions (LFs) for the emission at 5008 \AA\ and 88 $\mu$m. As expected, more galaxies populate the high luminosity end of the plot with decreasing redshift.
We also note that for every model of stellar feedback and at every redshift, the galaxies are 1-2 dex brighter in the optical than in the infrared, and this result is robust against the effects of dust attenuation in the optical. 

The impact of different stellar feedback models on \OIII\ emission varies with redshift. At $z = 10$, the \texttt{hyper-sn} model produces fewer luminous \OIII\ galaxies, as its strong HN feedback suppresses the accumulation of cold gas and delays metal enrichment. In particular, its \OIII\ LF reaches values up to $\sim$1 dex (1.5 dex) lower than those in \texttt{bursty-sn} (\texttt{smooth-sn}). As redshift decreases, the effect of HN feedback diminishes, and by $z = 7$ the \OIII\ luminosity function of the \texttt{hyper-sn} model closely resembles that of the \texttt{bursty-sn} model — consistent with the convergence in UV LFs reported by BH24 and \cite{Basu25}. In contrast, the \texttt{smooth-sn} model consistently hosts the brightest \OIII\ emitters across all redshifts, reaching luminosities up to $\sim$1 dex higher than the \texttt{bursty-sn} case by $z = 5$, owing to its efficiency in forming more massive galaxies with higher metallicities (see Figure~\ref{fig:MZ relation}). The differences in the amplitudes of the \OIII\ LFs between the three models, even after accounting for dust attenuation, make them valuable diagnostics of stellar feedback, in contrast to UV LFs which become indistinguishable between feedback models once dust absorption is included \citep[BH24;][]{Basu25}.

Despite differences in metal enrichment, by $z = 7$ and 5 all three feedback models produce similar maximum \LOIII\ values at 5008 \AA . These results highlight that \LOIII\ is not governed by metallicity alone, but also depends on the fraction of ionized gas. As such, numerical simulations are essential to disentangle the complex interplay between metallicity, gas temperature, density, and ionization state in shaping the observed \OIII\ luminosity functions. In the following sections, we explore in detail how factors like the neutral gas fraction and metal content influence \LOIII\ under different stellar feedback scenarios.

From a comparison of our results with JWST observations of the emission at $5008$ \AA\ \citep{Matthee23, Meyer24, Wold25}, we note that at $z = 7$ and 5 all models are consistent with the data. The \texttt{smooth-sn} model, however, seems to be slightly above the observational data at $z = 5$, indicating that at lower redshift galaxies with a smooth stellar feedback are affected by overcooling and therefore by an overproduction of metals. The fact that all three models are consistent with the data makes it particularly difficult to distinguish between them, especially in the high-luminosity bins. More sensitive and statistically robust observations from ALMA and JWST are needed — ideally extending to luminosities below $10^8$ L$_{\odot}$ and avoiding biases toward overdense regions — in order to constrain which stellar feedback mechanisms dominate at these redshifts.

\subsection{Dependence on neutral gas fraction}
\label{sect:Neutral gas fraction}

\begin{figure*}
\begin{center}
  \begin{tabular}{c}
    \includegraphics[width = 1\textwidth]{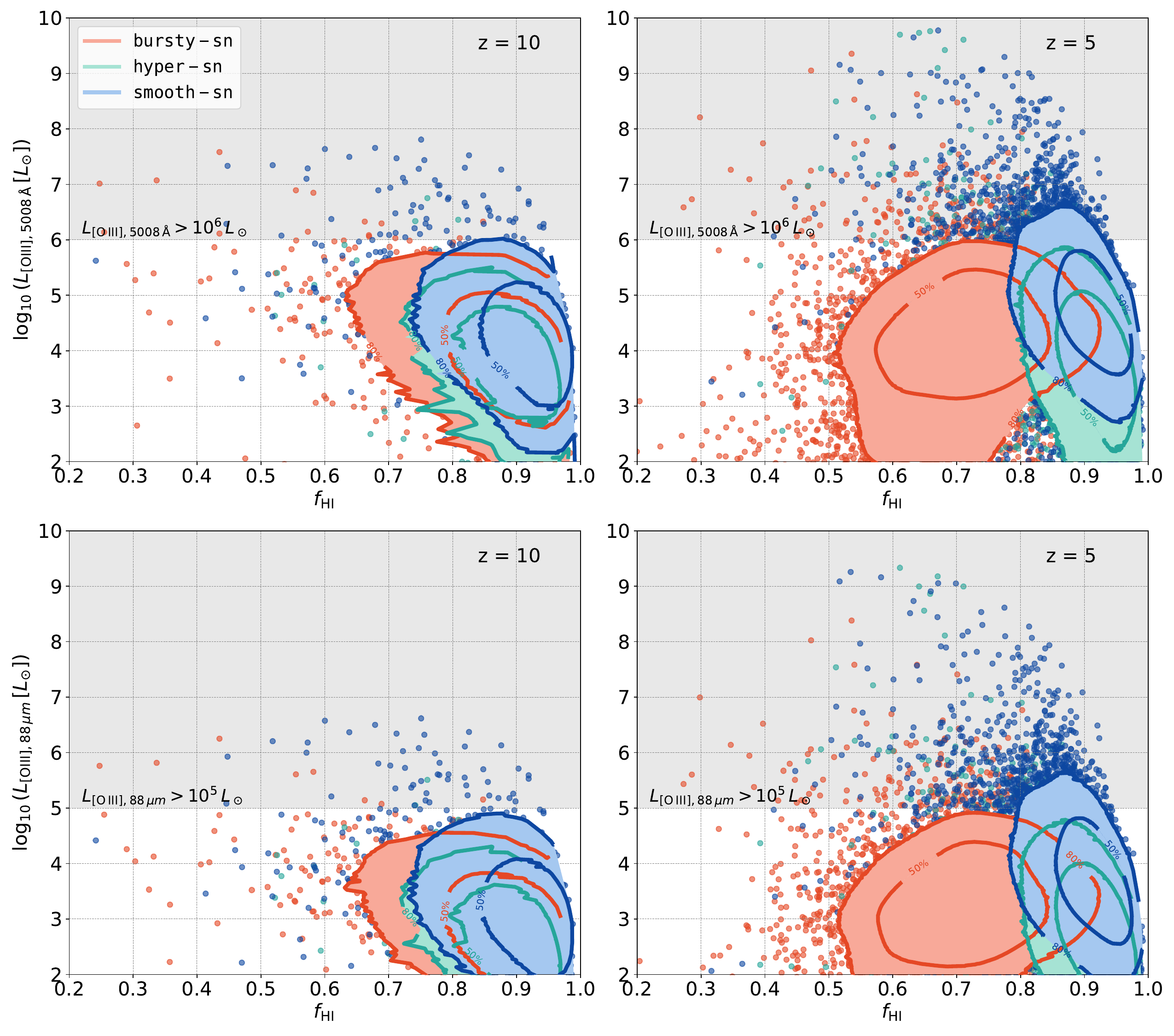}
  \end{tabular}
  \caption{Dust-corrected \LOIII\ at 5008 \AA\ (top) and 88 $\mu$m (bottom) as a function of the neutral gas fraction within the virial radius for galaxies at $z = 10$ (left) and $z = 5$ (right) in the \texttt{bursty-sn} (red), \texttt{hyper-sn} (green), and \texttt{smooth-sn} (blue) models. Contours enclose $50\%$ and $80\%$ of the data, while points indicate individual galaxies outside these contours. The gray-shaded areas mark the region where galaxies have $L_{\mathrm{[O \, III]},5008\text{\AA}} > 10^6 {\rm L}_\odot$ (top) and $L_{\mathrm{[O \, III]},88,\mu\mathrm{m}} > 10^4 {\rm L}_\odot$ (bottom). While models overlap at $z = 10$, at $z = 5$ \texttt{bursty-sn} yields lower neutral gas fractions. Despite having a higher neutral content, the \texttt{smooth-sn} model still hosts the most \OIII-luminous galaxies at both wavelengths.}
  \label{fig:neutral_gas}
\end{center}
\end{figure*}

In this section we study how the \LOIII\ of \texttt{SPICE} galaxies depends on their neutral gas content, where the fraction of neutral hydrogen is computed as the ratio between the volume occupied by \ion{H}{i} and the total gas volume, $f_{\mathrm{HI}} = V_{\mathrm{HI}}/V_{\rm gas}$. In Figure \ref{fig:neutral_gas} we show how dust-corrected \LOIII\ at 5008 \AA\ and 88 $\mu$m depends on $f_{\mathrm{HI}}$.

At $z = 10$, most \OIII\ emitters have a luminosity below $10^6$ L$_{\odot}$ at 5008 \AA, while below $10^4$ L$_{\odot}$ for 88 $\mu$m, and more than $80\%$ of their gas content is in neutral form. At higher luminosities (grey region of the plot), instead, while galaxies in the \texttt{smooth-sn} and \texttt{hyper-sn} model are still dominated by neutral gas, in the \texttt{bursty-sn} model they have a predominance of ionized gas (consistent with the results in Section \ref{sect:OIII origin}). The interplay between the different physical mechanisms responsible for \OIII\ emission and the way stellar feedback affects them result in larger differences at $z = 5$. In general, the \texttt{bursty-sn} model leads to higher ionized gas fractions, as a result of stronger and more frequent episodes of star formation that efficiently heat and ionize the gas. In contrast, the \texttt{smooth-sn} and \texttt{hyper-sn} models tend to preserve more neutral gas, with the majority of galaxies exhibiting neutral fractions above 0.75.
Despite the high fraction of neutral gas, as we will show in Section \ref{sect:LvsZ}, in the \texttt{smooth-sn} and \texttt{hyper-sn} models, the brightest \OIII\ emitters tend to have higher metallicities by $z = 5$, which boost line emission efficiency. In contrast, in the \texttt{bursty-sn} model the elevated \LOIII\ values are primarily driven by a higher fraction of ionized gas heated above $10^4$ K, despite lower metallicities. It should be noted, however, that excessively strong feedback — as in the \texttt{hyper-sn} case at $z=10$ — can suppress \OIII\ luminosity.

These results suggest that different physical conditions can produce similar \OIII\ luminosities. In Section~\ref{sect:LvsZ}, we explore in more detail how \LOIII\ depends on the gas metallicity, in order to test this interpretation.

\subsection{Dependence on metallicity}
\label{sect:LvsZ}

\begin{figure}
    \centering
    \includegraphics[width=\linewidth]{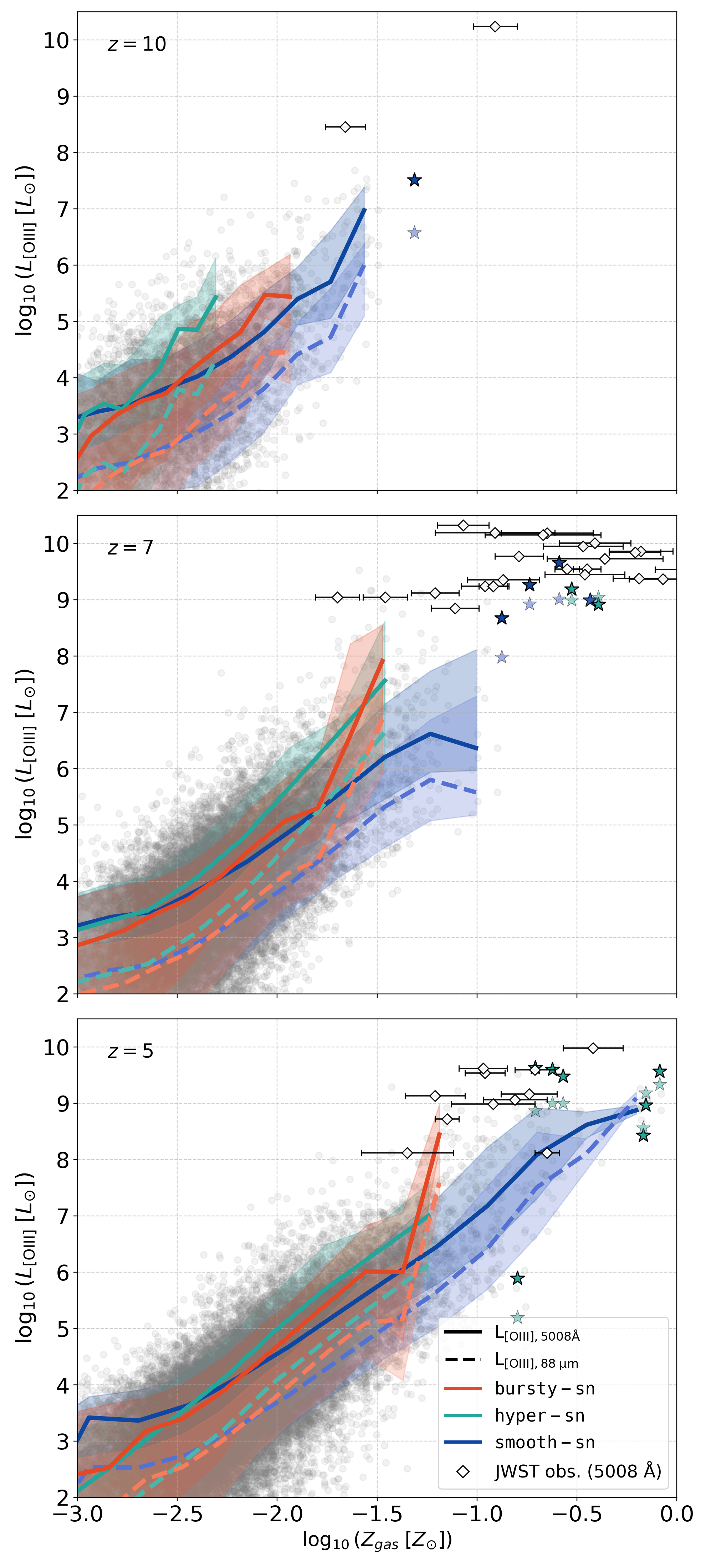}
    \caption{Dust-corrected \LOIII\ as a function of gas metallicity at $z = 10$ (top), $z = 7$ (middle) and $z = 5$ (bottom) for the \texttt{bursty-sn} (red), \texttt{hyper-sn} (green), and \texttt{smooth-sn} (blue) models. Lines indicate the median luminosities for the 5008 \AA\ (solid) and the 88 $\mu$m (dashed) transition, while shaded regions represent the 16th–84th percentile range around the medians. Stars indicate single galaxies in metallicity bins where there are less than 4 objects. Gray points represent the full \texttt{SPICE} galaxy sample for the three models. All feedback models exhibit a clear positive correlation. We compare our results with galaxies observed with JWST at $z > 4$ \citep[white diamonds, ][]{Curti23, Laseter24, Sanders24, Rowland26}. At $z = 5$, the \texttt{smooth-sn} and \texttt{hyper-sn} models reach both the highest luminosities and metallicities.}
    \label{fig:LvsZ}
\end{figure}

As shown in Figure~\ref{fig:LvsZ}, the luminosity–metallicity relation is approximately linear. By $z = 5$, galaxies across all three models extend to higher metallicities relative to $z = 10$, reflecting the progressive enrichment over cosmic time. At both $z = 10$ and $z = 5$, the luminosity–metallicity relation shows a broadly similar behavior: the models overlap at lower metallicities, while at higher metallicities the slope steepens. The difference among feedback scenarios lies in where this steepening occurs. Since the \texttt{hyper-sn} and \texttt{bursty-sn} models tend to reach metallicities lower than \texttt{smooth-sn}, the break in slope appears at smaller $Z_{\rm gas}$ values for them, while in the \texttt{smooth-sn} case it is shifted to higher $Z_{\rm gas}$. Interestingly, although the \texttt{hyper-sn} model initially produces galaxies with low metallicities and luminosities, close to \texttt{bursty-sn}, it evolves rapidly and eventually reaches metallicities and \LOIII\ comparable to those of the \texttt{smooth-sn} model of one order of magnitude higher than \texttt{bursty-sn}. This occurs because strong negative feedback from hypernovae at early times suppresses star formation by disrupting cold gas, and delaying metal enrichment. At lower redshifts, however, the impact of hypernovae weakens, as we have assumed that the probability of their occurrence decreases with increasing gas metallicity.

Both at $z = 10$ and $5$, we find that at fixed metallicity the optical \LOIII\ is systematically brighter than its infrared counterpart by approximately 2~dex. While this result has already been discussed in previous sections, we highlight here that the slope of the luminosity--metallicity relation is remarkably similar at both wavelengths. A deviation from the general trend is observed at $z = 5$ and for metallicities above $\log(Z/Z_\odot) > -0.5$, where the slope of the optical luminosity flattens for the \texttt{smooth-sn} feedback model because of the increased dust attenuation at higher metallicities, which has a stronger impact on the emission in the optical band than at the infrared wavelengths. 

The combined effect of metallicity evolution and neutral gas content—outlined in Section~\ref{sect:Neutral gas fraction}—can account for the differences in the Luminosity functions observed across the three feedback models and displayed in Figure~\ref{fig:LF_OIII}, with the \texttt{smooth-sn} model emerging as the most efficient in producing \OIII-bright galaxies. This model consistently forms more metal-rich systems even at early epochs (as early as $z = 10$), resulting in higher \OIII\ luminosities at both optical and infrared wavelengths. This milder feedback leads to a larger number of galaxies at $z = 5$ with \LOIII\ exceeding $10^8~{\rm L_{\odot}}$ (at 5008~\AA) and $10^6~{\rm L_{\odot}}$ (at 88~$\mu$m), compared to the other models. Although the \texttt{hyper-sn} model is eventually capable of forming similarly metal-rich and luminous galaxies, its rapid redshift evolution does not leave sufficient time for a large population of such systems to build up, resulting in LFs more similar to those of the \texttt{bursty-sn} model.

\subsection{\OIII\ halos size}
\label{sect:OIIIsize}

\begin{figure}
    \centering
    \includegraphics[width=\linewidth]{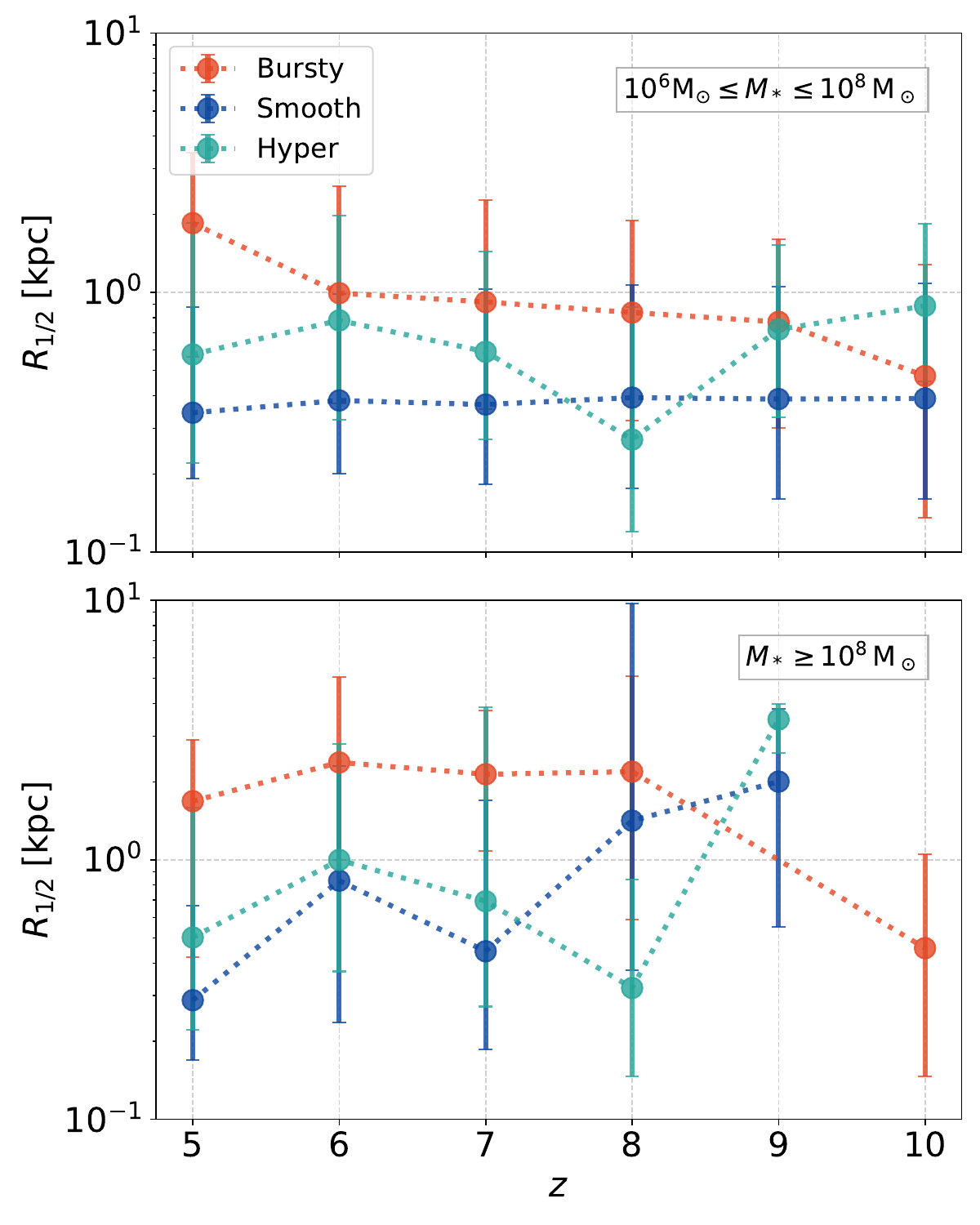}
    \caption{Redshift evolution of the $L_{\mathrm{[O\,III]},5008\,\text{\AA}}$ half-light radius, for the \texttt{bursty-sn} (red), \texttt{hyper-sn} (green) and \texttt{smooth-sn} (blue) stellar feedback models. Each point corresponds to the luminosity-weighted median value, and the error bars represent the 16th and 84th percentiles of the distribution. The top and bottom panels display the results for galaxies with $10^6\,\mathrm{M_\odot} \leq M_{\star} \leq 10^8\,\mathrm{M_\odot}$ and $M_{\star} > 10^8\,\mathrm{M_\odot}$, respectively. In both panels \texttt{bursty-sn} produces the most extended \OIII\ halos; although all models are consistent within the error bars.}
    \label{fig:Rvsz}
\end{figure}

In this section, we complement the analysis of the integrated \OIII\ luminosity presented in the previous sections by investigating spatially resolved properties of galaxies. In particular, combining the overall emission with the morphology and extent of the \OIII\ distribution can provide further constraints on the role of stellar feedback in shaping galaxy structure. Here, we analyze the redshift evolution of the median values for the \OIII\ half-light radius, $R_{1/2}$, defined as the projected radius enclosing 50$\%$ of the total \OIII\ emission. As the behavior of the 5008 \AA\,, 4960 \AA\ and the 88 $\mu$m lines is similar, in Figure \ref{fig:Rvsz} we show results only for the former and for two galaxy samples: one with $M_{\star} \leq 10^8 \mathrm{M}_\odot$, which includes the majority of halos in our simulations, and another with $M_{\star} > 10^8 \mathrm{M}_\odot$, which represents the systems we expect to be more easily detectable.

\texttt{Bursty-sn} galaxies with $10^6\,\mathrm{M}_\odot \leq M_\star \leq 10^8\,\mathrm{M}_\odot$ show a general increase in $R_{1/2}$ from $z = 10$ to $5$, which reflects the gradual growth of galaxy mass over cosmic time, naturally leading to larger emission regions \citep[see e.g.,][]{Buitrago08, vanDokkum10, Behroozi19, Costantin23, Wilkins23, Jia24, Ormerod24}. Across all redshifts, the three feedback models predict sizes with differences remaining within the error bars, although the \texttt{smooth-sn} model systematically tends to produce smaller half-light radii compared to the other two. This trend arises because a smoother stellar feedback does not drive outflows as strong as the other two models, allowing the gas to remain more tightly bound to the galaxy, which therefore appears more compact. For more massive galaxies there are no clear trends. Indeed, while these systems are not yet in place at $z = 10$, they begin to emerge at $z < 9$, when the median half-light radius exhibits a more stochastic evolution across redshift. Nonetheless, across all redshifts, \texttt{bursty-sn} tends to produce the most extended \OIII\ halos, while \texttt{smooth-sn} and \texttt{hyper-sn} generally yields the most compact ones, although the large scatter often makes these differences difficult to distinguish.

\begin{figure}
    \centering
    \includegraphics[width=0.95\linewidth]{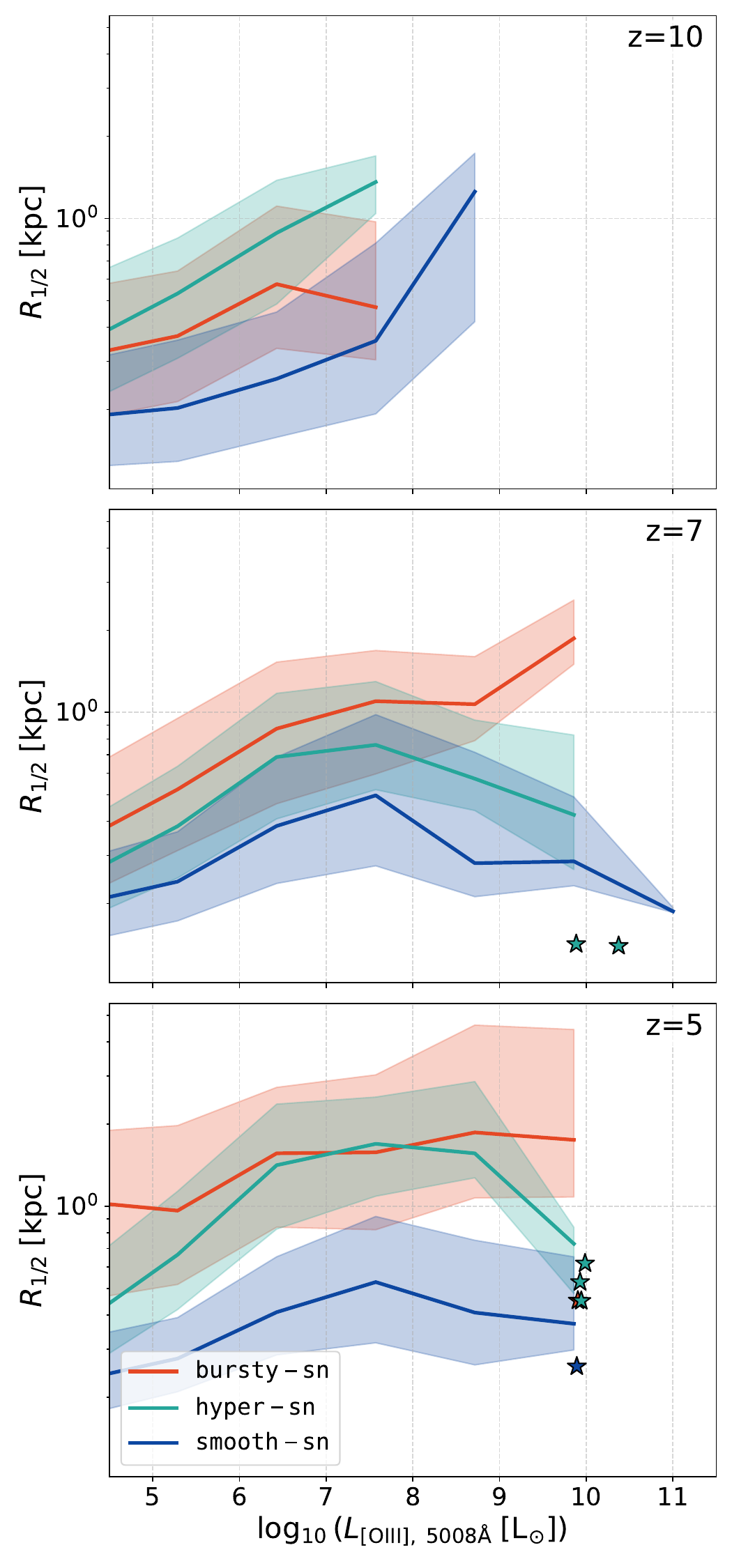}
    \caption{Half light radius $R_{1/2}$ as a function of \OIII\ luminosity at 5008 \AA. The solid lines represent the median values for \texttt{bursty-sn} (red), \texttt{hyper-sn} (green) and \texttt{smooth-sn} (blue), the shaded areas are the 16th and 84th percentiles, while the stars refers to individual galaxies in bins with fewer than four objects. $R_{1/2}$ increases with \OIII\ luminosity in \texttt{bursty-sn}, decreases in \texttt{smooth-sn}, and shows no clear trend in \texttt{hyper-sn}, which exhibits the most fluctuations.}
    \label{fig:RvsL32}
\end{figure}

To further explore this behavior, in Figure~\ref{fig:RvsL32} we present the half-light radius as a function of \OIII\ 5008\,\AA\ luminosity at three different redshifts. This plot illustrates how, at a fixed luminosity, the extent of the \OIII\ emission varies depending on the feedback implementation. At $z = 10$, all feedback models show an increasing trend of $R_{1/2}$ with \LOIII, with more luminous galaxies hosting more extended halos. At $z = 7$ and $5$, however, the \texttt{smooth-sn} feedback model shows a constant or even decreasing trend. 

This is a direct consequence of the more efficient gas cooling, which allows galaxies to grow more massive and develop deeper gravitational potential wells. Indeed, these galaxies become increasingly compact as their mass grows and stellar feedback becomes less effective at unbinding gas, which is quickly recycled into the galaxy. In contrast, the \texttt{bursty-sn} displays a consistently increasing $R_{1/2}$ with \LOIII\ at all redshifts, indicating an efficient heating and redistribution of the gas surrounding the galaxy. The \texttt{hyper-sn} model, instead, shows a no clear trend at $z = 7$ and 5, but with bigger fluctuations. This is because as the galaxies are dynamically disturbed, individual starburst events can significantly alter both their morphology and their \OIII\ spatial extent. Interestingly, at lower redshifts \texttt{hyper-sn} is able to produce a few luminous galaxies with \OIII\ luminosities comparable to the brightest systems in \texttt{smooth-sn}. Indeed, as the simulation evolves to lower redshift, the impact of hypernova feedback diminishes, causing the \texttt{hyper-sn} model to converge toward the behavior seen in \texttt{smooth-sn}.

\subsection{\OIII\ morphologies}

\begin{figure*}
    \centering
    \includegraphics[width=\linewidth]{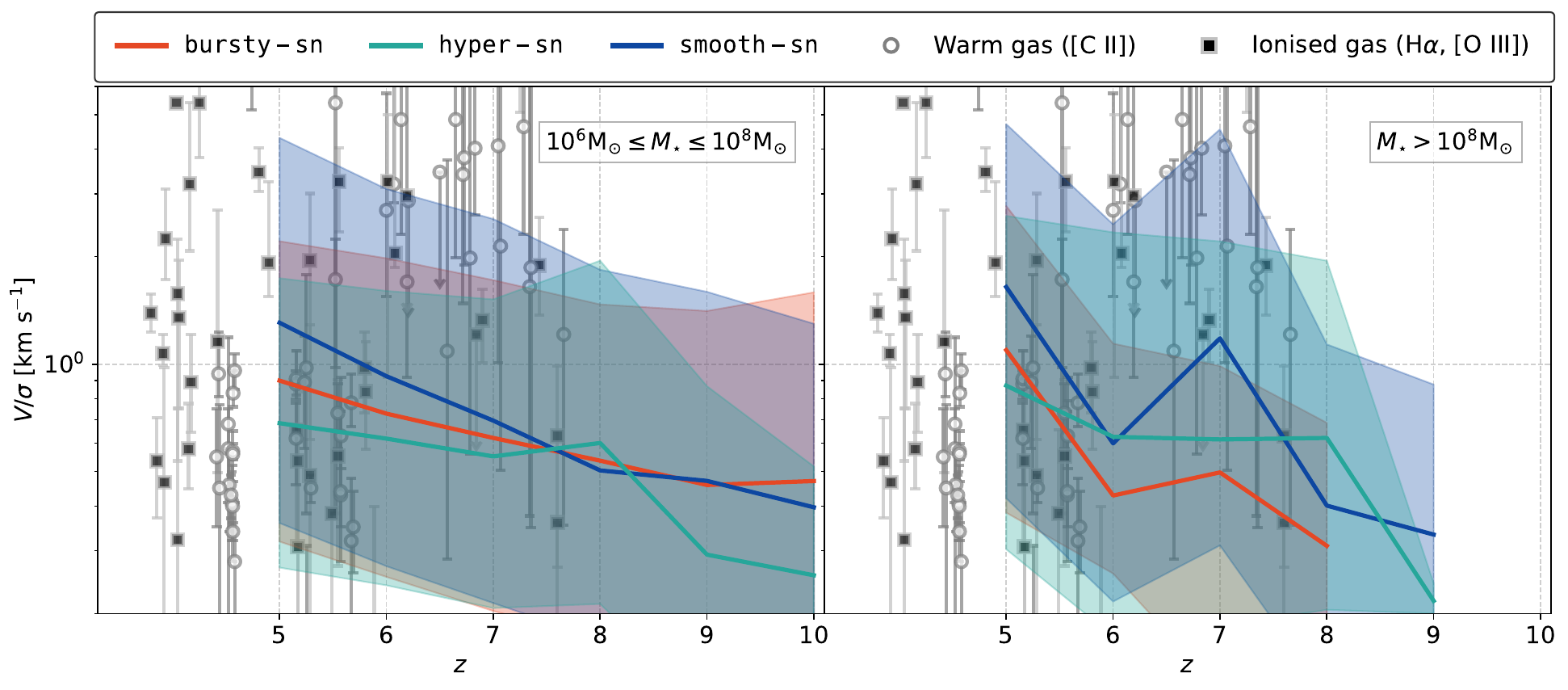}
    \caption{Redshift evolution of luminosity weighted $V/\sigma$ for galaxies with $10^6\,\mathrm{M_\odot} \leq M_{\star} \leq 10^8\,\mathrm{M_\odot}$ (top panel) and  $M_{\star} > 10^8\,\mathrm{M_\odot}$ (bottom). Solid lines displays the median values for the \texttt{bursty-sn} (red), \texttt{hyper-sn} (green) and \texttt{smooth-sn} (blue) supernova feedback models, while the shaded areas indicate the 16th and 84th percentiles. Grey points represent a compilation of ALMA and JWST observations based on ionized gas traced by H$\alpha$ and \OIII\ (filled points), and on warm gas detected in [\ion{C}{ii}] (empty points). The predicted redshift evolution of $V/\sigma$ is consistent with current observational constraints, and the differences among the three feedback models are within the scatter in both mass ranges, preventing a clear distinction between them.}
    \label{fig:vbysigma}
\end{figure*}

In this section, we investigate the time evolution of the morphologies of \texttt{SPICE} galaxies as traced by \OIII\ emission. We focus on the $V/\sigma$ ratio, a widely adopted diagnostic of galaxy morphology, where $V$ is the maximum value of the rotation curve and $\sigma$ the mean of the three-dimensional velocity dispersion, within twice $R_{1/2}$. 

Figure~\ref{fig:vbysigma} displays the redshift evolution of the luminosity-weighted median $V/\sigma$ values in the same mass ranges of Figure~\ref{fig:Rvsz}. For galaxies with $10^6\,\mathrm{M_\odot} \leq M_\star \leq 10^8 \mathrm{M_\odot}$, We find a slightly increasing trend across redshift, with the three feedback models indistinguishable within the scatter, indicating that differences in stellar feedback have little impact on the spatially resolved kinematics traced by \OIII\ in low-mass systems. Nonetheless, the \texttt{smooth-sn} model tends to lie above the others at $z \leq 7$, although still within the uncertainties. This suggests that even in low-mass galaxies, a smoother feedback is less efficient at driving turbulent motions, allowing the cold gas to remain more rotationally supported and leading to slightly higher $V/\sigma$ values. For galaxies with $M_\star > 10^8\mathrm{M_\odot}$, conclusions can only be drawn at lower redshifts, as no sufficiently luminous galaxies are present at $z=10$ due to the limited simulation volume. Here, the scatter around the median values in all models is bigger and there is not a clear redshift trend. Interestingly, despite the enhanced gas turbulence in the \texttt{bursty-sn} model—which would be expected to suppress the formation of rotationally supported structures—the resulting
$V/\sigma$ values remain comparable to those of the other models. This trend aligns well with both observational findings \citep[e.g.,][]{ForsterSchreiber09, Wisnioski15, Stott16, Ubler19, Genzel20} and other theoretical predictions \citep[e.g.,][]{Genel15, Pillepich19}, which suggest an increasing fraction of rotationally supported systems toward lower redshifts. Here, the \texttt{smooth-sn} model exhibits median $V/\sigma$ systematically higher than those found for less massive galaxies, suggesting that the more massive galaxies tend to be more evolved and capable of sustaining disk structures. On the contrary, the more massive galaxies in the \texttt{bursty-sn} and \texttt{hyper-sn} models are too dynamically disturbed to form as many disks as in the other feedback scenarios. Overall, these findings are consistent with the trends and values reported in BH24, who analyzed galaxy morphologies traced by H$\alpha$ emission.

We compare our results with a compilation of ALMA and JWST observations of both warm and ionized gas in early galaxies. Several studies based on ionized gas tracers, such as H$\alpha$ and \OIII\ have highlighted a broad diversity in kinematic properties. For instance, \cite{Parlanti23, Arribas24, Ubler24, Danhaive25} and \cite{Ivey25} find that while dynamically cold disks are present, many high-redshift galaxies exhibit significant turbulence and are predominantly dispersion-dominated. Their results suggest that rotational support becomes increasingly common with cosmic time, but that turbulent kinematics remain widespread at early epochs. Our simulation results align well with this emerging picture, as we find evidence for early disk formation, alongside a wide range of kinematic states across the galaxy population. On the other hand, investigations focusing on \CII\ emitting gas in very massive galaxies have reported the presence of dynamically cold disks at $z > 4$ \citep[e.g.,][]{Rizzo20, Lelli21, Parlanti23, DeGraaff24, Rowland24, Lee25}. These works generally report higher $V/\sigma$ values, indicative of more rotation-dominated systems, likely representative of the most luminous and extended disks at these epochs. These studies are primarily based on CO and \CII, colder gas tracers that typically show lower intrinsic velocity dispersions than \OIII. In contrast, CRISTAL measurements from \cite{Lee25} have lower $V/\sigma$ values, more consistent with high contribution of turbulence relative to rotational support. All three \texttt{SPICE} feedback models are consistent with ALMA and JWST data, given the large observational uncertainties and the significant scatter in our simulated galaxy sample. This agreement reinforces the idea that current models are capable of capturing the wide diversity of kinematic properties observed in early galaxies.

\section{Discussion}
\label{sect:Discussion}

\subsection{Main limitations}

While our results highlight robust trends in the \OIII\ emission properties across different stellar feedback models, several caveats must be considered when interpreting them. First, although \texttt{SPICE} achieves a physical resolution of $28$ pc in dense regions, this value can degrade to $84$ pc or more in the CGM, depending on location. As a result, we may not fully capture the multiphase structure of outflows, nor the small-scale clumping of the ISM. This limitation is particularly relevant for modeling \OIII\ emission, which originates from dense, compact, and highly ionized regions—typically \HII\ regions—with electron densities and temperatures in specific ranges where collisional excitation of O$^{++}$ is efficient \citep[e.g.,][]{Osterbrock06, Draine11}. At insufficient resolution, such regions may be artificially smoothed out, leading to underestimates of the emissivity due to lower gas densities and incorrect ionization structures \citep{Katz22, Yang24}. Furthermore, the geometry and filling factor of ionized gas—both critical for accurate line emission modeling—depend sensitively on resolving ISM substructure. Therefore, coarse spatial resolution can bias the predicted \LOIII, especially in low-mass or dynamically disturbed systems, where dense clumps are marginally resolved or missed entirely.

Moreover, we assume a fixed relative abundance of oxygen with respect to total metallicity, using a constant $\mathrm{O}/Z = 0.35$. While this is broadly consistent with solar-scaled compositions, it neglects the fact that oxygen yields depend on the stellar IMF, metallicity, and enrichment timescales. In particular, oxygen is primarily produced by short-lived massive stars (core-collapse SNe), whereas iron and other elements are released on longer timescales via Type Ia supernovae. As a result, galaxies with rapid or bursty star formation histories can exhibit enhanced [$\alpha$/Fe] ratios, and hence higher O/$Z$, at fixed total metallicity \citep[e.g.,][]{Vincenzo16,Vincenzo16b, Chruslinska24}. Based on the element-by-element chemodynamical simulations of \cite{Wiersma09}, the oxygen-to-metallicity ratio (O/$Z$) can vary by up to 0.3–0.5 dex, both across and within galaxies, particularly at early times, when the enrichment is dominated by Type II supernovae. These variations arise from differences in nucleosynthetic yields, the relative timing of SNe II/Ia \citep{Woosley95, Portinari98}, and inhomogeneous mixing in the ISM. As a result, assuming a fixed O/$Z$ can lead to systematic uncertainties of up to a factor 2–3 in predicted \OIII\ luminosities.

\subsection{Comparison with other numerical studies}

Connecting simulations and observations is becoming increasingly essential for interpreting the physics of early galaxies. Producing realistic observables from simulations typically involves post-processing pipelines that assign line luminosities based on precomputed grids of ISM parameters, using codes such as \textsc{CLOUDY} \citep{Ferland17} or \textsc{MAPPINGS} \citep{Binette85, Sutherland93, Sutherland18}. While powerful, such pipelines are computationally expensive and hinge on simplifying assumptions about the incident radiation field, the unresolved ISM structure, and metal abundances. In contrast, recent studies, including ours, model \OIII\ self-consistently by coupling radiative transfer and gas dynamics on the fly, so that the ionization state and emissivities emerge from the local stellar radiation and the simulated metal distribution \citep[e.g.][]{Costa22, Katz22RTZ, Smith22, Tacchella22, Bhagwat24b}. This setup captures the time-variable and spatially inhomogeneous conditions that regulate \OIII\ emission, and directly links observable line properties to the underlying feedback and enrichment physics.

Several high-redshift zoom-in and cosmological simulations have focused on modeling \OIII\ emission at 88 $\mu$m in detail. Conversely, the \OIII\ optical emission from such simulations remains less explored, and direct comparisons are still limited. Thanks to its design, \texttt{SPICE} combines the statistical power of a cosmological volume with a resolution sufficient to capture the physical scales that dominate \OIII\ emission, allowing us to test our predicted luminosities and emitting gas phases with the physical insights derived from recent zoom-in simulations. Overall, these studies have primarily aimed to disentangle the relative contributions of different galactic components (e.g. \HII\ regions, diffuse ionized gas, and the CGM) to the total \OIII\ emission at 88 $\mu$m, or to calibrate $L_{\OIII, 88 \mu m}$ as a tracer of star formation activity. For instance, \cite{Olsen17} find that in the SÍGAME simulations the \LOIII-SFR for the 88 $\mu$m line relation at $z \sim 6$ aligns well with local starburst galaxies \citep{DeLooze14}. Among all the numerical work, the SERRA simulations \citep{Pallottini22} show the biggest scatter in the \LOIII-SFR relation, revealing that stellar feedback drives strong time variability in both metal enrichment and ionization conditions, which directly impacts the visibility of \OIII. Similarly, in the FirstLight project, \cite{Nakazato23} find that young, metal-poor stellar populations can significantly boost \OIII\ luminosity, leading to $L_{\OIII, 88 \mu m}$/SFR ratios higher than observed in local analogs, in agreement with JWST observations. 
Overall, for galaxies in a similar mass and SFR regime, we find good agreement between the predicted \LOIII–SFR relations from these simulations and our \texttt{SPICE} results, as shown in Appendix~\ref{appendixA}.
Finally, the PONOS simulations from \cite{Schimek24b} demonstrate that different methods for estimating \OIII\ luminosities—particularly when omitting full radiative transfer—can lead to discrepancies of up to two orders of magnitude. This highlights the importance of a consistent treatment of ionizing radiation. In this regard, \texttt{SPICE} includes on-the-fly radiative transfer and in this paper we use \texttt{RASCAS} in post-processing, providing a more robust prediction for both the 88 $\mu$m and 5008 \AA\ \OIII\ emission.\\

Building on the zoom-in simulations discussed above—which collectively show that accurately modeling \OIII\ emission requires both high spatial resolution and a consistent treatment of radiative transfer—cosmological simulations provide a complementary perspective by predicting the statistical properties of \OIII\ emitters across much larger samples and volumes. Several studies have modeled \OIII\ emission in large-volume frameworks, such as IllustrisTNG \citep{Yang24}, which provides JWST luminosity–function forecasts, and EAGLE \citep{RamosPadilla23}, where post-processed galaxies reveal that 88 $\mu$m \OIII\ emission arises from both diffuse ionized gas and compact \HII\ regions, yielding \LOIII–SFR relations consistent with observations within $\sim0.5$ dex. Other cosmological simulations explicitly couple radiation–hydrodynamics to follow \HI\ ionization, as in THESAN \citep{Kannan22} and SPHINX \citep{Katz22}, while the MEGATRON \citep{Katz25} suite incorporates on-the-fly non-equilibrium chemical networks to track multiple ionic species and explore diverse feedback models at extremely high resolution, albeit within a relatively small volume of $\sim 20$ cMpc$^3/h$. All these cosmological models, including \texttt{SPICE}, successfully overlap with existing high-redshift observations, confirming the robustness of \OIII\ as a tracer of SFR. The largest discrepancies between simulations emerge at the low-mass and low-SFR end, where observational constraints remain scarce and where feedback prescriptions and numerical resolution exert the strongest influence on the predicted emission. In this regime, the \texttt{SPICE} \texttt{bursty-sn} model yields scaling relations comparable to those of \cite{Katz22}, while \cite{Kannan22} predict higher \LOIII\ at 88 $\mu$m. This excess may partly arise from resolution effects: when individual \HII\ regions are not fully resolved, the local ionization structure becomes smoothed out, leading to an overestimation of the \OIII\ emissivity per gas cell unless a correction is applied. Recent predictions of \OIII\ luminosity functions from large-volume simulations, such as \cite{Katz23} and \cite{Yang24b}, also show good agreement with early JWST measurements, though with a tendency to underpredict the brightest sources. This mild tension likely reflects both the limited current observational samples and the sensitivity of \OIII-based LF measurements to precise redshift estimates and to environmental effects such as cosmic variance. Building on these efforts, in this work we take a further step by explicitly isolating how different stellar-feedback models shape the predicted \OIII\ luminosity functions.


By combining high spatial resolution, a large cosmological box, on-the-fly radiative transfer, multiple stellar-feedback prescriptions, and by following galaxy evolution down to 
$z \sim 5$ \texttt{SPICE} resolves the physical scales most relevant to \OIII\ production, while retaining the statistical power of a big simulated volume. This makes \texttt{SPICE} uniquely positioned to bridge the gap between zoom-in and large-scale approaches, clarifying how feedback and resolution shape the diversity of \OIII-emitting galaxies in the early Universe. At the same time, the range of existing zoom-in and cosmological studies highlights the diversity of feedback implementations and post-processing methods currently adopted in the literature. Within this landscape, \texttt{SPICE} offers a uniquely controlled framework where we isolate the effects of stellar feedback by keeping all other initial conditions fixed. This allows us to attribute variations in \LOIII\ and morphology directly to differences in feedback strength and timing, providing valuable insight into the origin of the scatter in observed \OIII\ emission at high redshift.

\subsection{Testing Stellar Feedback in the EoR with \OIII\ Observations}

With the advent of powerful new facilities such as JWST and ALMA, it has become possible to characterise galaxies in the epoch of reionization with unprecedented detail. Among the many emission lines accessible at high redshift, \OIII\ is particularly valuable because it can be observed both in the rest-frame optical and FIR, providing complementary diagnostics of the ISM. Observations have now shown that strong \OIII\ emission is ubiquitous in early galaxies \citep[e.g.,][]{Inoue16, Carniani17, Carniani24, Hashimoto18, Trump23, Witstok25}.  

Recent measurements of the \OIII\ LF from deep JWST surveys have provided the first statistical constraints on the abundance and evolution of strong \OIII\ emitters in the reionization era \citep[e.g.,][]{Matthee23, Meyer24, Wold25}. These studies consistently find a steep faint-end slope and a rapid increase in number density towards lower luminosities, suggesting that low-mass galaxies contribute significantly to the ionizing photon budget. These are precisely the systems that \texttt{SPICE} is able to resolve within its cosmological volume, occupying the mass and SFR range where stellar feedback has the strongest dynamical and thermal impact on the ISM and on the resulting \OIII\ emission. In more massive galaxies, the deeper gravitational potential well tends to mitigate these effects, making the imprint of feedback less evident. Consequently, the galaxies captured by \texttt{SPICE} lie in a “sweet spot” where feedback-induced variations in ionization and enrichment are most directly observable. In this paper, we demonstrate that the shape and normalization of the \OIII\ LF provide some of the strongest observational constraints on stellar feedback models at high redshift.

Spatially resolved observations further reveal that \OIII\ emission can extend well beyond the stellar disk, forming ionized halos with radii of several kiloparsecs (e.g., \cite{Carniani20, Parlanti23}). These halos often exhibit complex morphologies and asymmetries. When combined with kinematic data —specifically velocity fields and dispersion measured in \OIII — in some cases, these maps show signs of massive bulges, clear rotation or turbulence (e.g., \cite{Lelli21, Parlanti23}). Such multi-dimensional observables provide strong constraints on feedback processes shaping the ionized ISM and CGM in the early Universe. In \texttt{SPICE}, all feedback models produce \OIII\ emission with spatial extents comparable to those observed, typically reaching a few kiloparsecs. Among them, the \texttt{bursty-sn} model is the one producing consistently the most extended and asymmetric morphologies, with ionized bubbles and filamentary features reaching several kiloparsecs into the circumgalactic medium, as repeated feedback episodes drive outflows and turbulence.

Measurements of \LOIII\ are widely used as tracers of SFR, often calibrated through empirical relations established in the local Universe \citep[e.g.,][]{DeLooze14, Cormier15, Villa-Velez21}. A wealth of recent ALMA and JWST observations have extended these relations to the reionization era, detecting \OIII\ emission across a wide range of galaxy masses and star formation rates \citep[e.g.,][]{Inoue16, Carniani17, Walter18, Hashimoto19, Harikane20, Witstok22, Wong22, Bakx24, Fujimoto24, Zavala24, Algera25, Carniani25, Schouws25, vanLeeuwen25}. \texttt{SPICE} simulations reproduce these empirical trends at both 5008 \AA\ and 88 $\mu$m, with predicted \LOIII–SFR relations that lie within the observed scatter of high-redshift samples (see Appendix \ref{appendixA}). This consistency across optical and far-infrared transitions supports the robustness of our modeling of the ionized gas and its coupling to star formation activity.

Line ratios involving \OIII, such as \OIII/H$\beta$ or \OIII/[C\,\textsc{ii}], provide powerful constraints on the gas-phase metallicity, ionization parameter, and the fraction of ionized to neutral gas. These diagnostics indicate that the ISM in many $z \gtrsim 6$ galaxies is already significantly enriched, with typical oxygen abundances reaching 10--50$\%$ of solar \citep[e.g.,][]{Jones20, Witstok22, Cameron23}. High-redshift galaxies display more extreme ionizing conditions and compact star-forming regions \citep[e.g.,][]{Bakx20, Witstok22}. Ionization parameters are often found to be high ($\log U \sim -2$ to $-1.5$), consistent with intense radiation fields from young, massive stars \citep[e.g.,][]{Nakajima14, Harikane20, Cameron23}. Furthermore, elevated \OIII/[C\,\textsc{ii}] ratios and spatially extended \OIII\ emission suggest that large fractions of the ISM are highly ionized, possibly due to strong burstiness and low dust content \citep[e.g.,][]{Inoue16, Carniani17, Carniani20, Katz23, Algera25}. Nevertheless, while most galaxies show relatively high values, some systems exhibit notably lower ratios, interpreted as signatures of a less bursty star formation history and a comparatively weaker ionizing radiation field \citep{Algera24}. Although our analysis focuses on \OIII, complementary diagnostics such as \CII, H$\alpha$, and other optical and infrared lines provide additional information on the ionization structure and feedback efficiency. Future work will combine multiple emission lines to jointly constrain the thermal, chemical, and dynamical impact of stellar feedback on the early interstellar medium.

In this observational framework, the spatial extent of \OIII\ emission and the luminosity function down to even fainter luminosities appear to be the most promising methods to constrain stellar feedback models according to the \texttt{SPICE} simulations. The size and kinematics of ionized halos provide complementary and sensitive probes of how feedback regulates the ionization state and distribution of gas in early galaxies. Looking ahead, we plan to extend this analysis by combining \OIII\ observations with other key emission lines—such as \CII, H$\alpha$, and UV metal lines—to achieve a more comprehensive understanding of the interplay between star formation, feedback, and ISM conditions in the epoch of reionization.

\section{Conclusions}
\label{sect:Conclusions}

The advent of powerful telescopes such as ALMA and JWST, has revolutionized our understanding of
cold and warm/ionized gas phases and the stellar component of galaxies at $z > 5$. Among the brightest emission lines, \OIII\ stands out as a particularly valuable tracer, observable both in the rest-frame optical (e.g., lines at 5008 and 4960 \AA) and in the far-infrared (e.g., the 88\,$\mu$m line). \OIII\ provides a robust tool to characterize how stellar feedback impacts the interstellar medium, specifically regarding the evolution of metallicity, gas ionization, and the size and morphology of ionized halos.

Modeling these processes within a cosmological context is challenging due to the wide range of physical scales involved, which are crucial to consistently reproduce the observed \OIII\ emission. In this work, we model the \OIII\ emission by treating the ion as a five-level atom, and we compute the resulting line intensities by accounting for the finite size of \HII\ regions, as well as dust attenuation and scattering effects. We present predictions for the 5008\,\AA, 4960\,\AA, and 88\,$\mu$m transitions within the \texttt{SPICE} simulations (BH24).
 \texttt{SPICE} consists of three cosmological simulation volumes that differ only in their stellar feedback prescriptions, including a bursty supernova feedback model (\texttt{bursty-sn}), a smoother supernova feedback model (\texttt{smooth-sn}), and a model incorporating the effects of hypernovae (\texttt{hyper-sn}).

Our main findings are as follows:
\begin{itemize}
    \item At $z=7$ and $z=5$, \OIII\ emission is predominantly contributed by gas ionized by radiative feedback across all models. At $z=10$, this dominance persists only in the \texttt{smooth-sn} model, whereas shock-heated gas dominates in both the \texttt{bursty-sn} and \texttt{hyper-sn} scenarios.
    
    \item The simulated mass–metallicity relations reproduce both the observed trends and scatter from JWST and ALMA datasets. However, the \texttt{smooth-sn} model reaches stellar masses (metallicities) up to $\sim1$\,dex  (0.5 dex) higher than \texttt{bursty-sn} by $z=5$, reflecting its more efficient cooling and metal retention. Despite this, the mass–metallicity relation alone remains insufficient to unambiguously discriminate between different feedback mechanisms at high redshift.

    \item The \OIII\ luminosity function exhibits a systematic increase in amplitude toward lower redshifts, mirroring the formation of progressively more luminous galaxies. Differences among feedback models can reach up to $\sim1$\,dex in amplitude, with the \texttt{smooth-sn} model consistently predicting the highest abundance of \OIII-bright galaxies, in good agreement with current observations. Because these differences persist even after accounting for dust attenuation, the \OIII\ LFs provide a powerful diagnostic of stellar feedback, in contrast to the UV LFs which can be degenerate between feedback models once dust absorption is included. Deeper or less overdensity-biased observations with ALMA and JWST will be essential to disentangle these feedback-driven variations in the epoch of reionization.

    \item While the three models converge at $z=10$, by $z=5$ the \texttt{bursty-sn} galaxies display systematically lower neutral gas fractions. In particular, about 80\% of galaxies in the \texttt{smooth-sn} and \texttt{hyper-sn} simulations have neutral fractions above 0.75, whereas in the \texttt{bursty-sn} case most systems lie between 0.5 and 0.9. Despite this higher neutral content, the \texttt{smooth-sn} model still hosts the most luminous \OIII\ emitters at both optical and FIR wavelengths.

    \item All feedback prescriptions show a positive correlation between \OIII\ luminosity and metallicity, with the \texttt{smooth-sn} model reaching almost 1.5 dex higher metallicities than \texttt{bursty-sn}.

    \item The three feedback models predict similar \OIII\ half-light radii across all redshifts and mass regimes, with differences remaining within the measurement uncertainties. Nonetheless, systematic differences emerge: \texttt{bursty-sn} produces more extended halos with $R_{1/2}$ increasing with luminosity, while the \texttt{smooth-sn} model forms more compact systems and shows a flatter or decreasing size–luminosity relation at $z \leq 7$. This behavior arises from its more efficient cooling, which allows galaxies to grow more massive and develop deeper gravitational potential wells. As a result, these galaxies become increasingly compact as their mass increases and stellar feedback becomes less effective at unbinding gas. The \texttt{hyper-sn} model exhibits no clear trend and displays larger scatter, reflecting its more stochastic nature.

    \item The kinematic analysis shows that all feedback models yield comparable $V/\sigma$ distributions, with no statistically significant differences across mass or redshift. This indicates that \OIII\ kinematics alone are not sufficient to distinguish between different feedback prescriptions within current observational uncertainties.

\end{itemize}

Overall, the \texttt{SPICE} simulations provide new theoretical insights into the origin of \OIII\ emission during the epoch of reionization and demonstrate that \OIII-based diagnostics can serve as powerful probes of stellar feedback in the early Universe. In particular, the normalization and shape of the \OIII\ luminosity function, together with the spatial extent of \OIII-emitting regions, are shown to be sensitive to the strength and timing of feedback episodes. Bursty feedback models suppress metal enrichment and \OIII\ luminosity, while smoother and more continuous feedback leads to more compact, metal-rich, and \OIII-bright systems. Indeed, upcoming JWST and ALMA campaigns could exploit these results to constrain feedback physics at high redshift.
In future work, we will extend this framework to include additional emission and absorption lines as well as continuum properties, enabling a more comprehensive multi-wavelength characterization of stellar feedback at $z > 5$.

\section*{Acknowledgements}
B Casavecchia is grateful to Ildar Khabibullin for insightful discussions on emission processes in high-$z$ galaxies. B Casavecchia and AB also thank Rüdiger Pakmor and Matteo Guardiani for their sweet encouragement and support throughout this work.
AB and B Ciardi acknowledge support from the ERC Synergy Grant 101166930 - RECAP. The numerical calculations presented in this work were performed on the machines of the Max Planck Computing and Data Facility of the Max Planck Society, Germany. This research was supported by the International Space Science Institute (ISSI) in Bern (Switzerland), through ISSI International Team project no. 564 (The Cosmic Baryon Cycle from Space). We acknowledge the NASA Astrophysics Data System for providing access to their bibliographic research tools. We acknowledge the use of OpenAI’s ChatGPT language model, which assisted in debugging and refining parts of the analysis and visualization code used in this work.

\section*{Data Availability}

 The data underlying this article is available upon reasonable request to the corresponding author.



\bibliographystyle{mnras}
\bibliography{main} 

@BOOK{Draine11,
       author = {{Draine}, Bruce T.},
        title = "{Physics of the Interstellar and Intergalactic Medium}",
         year = 2011,
       adsurl = {https://ui.adsabs.harvard.edu/abs/2011piim.book.....D}
}

@ARTICLE{Katz22RTZ,
       author = {{Katz}, Harley},
        title = "{RAMSES-RTZ: non-equilibrium metal chemistry and cooling coupled to on-the-fly radiation hydrodynamics}",
      journal = {\mnras},
         year = 2022,
        month = may,
       volume = {512},
       number = {1},
        pages = {348-365},
          doi = {10.1093/mnras/stac423},
archivePrefix = {arXiv},
       eprint = {2202.04083},
 primaryClass = {astro-ph.GA},
       adsurl = {https://ui.adsabs.harvard.edu/abs/2022MNRAS.512..348K}
}

@ARTICLE{Oppenheimer13,
       author = {{Oppenheimer}, Benjamin D. and {Schaye}, Joop},
        title = "{Non-equilibrium ionization and cooling of metal-enriched gas in the presence of a photoionization background}",
      journal = {\mnras},
         year = 2013,
        month = sep,
       volume = {434},
       number = {2},
        pages = {1043-1062},
          doi = {10.1093/mnras/stt1043},
archivePrefix = {arXiv},
       eprint = {1302.5710},
 primaryClass = {astro-ph.CO},
       adsurl = {https://ui.adsabs.harvard.edu/abs/2013MNRAS.434.1043O}
}

@ARTICLE{Voronov97,
       author = {{Voronov}, G.~S.},
        title = "{A Practical Fit Formula for Ionization Rate Coefficients of Atoms and Ions by Electron Impact: Z = 1-28}",
      journal = {Atomic Data and Nuclear Data Tables},
         year = 1997,
        month = jan,
       volume = {65},
        pages = {1},
          doi = {10.1006/adnd.1997.0732},
       adsurl = {https://ui.adsabs.harvard.edu/abs/1997ADNDT..65....1V}
}

@ARTICLE{Badnell06,
       author = {{Badnell}, N.~R.},
        title = "{Radiative Recombination Data for Modeling Dynamic Finite-Density Plasmas}",
      journal = {\apjs},
         year = 2006,
        month = dec,
       volume = {167},
       number = {2},
        pages = {334-342},
          doi = {10.1086/508465},
archivePrefix = {arXiv},
       eprint = {astro-ph/0604144},
 primaryClass = {astro-ph},
       adsurl = {https://ui.adsabs.harvard.edu/abs/2006ApJS..167..334B}
}

@ARTICLE{Verner96,
       author = {{Verner}, D.~A. and {Ferland}, G.~J. and {Korista}, K.~T. and {Yakovlev}, D.~G.},
        title = "{Atomic Data for Astrophysics. II. New Analytic Fits for Photoionization Cross Sections of Atoms and Ions}",
      journal = {\apj},
         year = 1996,
        month = jul,
       volume = {465},
        pages = {487},
          doi = {10.1086/177435},
archivePrefix = {arXiv},
       eprint = {astro-ph/9601009},
 primaryClass = {astro-ph},
       adsurl = {https://ui.adsabs.harvard.edu/abs/1996ApJ...465..487V}
}

@ARTICLE{Yang24,
       author = {{Yang}, Shengqi and {Lidz}, Adam and {Benson}, Andrew and {Chauhan}, Swathya Singh and {Smith}, Aaron and {Li}, Hui},
        title = "{Analytical strong line diagnostics and their redshift evolution}",
      journal = {\mnras},
         year = 2024,
        month = nov,
       volume = {534},
       number = {4},
        pages = {3665-3675},
          doi = {10.1093/mnras/stae2337},
archivePrefix = {arXiv},
       eprint = {2312.09213},
 primaryClass = {astro-ph.GA},
       adsurl = {https://ui.adsabs.harvard.edu/abs/2024MNRAS.534.3665Y}
}

@ARTICLE{Yang24b,
       author = {{Yang}, Shengqi and {Lidz}, Adam and {Benson}, Andrew and {Zhao}, Yizhou and {Li}, Hui and {Zhao}, Amelia and {Smith}, Aaron and {Zhang}, Yucheng and {Somerville}, Rachel and {Pullen}, Anthony and {Li}, Hui},
        title = "{A New Framework for ISM Emission Line Models: Connecting Multi-Scale Simulations Across Cosmological Volumes}",
      journal = {arXiv e-prints},
         year = 2024,
        month = sep,
          eid = {arXiv:2409.03997},
        pages = {arXiv:2409.03997},
          doi = {10.48550/arXiv.2409.03997},
archivePrefix = {arXiv},
       eprint = {2409.03997},
 primaryClass = {astro-ph.GA},
       adsurl = {https://ui.adsabs.harvard.edu/abs/2024arXiv240903997Y}
}

@ARTICLE{Harikane23,
       author = {{Harikane}, Yuichi and {Ouchi}, Masami and {Oguri}, Masamune and {Ono}, Yoshiaki and {Nakajima}, Kimihiko and {Isobe}, Yuki and {Umeda}, Hiroya and {Mawatari}, Ken and {Zhang}, Yechi},
        title = "{A Comprehensive Study of Galaxies at z   9-16 Found in the Early JWST Data: Ultraviolet Luminosity Functions and Cosmic Star Formation History at the Pre-reionization Epoch}",
      journal = {\apjs},
         year = 2023,
        month = mar,
       volume = {265},
       number = {1},
          eid = {5},
        pages = {5},
          doi = {10.3847/1538-4365/acaaa9},
archivePrefix = {arXiv},
       eprint = {2208.01612},
 primaryClass = {astro-ph.GA},
       adsurl = {https://ui.adsabs.harvard.edu/abs/2023ApJS..265....5H}
}

@ARTICLE{Naidu22,
       author = {{Naidu}, Rohan P. and {Oesch}, Pascal A. and {van Dokkum}, Pieter and {Nelson}, Erica J. and {Suess}, Katherine A. and {Brammer}, Gabriel and {Whitaker}, Katherine E. and {Illingworth}, Garth and {Bouwens}, Rychard and {Tacchella}, Sandro and {Matthee}, Jorryt and {Allen}, Natalie and {Bezanson}, Rachel and {Conroy}, Charlie and {Labbe}, Ivo and {Leja}, Joel and {Leonova}, Ecaterina and {Magee}, Dan and {Price}, Sedona H. and {Setton}, David J. and {Strait}, Victoria and {Stefanon}, Mauro and {Toft}, Sune and {Weaver}, John R. and {Weibel}, Andrea},
        title = "{Two Remarkably Luminous Galaxy Candidates at z {\ensuremath{\approx}} 10-12 Revealed by JWST}",
      journal = {\apjl},
         year = 2022,
        month = nov,
       volume = {940},
       number = {1},
          eid = {L14},
        pages = {L14},
          doi = {10.3847/2041-8213/ac9b22},
archivePrefix = {arXiv},
       eprint = {2207.09434},
 primaryClass = {astro-ph.GA},
       adsurl = {https://ui.adsabs.harvard.edu/abs/2022ApJ...940L..14N}
}

@ARTICLE{Bouwens23a,
       author = {{Bouwens}, Rychard and {Illingworth}, Garth and {Oesch}, Pascal and {Stefanon}, Mauro and {Naidu}, Rohan and {van Leeuwen}, Ivana and {Magee}, Dan},
        title = "{UV luminosity density results at z > 8 from the first JWST/NIRCam fields: limitations of early data sets and the need for spectroscopy}",
      journal = {\mnras},
         year = 2023,
        month = jul,
       volume = {523},
       number = {1},
        pages = {1009-1035},
          doi = {10.1093/mnras/stad1014},
archivePrefix = {arXiv},
       eprint = {2212.06683},
 primaryClass = {astro-ph.CO},
       adsurl = {https://ui.adsabs.harvard.edu/abs/2023MNRAS.523.1009B}
}

@ARTICLE{Bouwens23b,
       author = {{Bouwens}, Rychard J. and {Stefanon}, Mauro and {Brammer}, Gabriel and {Oesch}, Pascal A. and {Herard-Demanche}, Thomas and {Illingworth}, Garth D. and {Matthee}, Jorryt and {Naidu}, Rohan P. and {van Dokkum}, Pieter G. and {van Leeuwen}, Ivana F.},
        title = "{Evolution of the UV LF from z   15 to z   8 using new JWST NIRCam medium-band observations over the HUDF/XDF}",
      journal = {\mnras},
         year = 2023,
        month = jul,
       volume = {523},
       number = {1},
        pages = {1036-1055},
          doi = {10.1093/mnras/stad1145},
archivePrefix = {arXiv},
       eprint = {2211.02607},
 primaryClass = {astro-ph.GA},
       adsurl = {https://ui.adsabs.harvard.edu/abs/2023MNRAS.523.1036B}
}

@ARTICLE{Robertson23,
       author = {{Robertson}, B.~E. and {Tacchella}, S. and {Johnson}, B.~D. and {Hainline}, K. and {Whitler}, L. and {Eisenstein}, D.~J. and {Endsley}, R. and {Rieke}, M. and {Stark}, D.~P. and {Alberts}, S. and {Dressler}, A. and {Egami}, E. and {Hausen}, R. and {Rieke}, G. and {Shivaei}, I. and {Williams}, C.~C. and {Willmer}, C.~N.~A. and {Arribas}, S. and {Bonaventura}, N. and {Bunker}, A. and {Cameron}, A.~J. and {Carniani}, S. and {Charlot}, S. and {Chevallard}, J. and {Curti}, M. and {Curtis-Lake}, E. and {D'Eugenio}, F. and {Jakobsen}, P. and {Looser}, T.~J. and {L{\"u}tzgendorf}, N. and {Maiolino}, R. and {Maseda}, M.~V. and {Rawle}, T. and {Rix}, H. -W. and {Smit}, R. and {{\"U}bler}, H. and {Willott}, C. and {Witstok}, J. and {Baum}, S. and {Bhatawdekar}, R. and {Boyett}, K. and {Chen}, Z. and {de Graaff}, A. and {Florian}, M. and {Helton}, J.~M. and {Hviding}, R.~E. and {Ji}, Z. and {Kumari}, N. and {Lyu}, J. and {Nelson}, E. and {Sandles}, L. and {Saxena}, A. and {Suess}, K.~A. and {Sun}, F. and {Topping}, M. and {Wallace}, I.~E.~B.},
        title = "{Identification and properties of intense star-forming galaxies at redshifts z > 10}",
      journal = {Nature Astronomy},
         year = 2023,
        month = may,
       volume = {7},
        pages = {611-621},
          doi = {10.1038/s41550-023-01921-1},
archivePrefix = {arXiv},
       eprint = {2212.04480},
 primaryClass = {astro-ph.GA},
       adsurl = {https://ui.adsabs.harvard.edu/abs/2023NatAs...7..611R}
}

@ARTICLE{Zavala24,
       author = {{Zavala}, Jorge A. and {Bakx}, Tom and {Mitsuhashi}, Ikki and {Castellano}, Marco and {Calabro}, Antonello and {Akins}, Hollis and {Buat}, Veronique and {Casey}, Caitlin M. and {Fernandez-Arenas}, David and {Franco}, Maximilien and {Fontana}, Adriano and {Hatsukade}, Bunyo and {Ho}, Luis C. and {Ikeda}, Ryota and {Kartaltepe}, Jeyhan and {Koekemoer}, Anton M. and {McKinney}, Jed and {Napolitano}, Lorenzo and {P{\'e}rez-Gonz{\'a}lez}, Pablo G. and {Santini}, Paola and {Serjeant}, Stephen and {Terlevich}, Elena and {Terlevich}, Roberto and {Yung}, L.~Y. Aaron},
        title = "{ALMA Detection of [O III] 88 {\ensuremath{\mu}}m at z = 12.33: Exploring the Nature and Evolution of GHZ2 as a Massive Compact Stellar System}",
      journal = {\apjl},
         year = 2024,
        month = dec,
       volume = {977},
       number = {1},
          eid = {L9},
        pages = {L9},
          doi = {10.3847/2041-8213/ad8f38},
archivePrefix = {arXiv},
       eprint = {2411.03593},
 primaryClass = {astro-ph.GA},
       adsurl = {https://ui.adsabs.harvard.edu/abs/2024ApJ...977L...9Z}
}

@ARTICLE{Hashimoto18,
       author = {{Hashimoto}, Takuya and {Laporte}, Nicolas and {Mawatari}, Ken and {Ellis}, Richard S. and {Inoue}, Akio K. and {Zackrisson}, Erik and {Roberts-Borsani}, Guido and {Zheng}, Wei and {Tamura}, Yoichi and {Bauer}, Franz E. and {Fletcher}, Thomas and {Harikane}, Yuichi and {Hatsukade}, Bunyo and {Hayatsu}, Natsuki H. and {Matsuda}, Yuichi and {Matsuo}, Hiroshi and {Okamoto}, Takashi and {Ouchi}, Masami and {Pell{\'o}}, Roser and {Rydberg}, Claes-Erik and {Shimizu}, Ikkoh and {Taniguchi}, Yoshiaki and {Umehata}, Hideki and {Yoshida}, Naoki},
        title = "{The onset of star formation 250 million years after the Big Bang}",
      journal = {\nat},
         year = 2018,
        month = may,
       volume = {557},
       number = {7705},
        pages = {392-395},
          doi = {10.1038/s41586-018-0117-z},
archivePrefix = {arXiv},
       eprint = {1805.05966},
 primaryClass = {astro-ph.GA},
       adsurl = {https://ui.adsabs.harvard.edu/abs/2018Natur.557..392H}
}

@ARTICLE{Hashimoto19,
       author = {{Hashimoto}, Takuya and {Inoue}, Akio K. and {Mawatari}, Ken and {Tamura}, Yoichi and {Matsuo}, Hiroshi and {Furusawa}, Hisanori and {Harikane}, Yuichi and {Shibuya}, Takatoshi and {Knudsen}, Kirsten K. and {Kohno}, Kotaro and {Ono}, Yoshiaki and {Zackrisson}, Erik and {Okamoto}, Takashi and {Kashikawa}, Nobunari and {Oesch}, Pascal A. and {Ouchi}, Masami and {Ota}, Kazuaki and {Shimizu}, Ikkoh and {Taniguchi}, Yoshiaki and {Umehata}, Hideki and {Watson}, Darach},
        title = "{Big Three Dragons: A z = 7.15 Lyman-break galaxy detected in [O III] 88 {\ensuremath{\mu}}m, [C II] 158 {\ensuremath{\mu}}m, and dust continuum with ALMA}",
      journal = {\pasj},
         year = 2019,
        month = aug,
       volume = {71},
       number = {4},
          eid = {71},
        pages = {71},
          doi = {10.1093/pasj/psz049},
archivePrefix = {arXiv},
       eprint = {1806.00486},
 primaryClass = {astro-ph.GA},
       adsurl = {https://ui.adsabs.harvard.edu/abs/2019PASJ...71...71H}
}

@ARTICLE{LeFevre20,
       author = {{Le F{\`e}vre}, O. and {B{\'e}thermin}, M. and {Faisst}, A. and {Jones}, G.~C. and {Capak}, P. and {Cassata}, P. and {Silverman}, J.~D. and {Schaerer}, D. and {Yan}, L. and {Amorin}, R. and {Bardelli}, S. and {Boquien}, M. and {Cimatti}, A. and {Dessauges-Zavadsky}, M. and {Giavalisco}, M. and {Hathi}, N.~P. and {Fudamoto}, Y. and {Fujimoto}, S. and {Ginolfi}, M. and {Gruppioni}, C. and {Hemmati}, S. and {Ibar}, E. and {Koekemoer}, A. and {Khusanova}, Y. and {Lagache}, G. and {Lemaux}, B.~C. and {Loiacono}, F. and {Maiolino}, R. and {Mancini}, C. and {Narayanan}, D. and {Morselli}, L. and {M{\'e}ndez-Hern{\`a}ndez}, Hugo and {Oesch}, P.~A. and {Pozzi}, F. and {Romano}, M. and {Riechers}, D. and {Scoville}, N. and {Talia}, M. and {Tasca}, L.~A.~M. and {Thomas}, R. and {Toft}, S. and {Vallini}, L. and {Vergani}, D. and {Walter}, F. and {Zamorani}, G. and {Zucca}, E.},
        title = "{The ALPINE-ALMA [CII] survey. Survey strategy, observations, and sample properties of 118 star-forming galaxies at 4 < z < 6}",
      journal = {\aap},
         year = 2020,
        month = nov,
       volume = {643},
          eid = {A1},
        pages = {A1},
          doi = {10.1051/0004-6361/201936965},
archivePrefix = {arXiv},
       eprint = {1910.09517},
 primaryClass = {astro-ph.CO},
       adsurl = {https://ui.adsabs.harvard.edu/abs/2020A&A...643A...1L}
}

@ARTICLE{Finkelstein22,
       author = {{Finkelstein}, Steven L. and {Bagley}, Micaela B. and {Arrabal Haro}, Pablo and {Dickinson}, Mark and {Ferguson}, Henry C. and {Kartaltepe}, Jeyhan S. and {Papovich}, Casey and {Burgarella}, Denis and {Kocevski}, Dale D. and {Huertas-Company}, Marc and {Iyer}, Kartheik G. and {Koekemoer}, Anton M. and {Larson}, Rebecca L. and {P{\'e}rez-Gonz{\'a}lez}, Pablo G. and {Rose}, Caitlin and {Tacchella}, Sandro and {Wilkins}, Stephen M. and {Chworowsky}, Katherine and {Medrano}, Aubrey and {Morales}, Alexa M. and {Somerville}, Rachel S. and {Yung}, L.~Y. Aaron and {Fontana}, Adriano and {Giavalisco}, Mauro and {Grazian}, Andrea and {Grogin}, Norman A. and {Kewley}, Lisa J. and {Kirkpatrick}, Allison and {Kurczynski}, Peter and {Lotz}, Jennifer M. and {Pentericci}, Laura and {Pirzkal}, Nor and {Ravindranath}, Swara and {Ryan}, Russell E. and {Trump}, Jonathan R. and {Yang}, Guang and {Almaini}, Omar and {Amor{\'\i}n}, Ricardo O. and {Annunziatella}, Marianna and {Backhaus}, Bren E. and {Barro}, Guillermo and {Behroozi}, Peter and {Bell}, Eric F. and {Bhatawdekar}, Rachana and {Bisigello}, Laura and {Bromm}, Volker and {Buat}, V{\'e}ronique and {Buitrago}, Fernando and {Calabr{\`o}}, Antonello and {Casey}, Caitlin M. and {Castellano}, Marco and {Ch{\'a}vez Ortiz}, {\'O}scar A. and {Ciesla}, Laure and {Cleri}, Nikko J. and {Cohen}, Seth H. and {Cole}, Justin W. and {Cooke}, Kevin C. and {Cooper}, M.~C. and {Cooray}, Asantha R. and {Costantin}, Luca and {Cox}, Isabella G. and {Croton}, Darren and {Daddi}, Emanuele and {Dav{\'e}}, Romeel and {de La Vega}, Alexander and {Dekel}, Avishai and {Elbaz}, David and {Estrada-Carpenter}, Vicente and {Faber}, Sandra M. and {Fern{\'a}ndez}, Vital and {Finkelstein}, Keely D. and {Freundlich}, Jonathan and {Fujimoto}, Seiji and {Garc{\'\i}a-Argum{\'a}nez}, {\'A}ngela and {Gardner}, Jonathan P. and {Gawiser}, Eric and {G{\'o}mez-Guijarro}, Carlos and {Guo}, Yuchen and {Hamblin}, Kurt and {Hamilton}, Timothy S. and {Hathi}, Nimish P. and {Holwerda}, Benne W. and {Hirschmann}, Michaela and {Hutchison}, Taylor A. and {Jaskot}, Anne E. and {Jha}, Saurabh W. and {Jogee}, Shardha and {Juneau}, St{\'e}phanie and {Jung}, Intae and {Kassin}, Susan A. and {Le Bail}, Aur{\'e}lien and {Leung}, Gene C.~K. and {Lucas}, Ray A. and {Magnelli}, Benjamin and {Mantha}, Kameswara Bharadwaj and {Matharu}, Jasleen and {McGrath}, Elizabeth J. and {McIntosh}, Daniel H. and {Merlin}, Emiliano and {Mobasher}, Bahram and {Newman}, Jeffrey A. and {Nicholls}, David C. and {Pandya}, Viraj and {Rafelski}, Marc and {Ronayne}, Kaila and {Santini}, Paola and {Seill{\'e}}, Lise-Marie and {Shah}, Ekta A. and {Shen}, Lu and {Simons}, Raymond C. and {Snyder}, Gregory F. and {Stanway}, Elizabeth R. and {Straughn}, Amber N. and {Teplitz}, Harry I. and {Vanderhoof}, Brittany N. and {Vega-Ferrero}, Jes{\'u}s and {Wang}, Weichen and {Weiner}, Benjamin J. and {Willmer}, Christopher N.~A. and {Wuyts}, Stijn and {Zavala}, Jorge A. and {Ceers Team}},
        title = "{A Long Time Ago in a Galaxy Far, Far Away: A Candidate z {\ensuremath{\sim}} 12 Galaxy in Early JWST CEERS Imaging}",
      journal = {\apjl},
         year = 2022,
        month = dec,
       volume = {940},
       number = {2},
          eid = {L55},
        pages = {L55},
          doi = {10.3847/2041-8213/ac966e},
archivePrefix = {arXiv},
       eprint = {2207.12474},
 primaryClass = {astro-ph.GA},
       adsurl = {https://ui.adsabs.harvard.edu/abs/2022ApJ...940L..55F}
}

@ARTICLE{Castellano24,
       author = {{Castellano}, Marco and {Napolitano}, Lorenzo and {Fontana}, Adriano and {Roberts-Borsani}, Guido and {Treu}, Tommaso and {Vanzella}, Eros and {Zavala}, Jorge A. and {Arrabal Haro}, Pablo and {Calabr{\`o}}, Antonello and {Llerena}, Mario and {Mascia}, Sara and {Merlin}, Emiliano and {Paris}, Diego and {Pentericci}, Laura and {Santini}, Paola and {Bakx}, Tom J.~L.~C. and {Bergamini}, Pietro and {Cupani}, Guido and {Dickinson}, Mark and {Filippenko}, Alexei V. and {Glazebrook}, Karl and {Grillo}, Claudio and {Kelly}, Patrick L. and {Malkan}, Matthew A. and {Mason}, Charlotte A. and {Morishita}, Takahiro and {Nanayakkara}, Themiya and {Rosati}, Piero and {Sani}, Eleonora and {Wang}, Xin and {Yoon}, Ilsang},
        title = "{JWST NIRSpec Spectroscopy of the Remarkable Bright Galaxy GHZ2/GLASS-z12 at Redshift 12.34}",
      journal = {\apj},
         year = 2024,
        month = sep,
       volume = {972},
       number = {2},
          eid = {143},
        pages = {143},
          doi = {10.3847/1538-4357/ad5f88},
archivePrefix = {arXiv},
       eprint = {2403.10238},
 primaryClass = {astro-ph.GA},
       adsurl = {https://ui.adsabs.harvard.edu/abs/2024ApJ...972..143C}
}

@ARTICLE{Carniani24,
       author = {{Carniani}, Stefano and {Hainline}, Kevin and {D'Eugenio}, Francesco and {Eisenstein}, Daniel J. and {Jakobsen}, Peter and {Witstok}, Joris and {Johnson}, Benjamin D. and {Chevallard}, Jacopo and {Maiolino}, Roberto and {Helton}, Jakob M. and {Willott}, Chris and {Robertson}, Brant and {Alberts}, Stacey and {Arribas}, Santiago and {Baker}, William M. and {Bhatawdekar}, Rachana and {Boyett}, Kristan and {Bunker}, Andrew J. and {Cameron}, Alex J. and {Cargile}, Phillip A. and {Charlot}, St{\'e}phane and {Curti}, Mirko and {Curtis-Lake}, Emma and {Egami}, Eiichi and {Giardino}, Giovanna and {Isaak}, Kate and {Ji}, Zhiyuan and {Jones}, Gareth C. and {Kumari}, Nimisha and {Maseda}, Michael V. and {Parlanti}, Eleonora and {P{\'e}rez-Gonz{\'a}lez}, Pablo G. and {Rawle}, Tim and {Rieke}, George and {Rieke}, Marcia and {Del Pino}, Bruno Rodr{\'\i}guez and {Saxena}, Aayush and {Scholtz}, Jan and {Smit}, Renske and {Sun}, Fengwu and {Tacchella}, Sandro and {{\"U}bler}, Hannah and {Venturi}, Giacomo and {Williams}, Christina C. and {Willmer}, Christopher N.~A.},
        title = "{Spectroscopic confirmation of two luminous galaxies at a redshift of 14}",
      journal = {\nat},
         year = 2024,
        month = sep,
       volume = {633},
       number = {8029},
        pages = {318-322},
          doi = {10.1038/s41586-024-07860-9},
archivePrefix = {arXiv},
       eprint = {2405.18485},
 primaryClass = {astro-ph.GA},
       adsurl = {https://ui.adsabs.harvard.edu/abs/2024Natur.633..318C}
}

@ARTICLE{Carniani25,
       author = {{Carniani}, Stefano and {D'Eugenio}, Francesco and {Ji}, Xihan and {Parlanti}, Eleonora and {Scholtz}, Jan and {Sun}, Fengwu and {Venturi}, Giacomo and {Bakx}, Tom J.~L.~C. and {Curti}, Mirko and {Maiolino}, Roberto and {Tacchella}, Sandro and {Zavala}, Jorge A. and {Hainline}, Kevin and {Witstok}, Joris and {Johnson}, Benjamin D. and {Alberts}, Stacey and {Bunker}, Andrew J. and {Charlot}, St{\'e}phane and {Eisenstein}, Daniel J. and {Helton}, Jakob M. and {Jakobsen}, Peter and {Kumari}, Nimisha and {Robertson}, Brant and {Saxena}, Aayush and {{\"U}bler}, Hannah and {Williams}, Christina C. and {Willmer}, Christopher N.~A. and {Willott}, Chris},
        title = "{The eventful life of a luminous galaxy at z = 14: metal enrichment, feedback, and low gas fraction?}",
      journal = {\aap},
         year = 2025,
        month = apr,
       volume = {696},
          eid = {A87},
        pages = {A87},
          doi = {10.1051/0004-6361/202452451},
archivePrefix = {arXiv},
       eprint = {2409.20533},
 primaryClass = {astro-ph.GA},
       adsurl = {https://ui.adsabs.harvard.edu/abs/2025A&A...696A..87C}
}

@ARTICLE{Napolitano24,
       author = {{Napolitano}, Lorenzo and {Castellano}, Marco and {Pentericci}, Laura and {Vignali}, Cristian and {Gilli}, Roberto and {Fontana}, Adriano and {Santini}, Paola and {Treu}, Tommaso and {Calabr{\`o}}, Antonello and {Llerena}, Mario and {Piconcelli}, Enrico and {Zappacosta}, Luca and {Mascia}, Sara and {Bergamini}, Pietro and {Bakx}, Tom J.~L.~C. and {Dickinson}, Mark and {Glazebrook}, Karl and {Henry}, Alaina and {Leethochawalit}, Nicha and {Mazzolari}, Giovanni and {Merlin}, Emiliano and {Morishita}, Takahiro and {Nanayakkara}, Themiya and {Paris}, Diego and {Puccetti}, Simonetta and {Roberts-Borsani}, Guido and {Rojas Ruiz}, Sofia and {Vanzella}, Eros and {Vito}, Fabio and {Vulcani}, Benedetta and {Wang}, Xin and {Yoon}, Ilsang and {Zavala}, Jorge A.},
        title = "{The dual nature of GHZ9: coexisting AGN and star formation activity in a remote X-ray source at z=10.145}",
      journal = {arXiv e-prints},
         year = 2024,
        month = oct,
          eid = {arXiv:2410.18763},
        pages = {arXiv:2410.18763},
          doi = {10.48550/arXiv.2410.18763},
archivePrefix = {arXiv},
       eprint = {2410.18763},
 primaryClass = {astro-ph.GA},
       adsurl = {https://ui.adsabs.harvard.edu/abs/2024arXiv241018763N}
}

@ARTICLE{Donnan23b,
       author = {{Donnan}, C.~T. and {McLeod}, D.~J. and {McLure}, R.~J. and {Dunlop}, J.~S. and {Carnall}, A.~C. and {Cullen}, F. and {Magee}, D.},
        title = "{The abundance of z {\ensuremath{\gtrsim}} 10 galaxy candidates in the HUDF using deep JWST NIRCam medium-band imaging}",
      journal = {\mnras},
         year = 2023,
        month = apr,
       volume = {520},
       number = {3},
        pages = {4554-4561},
          doi = {10.1093/mnras/stad471},
archivePrefix = {arXiv},
       eprint = {2212.10126},
 primaryClass = {astro-ph.GA},
       adsurl = {https://ui.adsabs.harvard.edu/abs/2023MNRAS.520.4554D}
}

@ARTICLE{Donnan23a,
       author = {{Donnan}, C.~T. and {McLeod}, D.~J. and {Dunlop}, J.~S. and {McLure}, R.~J. and {Carnall}, A.~C. and {Begley}, R. and {Cullen}, F. and {Hamadouche}, M.~L. and {Bowler}, R.~A.~A. and {Magee}, D. and {McCracken}, H.~J. and {Milvang-Jensen}, B. and {Moneti}, A. and {Targett}, T.},
        title = "{The evolution of the galaxy UV luminosity function at redshifts z ≃ 8 - 15 from deep JWST and ground-based near-infrared imaging}",
      journal = {\mnras},
         year = 2023,
        month = feb,
       volume = {518},
       number = {4},
        pages = {6011-6040},
          doi = {10.1093/mnras/stac3472},
archivePrefix = {arXiv},
       eprint = {2207.12356},
 primaryClass = {astro-ph.GA},
       adsurl = {https://ui.adsabs.harvard.edu/abs/2023MNRAS.518.6011D}
}

@ARTICLE{Whitler25,
       author = {{Whitler}, Lily and {Stark}, Daniel P. and {Topping}, Michael W. and {Robertson}, Brant and {Rieke}, Marcia and {Hainline}, Kevin N. and {Endsley}, Ryan and {Chen}, Zuyi and {Baker}, William M. and {Bhatawdekar}, Rachana and {Bunker}, Andrew J. and {Carniani}, Stefano and {Charlot}, St{\'e}phane and {Chevallard}, Jacopo and {Curtis-Lake}, Emma and {Egami}, Eiichi and {Eisenstein}, Daniel J. and {Helton}, Jakob M. and {Ji}, Zhiyuan and {Johnson}, Benjamin D. and {P{\'e}rez-Gonz{\'a}lez}, Pablo G. and {Rinaldi}, Pierluigi and {Tacchella}, Sandro and {Williams}, Christina C. and {Willmer}, Christopher N.~A. and {Willott}, Chris and {Witstok}, Joris},
        title = "{The $z rsim 9$ galaxy UV luminosity function from the JWST Advanced Deep Extragalactic Survey: insights into early galaxy evolution and reionization}",
      journal = {arXiv e-prints},
         year = 2025,
        month = jan,
          eid = {arXiv:2501.00984},
        pages = {arXiv:2501.00984},
          doi = {10.48550/arXiv.2501.00984},
archivePrefix = {arXiv},
       eprint = {2501.00984},
 primaryClass = {astro-ph.GA},
       adsurl = {https://ui.adsabs.harvard.edu/abs/2025arXiv250100984W}
}

@ARTICLE{Trinca24,
       author = {{Trinca}, Alessandro and {Schneider}, Raffaella and {Valiante}, Rosa and {Graziani}, Luca and {Ferrotti}, Arianna and {Omukai}, Kazuyuki and {Chon}, Sunmyon},
        title = "{Exploring the nature of UV-bright z {\ensuremath{\gtrsim}} 10 galaxies detected by JWST: star formation, black hole accretion, or a non-universal IMF?}",
      journal = {\mnras},
         year = 2024,
        month = apr,
       volume = {529},
       number = {4},
        pages = {3563-3581},
          doi = {10.1093/mnras/stae651},
archivePrefix = {arXiv},
       eprint = {2305.04944},
 primaryClass = {astro-ph.GA},
       adsurl = {https://ui.adsabs.harvard.edu/abs/2024MNRAS.529.3563T}
}

@ARTICLE{Napolitano25,
       author = {{Napolitano}, L. and {Castellano}, M. and {Pentericci}, L. and {Arrabal Haro}, P. and {Fontana}, A. and {Treu}, T. and {Bergamini}, P. and {Calabr{\`o}}, A. and {Mascia}, S. and {Morishita}, T. and {Roberts-Borsani}, G. and {Santini}, P. and {Vanzella}, E. and {Vulcani}, B. and {Zakharova}, D. and {Bakx}, T. and {Dickinson}, M. and {Grillo}, C. and {Leethochawalit}, N. and {Llerena}, M. and {Merlin}, E. and {Paris}, D. and {Rojas-Ruiz}, S. and {Rosati}, P. and {Wang}, X. and {Yoon}, I. and {Zavala}, J.},
        title = "{Seven wonders of Cosmic Dawn: JWST confirms a high abundance of galaxies and AGN at z ≃ 9{\textendash}11 in the GLASS field}",
      journal = {\aap},
         year = 2025,
        month = jan,
       volume = {693},
          eid = {A50},
        pages = {A50},
          doi = {10.1051/0004-6361/202452090},
archivePrefix = {arXiv},
       eprint = {2410.10967},
 primaryClass = {astro-ph.GA},
       adsurl = {https://ui.adsabs.harvard.edu/abs/2025A&A...693A..50N}
}

@ARTICLE{Ferrara23,
       author = {{Ferrara}, Andrea and {Pallottini}, Andrea and {Dayal}, Pratika},
        title = "{On the stunning abundance of super-early, luminous galaxies revealed by JWST}",
      journal = {\mnras},
         year = 2023,
        month = jul,
       volume = {522},
       number = {3},
        pages = {3986-3991},
          doi = {10.1093/mnras/stad1095},
archivePrefix = {arXiv},
       eprint = {2208.00720},
 primaryClass = {astro-ph.GA},
       adsurl = {https://ui.adsabs.harvard.edu/abs/2023MNRAS.522.3986F}
}

@ARTICLE{Ferrara24,
       author = {{Ferrara}, A.},
        title = "{The eventful life of GS-z14-0, the most distant galaxy at redshift z = 14.32}",
      journal = {\aap},
         year = 2024,
        month = sep,
       volume = {689},
          eid = {A310},
        pages = {A310},
          doi = {10.1051/0004-6361/202450944},
archivePrefix = {arXiv},
       eprint = {2405.20370},
 primaryClass = {astro-ph.GA},
       adsurl = {https://ui.adsabs.harvard.edu/abs/2024A&A...689A.310F}
}

@ARTICLE{Inayoshi22,
       author = {{Inayoshi}, Kohei and {Harikane}, Yuichi and {Inoue}, Akio K. and {Li}, Wenxiu and {Ho}, Luis C.},
        title = "{A Lower Bound of Star Formation Activity in Ultra-high-redshift Galaxies Detected with JWST: Implications for Stellar Populations and Radiation Sources}",
      journal = {\apjl},
         year = 2022,
        month = oct,
       volume = {938},
       number = {2},
          eid = {L10},
        pages = {L10},
          doi = {10.3847/2041-8213/ac9310},
archivePrefix = {arXiv},
       eprint = {2208.06872},
 primaryClass = {astro-ph.GA},
       adsurl = {https://ui.adsabs.harvard.edu/abs/2022ApJ...938L..10I}
}

@ARTICLE{Dekel23,
       author = {{Dekel}, Avishai and {Sarkar}, Kartick C. and {Birnboim}, Yuval and {Mandelker}, Nir and {Li}, Zhaozhou},
        title = "{Efficient formation of massive galaxies at cosmic dawn by feedback-free starbursts}",
      journal = {\mnras},
         year = 2023,
        month = aug,
       volume = {523},
       number = {3},
        pages = {3201-3218},
          doi = {10.1093/mnras/stad1557},
archivePrefix = {arXiv},
       eprint = {2303.04827},
 primaryClass = {astro-ph.GA},
       adsurl = {https://ui.adsabs.harvard.edu/abs/2023MNRAS.523.3201D}
}

@ARTICLE{Li24,
       author = {{Li}, Zhaozhou and {Dekel}, Avishai and {Sarkar}, Kartick C. and {Aung}, Han and {Giavalisco}, Mauro and {Mandelker}, Nir and {Tacchella}, Sandro},
        title = "{Feedback-free starbursts at cosmic dawn: Observable predictions for JWST}",
      journal = {\aap},
         year = 2024,
        month = oct,
       volume = {690},
          eid = {A108},
        pages = {A108},
          doi = {10.1051/0004-6361/202348727},
archivePrefix = {arXiv},
       eprint = {2311.14662},
 primaryClass = {astro-ph.GA},
       adsurl = {https://ui.adsabs.harvard.edu/abs/2024A&A...690A.108L}
}

@ARTICLE{Ceverino24,
       author = {{Ceverino}, D. and {Nakazato}, Y. and {Yoshida}, N. and {Klessen}, R.~S. and {Glover}, S.~C.~O.},
        title = "{Redshift-dependent galaxy formation efficiency at z = 5 ‑ 13 in the FirstLight Simulations}",
      journal = {\aap},
         year = 2024,
        month = sep,
       volume = {689},
          eid = {A244},
        pages = {A244},
          doi = {10.1051/0004-6361/202450224},
archivePrefix = {arXiv},
       eprint = {2404.02537},
 primaryClass = {astro-ph.GA},
       adsurl = {https://ui.adsabs.harvard.edu/abs/2024A&A...689A.244C}
}

@ARTICLE{Feldmann25,
       author = {{Feldmann}, Robert and {Boylan-Kolchin}, Michael and {Bullock}, James S. and {{\c{C}}atmabacak}, Onur and {Faucher-Gigu{\`e}re}, Claude-Andr{\'e} and {Hayward}, Christopher C. and {Kere{\v{s}}}, Du{\v{s}}an and {Lazar}, Alexandres and {Liang}, Lichen and {Moreno}, Jorge and {Oesch}, Pascal A. and {Quataert}, Eliot and {Shen}, Xuejian and {Sun}, Guochao},
        title = "{Elevated UV luminosity density at Cosmic Dawn explained by non-evolving, weakly mass-dependent star formation efficiency}",
      journal = {\mnras},
         year = 2025,
        month = jan,
       volume = {536},
       number = {1},
        pages = {988-1016},
          doi = {10.1093/mnras/stae2633},
archivePrefix = {arXiv},
       eprint = {2407.02674},
 primaryClass = {astro-ph.CO},
       adsurl = {https://ui.adsabs.harvard.edu/abs/2025MNRAS.536..988F}
}

@ARTICLE{Schouws24,
       author = {{Schouws}, Sander and {Bouwens}, Rychard J. and {Ormerod}, Katherine and {Smit}, Renske and {Algera}, Hiddo and {Sommovigo}, Laura and {Hodge}, Jacqueline and {Ferrara}, Andrea and {Oesch}, Pascal A. and {Rowland}, Lucie E. and {van Leeuwen}, Ivana and {Stefanon}, Mauro and {Herard-Demanche}, Thomas and {Fudamoto}, Yoshinobu and {R{\"o}ttgering}, Huub and {van der Werf}, Paul},
        title = "{Detection of [OIII]88$\mu$m in JADES-GS-z14-0 at z=14.1793}",
      journal = {arXiv e-prints},
         year = 2024,
        month = sep,
          eid = {arXiv:2409.20549},
        pages = {arXiv:2409.20549},
          doi = {10.48550/arXiv.2409.20549},
archivePrefix = {arXiv},
       eprint = {2409.20549},
 primaryClass = {astro-ph.GA},
       adsurl = {https://ui.adsabs.harvard.edu/abs/2024arXiv240920549S}
}

@ARTICLE{Algera24,
       author = {{Algera}, Hiddo S.~B. and {Inami}, Hanae and {Sommovigo}, Laura and {Fudamoto}, Yoshinobu and {Schneider}, Raffaella and {Graziani}, Luca and {Dayal}, Pratika and {Bouwens}, Rychard and {Aravena}, Manuel and {da Cunha}, Elisabete and {Ferrara}, Andrea and {Hygate}, Alexander P.~S. and {van Leeuwen}, Ivana and {De Looze}, Ilse and {Palla}, Marco and {Pallottini}, Andrea and {Smit}, Renske and {Stefanon}, Mauro and {Topping}, Michael and {van der Werf}, Paul P.},
        title = "{Cold dust and low [O III]/[C II] ratios: an evolved star-forming population at redshift 7}",
      journal = {\mnras},
         year = 2024,
        month = jan,
       volume = {527},
       number = {3},
        pages = {6867-6887},
          doi = {10.1093/mnras/stad3111},
archivePrefix = {arXiv},
       eprint = {2301.09659},
 primaryClass = {astro-ph.GA},
       adsurl = {https://ui.adsabs.harvard.edu/abs/2024MNRAS.527.6867A}
}

@ARTICLE{Fujimoto24,
       author = {{Fujimoto}, Seiji and {Ouchi}, Masami and {Nakajima}, Kimihiko and {Harikane}, Yuichi and {Isobe}, Yuki and {Brammer}, Gabriel and {Oguri}, Masamune and {Gim{\'e}nez-Arteaga}, Clara and {Heintz}, Kasper E. and {Kokorev}, Vasily and {Bauer}, Franz E. and {Ferrara}, Andrea and {Kojima}, Takashi and {Lagos}, Claudia del P. and {Laura}, Sommovigo and {Schaerer}, Daniel and {Shimasaku}, Kazuhiro and {Hatsukade}, Bunyo and {Kohno}, Kotaro and {Sun}, Fengwu and {Valentino}, Francesco and {Watson}, Darach and {Fudamoto}, Yoshinobu and {Inoue}, Akio K. and {Gonz{\'a}lez-L{\'o}pez}, Jorge and {Koekemoer}, Anton M. and {Knudsen}, Kirsten and {Lee}, Minju M. and {Magdis}, Georgios E. and {Richard}, Johan and {Strait}, Victoria B. and {Sugahara}, Yuma and {Tamura}, Yoichi and {Toft}, Sune and {Umehata}, Hideki and {Walth}, Gregory},
        title = "{JWST and ALMA Multiple-line Study in and around a Galaxy at z = 8.496: Optical to Far-Infrared Line Ratios and the Onset of an Outflow Promoting Ionizing Photon Escape}",
      journal = {\apj},
         year = 2024,
        month = apr,
       volume = {964},
       number = {2},
          eid = {146},
        pages = {146},
          doi = {10.3847/1538-4357/ad235c},
archivePrefix = {arXiv},
       eprint = {2212.06863},
 primaryClass = {astro-ph.GA},
       adsurl = {https://ui.adsabs.harvard.edu/abs/2024ApJ...964..146F}
}

@ARTICLE{Katz22,
       author = {{Katz}, Harley and {Rosdahl}, Joakim and {Kimm}, Taysun and {Garel}, Thibault and {Blaizot}, J{\'e}r{\'e}my and {Haehnelt}, Martin G. and {Michel-Dansac}, L{\'e}o and {Martin-Alvarez}, Sergio and {Devriendt}, Julien and {Slyz}, Adrianne and {Teyssier}, Romain and {Ocvirk}, Pierre and {Laporte}, Nicolas and {Ellis}, Richard},
        title = "{The nature of high [O III]$_{88 {\ensuremath{\mu}} m}$/[C II]$_{158 {\ensuremath{\mu}}m}$ galaxies in the epoch of reionization: Low carbon abundance and a top-heavy IMF?}",
      journal = {\mnras},
         year = 2022,
        month = mar,
       volume = {510},
       number = {4},
        pages = {5603-5622},
          doi = {10.1093/mnras/stac028},
archivePrefix = {arXiv},
       eprint = {2108.01074},
 primaryClass = {astro-ph.GA},
       adsurl = {https://ui.adsabs.harvard.edu/abs/2022MNRAS.510.5603K}
}

@ARTICLE{Nyhagen24,
       author = {{Nyhagen}, Camilla T. and {Schimek}, Alice and {Cicone}, Claudia and {Decataldo}, Davide and {Shen}, Sijing},
        title = "{A theoretical investigation of far-infrared fine structure lines at $z>6$ and of the origin of the [OIII]88/[CII]158 enhancement}",
      journal = {arXiv e-prints},
         year = 2024,
        month = oct,
          eid = {arXiv:2410.18471},
        pages = {arXiv:2410.18471},
          doi = {10.48550/arXiv.2410.18471},
archivePrefix = {arXiv},
       eprint = {2410.18471},
 primaryClass = {astro-ph.GA},
       adsurl = {https://ui.adsabs.harvard.edu/abs/2024arXiv241018471N}
}

@ARTICLE{Schouws25,
       author = {{Schouws}, Sander and {Bouwens}, Rychard J. and {Algera}, Hiddo and {Smit}, Renske and {Kumari}, Nimisha and {Rowland}, Lucie E. and {van Leeuwen}, Ivana and {Sommovigo}, Laura and {Ferrara}, Andrea and {Oesch}, Pascal A. and {Ormerod}, Katherine and {Stefanon}, Mauro and {Herard-Demanche}, Thomas and {Hodge}, Jacqueline and {Fudamoto}, Yoshinobu and {R{\"o}ttgering}, Huub and {van der Werf}, Paul},
        title = "{Deep Constraints on [CII]158$\mu$m in JADES-GS-z14-0: Further Evidence for a Galaxy with Low Gas Content at z=14.2}",
      journal = {arXiv e-prints},
         year = 2025,
        month = feb,
          eid = {arXiv:2502.01610},
        pages = {arXiv:2502.01610},
          doi = {10.48550/arXiv.2502.01610},
archivePrefix = {arXiv},
       eprint = {2502.01610},
 primaryClass = {astro-ph.GA},
       adsurl = {https://ui.adsabs.harvard.edu/abs/2025arXiv250201610S}
}

@ARTICLE{Sun22a,
       author = {{Sun}, Fengwu and {Egami}, Eiichi and {Fujimoto}, Seiji and {Rawle}, Timothy and {Bauer}, Franz E. and {Kohno}, Kotaro and {Smail}, Ian and {P{\'e}rez-Gonz{\'a}lez}, Pablo G. and {Ao}, Yiping and {Chapman}, Scott C. and {Combes}, Francoise and {Dessauges-Zavadsky}, Miroslava and {Espada}, Daniel and {Gonz{\'a}lez-L{\'o}pez}, Jorge and {Koekemoer}, Anton M. and {Kokorev}, Vasily and {Lee}, Minju M. and {Morokuma-Matsui}, Kana and {Mu{\~n}oz Arancibia}, Alejandra M. and {Oguri}, Masamune and {Pell{\'o}}, Roser and {Ueda}, Yoshihiro and {Uematsu}, Ryosuke and {Valentino}, Francesco and {Van der Werf}, Paul and {Walth}, Gregory L. and {Zemcov}, Michael and {Zitrin}, Adi},
        title = "{ALMA Lensing Cluster Survey: ALMA-Herschel Joint Study of Lensed Dusty Star-forming Galaxies across z ≃ 0.5 - 6}",
      journal = {\apj},
         year = 2022,
        month = jun,
       volume = {932},
       number = {2},
          eid = {77},
        pages = {77},
          doi = {10.3847/1538-4357/ac6e3f},
archivePrefix = {arXiv},
       eprint = {2204.07187},
 primaryClass = {astro-ph.GA},
       adsurl = {https://ui.adsabs.harvard.edu/abs/2022ApJ...932...77S}
}

@ARTICLE{Rigby23,
       author = {{Rigby}, Jane and {Perrin}, Marshall and {McElwain}, Michael and {Kimble}, Randy and {Friedman}, Scott and {Lallo}, Matt and {Doyon}, Ren{\'e} and {Feinberg}, Lee and {Ferruit}, Pierre and {Glasse}, Alistair and {Rieke}, Marcia and {Rieke}, George and {Wright}, Gillian and {Willott}, Chris and {Colon}, Knicole and {Milam}, Stefanie and {Neff}, Susan and {Stark}, Christopher and {Valenti}, Jeff and {Abell}, Jim and {Abney}, Faith and {Abul-Huda}, Yasin and {Acton}, D. Scott and {Adams}, Evan and {Adler}, David and {Aguilar}, Jonathan and {Ahmed}, Nasif and {Albert}, Lo{\"\i}c and {Alberts}, Stacey and {Aldridge}, David and {Allen}, Marsha and {Altenburg}, Martin and {{\'A}lvarez-M{\'a}rquez}, Javier and {Alves de Oliveira}, Catarina and {Andersen}, Greg and {Anderson}, Harry and {Anderson}, Sara and {Argyriou}, Ioannis and {Armstrong}, Amber and {Arribas}, Santiago and {Artigau}, Etienne and {Arvai}, Amanda and {Atkinson}, Charles and {Bacon}, Gregory and {Bair}, Thomas and {Banks}, Kimberly and {Barrientes}, Jaclyn and {Barringer}, Bruce and {Bartosik}, Peter and {Bast}, William and {Baudoz}, Pierre and {Beatty}, Thomas and {Bechtold}, Katie and {Beck}, Tracy and {Bergeron}, Eddie and {Bergkoetter}, Matthew and {Bhatawdekar}, Rachana and {Birkmann}, Stephan and {Blazek}, Ronald and {Blome}, Claire and {Boccaletti}, Anthony and {B{\"o}ker}, Torsten and {Boia}, John and {Bonaventura}, Nina and {Bond}, Nicholas and {Bosley}, Kari and {Boucarut}, Ray and {Bourque}, Matthew and {Bouwman}, Jeroen and {Bower}, Gary and {Bowers}, Charles and {Boyer}, Martha and {Bradley}, Larry and {Brady}, Greg and {Braun}, Hannah and {Breda}, David and {Bresnahan}, Pamela and {Bright}, Stacey and {Britt}, Christopher and {Bromenschenkel}, Asa and {Brooks}, Brian and {Brooks}, Keira and {Brown}, Bob and {Brown}, Matthew and {Brown}, Patricia and {Bunker}, Andy and {Burger}, Matthew and {Bushouse}, Howard and {Cale}, Steven and {Cameron}, Alex and {Cameron}, Peter and {Canipe}, Alicia and {Caplinger}, James and {Caputo}, Francis and {Cara}, Mihai and {Carey}, Larkin and {Carniani}, Stefano and {Carrasquilla}, Maria and {Carruthers}, Margaret and {Case}, Michael and {Catherine}, Riggs and {Chance}, Don and {Chapman}, George and {Charlot}, St{\'e}phane and {Charlow}, Brian and {Chayer}, Pierre and {Chen}, Bin and {Cherinka}, Brian and {Chichester}, Sarah and {Chilton}, Zack and {Chonis}, Taylor and {Clampin}, Mark and {Clark}, Charles and {Clark}, Kerry and {Coe}, Dan and {Coleman}, Benee and {Comber}, Brian and {Comeau}, Tom and {Connolly}, Dennis and {Cooper}, James and {Cooper}, Rachel and {Coppock}, Eric and {Correnti}, Matteo and {Cossou}, Christophe and {Coulais}, Alain and {Coyle}, Laura and {Cracraft}, Misty and {Curti}, Mirko and {Cuturic}, Steven and {Davis}, Katherine and {Davis}, Michael and {Dean}, Bruce and {DeLisa}, Amy and {deMeester}, Wim and {Dencheva}, Nadia and {Dencheva}, Nadezhda and {DePasquale}, Joseph and {Deschenes}, Jeremy and {Hunor Detre}, {\"O}rs and {Diaz}, Rosa and {Dicken}, Dan and {DiFelice}, Audrey and {Dillman}, Matthew and {Dixon}, William and {Doggett}, Jesse and {Donaldson}, Tom and {Douglas}, Rob and {DuPrie}, Kimberly and {Dupuis}, Jean and {Durning}, John and {Easmin}, Nilufar and {Eck}, Weston and {Edeani}, Chinwe and {Egami}, Eiichi and {Ehrenwinkler}, Ralf and {Eisenhamer}, Jonathan and {Eisenhower}, Michael and {Elie}, Michelle and {Elliott}, James and {Elliott}, Kyle and {Ellis}, Tracy and {Engesser}, Michael and {Espinoza}, Nestor and {Etienne}, Odessa and {Etxaluze}, Mireya and {Falini}, Patrick and {Feeney}, Matthew and {Ferry}, Malcolm and {Filippazzo}, Joseph and {Fincham}, Brian and {Fix}, Mees and {Flagey}, Nicolas and {Florian}, Michael and {Flynn}, Jim and {Fontanella}, Erin and {Ford}, Terrance and {Forshay}, Peter and {Fox}, Ori and {Franz}, David and {Fu}, Henry and {Fullerton}, Alexander and {Galkin}, Sergey and {Galyer}, Anthony and {Garc{\'\i}a Mar{\'\i}n}, Macarena and {Gardner}, Jonathan P. and {Gardner}, Lisa and {Garland}, Dennis and {Garrett}, Bruce and {Gasman}, Danny and {Gaspar}, Andras and {Gaudreau}, Daniel and {Gauthier}, Peter and {Geers}, Vincent and {Geithner}, Paul and {Gennaro}, Mario and {Giardino}, Giovanna and {Girard}, Julien and {Giuliano}, Mark and {Glassmire}, Kirk and {Glauser}, Adrian},
        title = "{The Science Performance of JWST as Characterized in Commissioning}",
      journal = {\pasp},
         year = 2023,
        month = apr,
       volume = {135},
       number = {1046},
          eid = {048001},
        pages = {048001},
          doi = {10.1088/1538-3873/acb293},
archivePrefix = {arXiv},
       eprint = {2207.05632},
 primaryClass = {astro-ph.IM},
       adsurl = {https://ui.adsabs.harvard.edu/abs/2023PASP..135d8001R}
}

@ARTICLE{Brinchmann23,
       author = {{Brinchmann}, Jarle},
        title = "{High-z galaxies with JWST and local analogues - it is not only star formation}",
      journal = {\mnras},
         year = 2023,
        month = oct,
       volume = {525},
       number = {2},
        pages = {2087-2106},
          doi = {10.1093/mnras/stad1704},
archivePrefix = {arXiv},
       eprint = {2208.07467},
 primaryClass = {astro-ph.GA},
       adsurl = {https://ui.adsabs.harvard.edu/abs/2023MNRAS.525.2087B}
}

@ARTICLE{Carnall23,
       author = {{Carnall}, A.~C. and {Begley}, R. and {McLeod}, D.~J. and {Hamadouche}, M.~L. and {Donnan}, C.~T. and {McLure}, R.~J. and {Dunlop}, J.~S. and {Milvang-Jensen}, B. and {Bondestam}, C.~L. and {Cullen}, F. and {Jewell}, S.~M. and {Pollock}, C.~L.},
        title = "{A first look at the SMACS0723 JWST ERO: spectroscopic redshifts, stellar masses, and star-formation histories}",
      journal = {\mnras},
         year = 2023,
        month = jan,
       volume = {518},
       number = {1},
        pages = {L45-L50},
          doi = {10.1093/mnrasl/slac136},
archivePrefix = {arXiv},
       eprint = {2207.08778},
 primaryClass = {astro-ph.GA},
       adsurl = {https://ui.adsabs.harvard.edu/abs/2023MNRAS.518L..45C}
}

@ARTICLE{Sanders24,
       author = {{Sanders}, Ryan L. and {Shapley}, Alice E. and {Topping}, Michael W. and {Reddy}, Naveen A. and {Brammer}, Gabriel B.},
        title = "{Direct T $_{e}$-based Metallicities of z = 2─9 Galaxies with JWST/NIRSpec: Empirical Metallicity Calibrations Applicable from Reionization to Cosmic Noon}",
      journal = {\apj},
         year = 2024,
        month = feb,
       volume = {962},
       number = {1},
          eid = {24},
        pages = {24},
          doi = {10.3847/1538-4357/ad15fc},
archivePrefix = {arXiv},
       eprint = {2303.08149},
 primaryClass = {astro-ph.GA},
       adsurl = {https://ui.adsabs.harvard.edu/abs/2024ApJ...962...24S}
}

@ARTICLE{Laseter24,
       author = {{Laseter}, Isaac H. and {Maseda}, Michael V. and {Curti}, Mirko and {Maiolino}, Roberto and {D'Eugenio}, Francesco and {Cameron}, Alex J. and {Looser}, Tobias J. and {Arribas}, Santiago and {Baker}, William M. and {Bhatawdekar}, Rachana and {Boyett}, Kristan and {Bunker}, Andrew J. and {Carniani}, Stefano and {Charlot}, Stephane and {Chevallard}, Jacopo and {Curtis-lake}, Emma and {Egami}, Eiichi and {Eisenstein}, Daniel J. and {Hainline}, Kevin and {Hausen}, Ryan and {Ji}, Zhiyuan and {Kumari}, Nimisha and {Perna}, Michele and {Rawle}, Tim and {Rix}, Hans-Walter and {Robertson}, Brant and {Rodr{\'\i}guez Del Pino}, Bruno and {Sandles}, Lester and {Scholtz}, Jan and {Smit}, Renske and {Tacchella}, Sandro and {{\"U}bler}, Hannah and {Williams}, Christina C. and {Willott}, Chris and {Witstok}, Joris},
        title = "{JADES: Detecting [OIII]{\ensuremath{\lambda}}4363 emitters and testing strong line calibrations in the high-z Universe with ultra-deep JWST/NIRSpec spectroscopy up to z {\ensuremath{\sim}} 9.5}",
      journal = {\aap},
         year = 2024,
        month = jan,
       volume = {681},
          eid = {A70},
        pages = {A70},
          doi = {10.1051/0004-6361/202347133},
archivePrefix = {arXiv},
       eprint = {2306.03120},
 primaryClass = {astro-ph.GA},
       adsurl = {https://ui.adsabs.harvard.edu/abs/2024A&A...681A..70L}
}

@ARTICLE{Curti23,
       author = {{Curti}, Mirko and {D'Eugenio}, Francesco and {Carniani}, Stefano and {Maiolino}, Roberto and {Sandles}, Lester and {Witstok}, Joris and {Baker}, William M. and {Bennett}, Jake S. and {Piotrowska}, Joanna M. and {Tacchella}, Sandro and {Charlot}, Stephane and {Nakajima}, Kimihiko and {Maheson}, Gabriel and {Mannucci}, Filippo and {Amiri}, Amirnezam and {Arribas}, Santiago and {Belfiore}, Francesco and {Bonaventura}, Nina R. and {Bunker}, Andrew J. and {Chevallard}, Jacopo and {Cresci}, Giovanni and {Curtis-Lake}, Emma and {Hayden-Pawson}, Connor and {Jones}, Gareth C. and {Kumari}, Nimisha and {Laseter}, Isaac and {Looser}, Tobias J. and {Marconi}, Alessandro and {Maseda}, Michael V. and {Scholtz}, Jan and {Smit}, Renske and {{\"U}bler}, Hannah and {Wallace}, Imaan E.~B.},
        title = "{The chemical enrichment in the early Universe as probed by JWST via direct metallicity measurements at z {\ensuremath{\sim}} 8}",
      journal = {\mnras},
         year = 2023,
        month = jan,
       volume = {518},
       number = {1},
        pages = {425-438},
          doi = {10.1093/mnras/stac2737},
archivePrefix = {arXiv},
       eprint = {2207.12375},
 primaryClass = {astro-ph.GA},
       adsurl = {https://ui.adsabs.harvard.edu/abs/2023MNRAS.518..425C}
}

@ARTICLE{Katz23,
       author = {{Katz}, Harley and {Saxena}, Aayush and {Cameron}, Alex J. and {Carniani}, Stefano and {Bunker}, Andrew J. and {Arribas}, Santiago and {Bhatawdekar}, Rachana and {Bowler}, Rebecca A.~A. and {Boyett}, Kristan N.~K. and {Cresci}, Giovanni and {Curtis-Lake}, Emma and {D'Eugenio}, Francesco and {Kumari}, Nimisha and {Looser}, Tobias J. and {Maiolino}, Roberto and {{\"U}bler}, Hannah and {Willott}, Chris and {Witstok}, Joris},
        title = "{First insights into the ISM at z > 8 with JWST: possible physical implications of a high [O III] {\ensuremath{\lambda}}4363/[O III] {\ensuremath{\lambda}}5007}",
      journal = {\mnras},
         year = 2023,
        month = jan,
       volume = {518},
       number = {1},
        pages = {592-603},
          doi = {10.1093/mnras/stac2657},
archivePrefix = {arXiv},
       eprint = {2207.13693},
 primaryClass = {astro-ph.GA},
       adsurl = {https://ui.adsabs.harvard.edu/abs/2023MNRAS.518..592K}
}

@ARTICLE{Rhoads23,
       author = {{Rhoads}, James E. and {Wold}, Isak G.~B. and {Harish}, Santosh and {Kim}, Keunho J. and {Pharo}, John and {Malhotra}, Sangeeta and {Gabrielpillai}, Austen and {Jiang}, Tianxing and {Yang}, Huan},
        title = "{Finding Peas in the Early Universe with JWST}",
      journal = {\apjl},
         year = 2023,
        month = jan,
       volume = {942},
       number = {1},
          eid = {L14},
        pages = {L14},
          doi = {10.3847/2041-8213/acaaaf},
archivePrefix = {arXiv},
       eprint = {2207.13020},
 primaryClass = {astro-ph.GA},
       adsurl = {https://ui.adsabs.harvard.edu/abs/2023ApJ...942L..14R}
}

@ARTICLE{Schaerer22,
       author = {{Schaerer}, D. and {Marques-Chaves}, R. and {Barrufet}, L. and {Oesch}, P. and {Izotov}, Y.~I. and {Naidu}, R. and {Guseva}, N.~G. and {Brammer}, G.},
        title = "{First look with JWST spectroscopy: Resemblance among z {\ensuremath{\sim}} 8 galaxies and local analogs}",
      journal = {\aap},
         year = 2022,
        month = sep,
       volume = {665},
          eid = {L4},
        pages = {L4},
          doi = {10.1051/0004-6361/202244556},
archivePrefix = {arXiv},
       eprint = {2207.10034},
 primaryClass = {astro-ph.GA},
       adsurl = {https://ui.adsabs.harvard.edu/abs/2022A&A...665L...4S}
}

@ARTICLE{Tacchella22,
       author = {{Tacchella}, Sandro and {Finkelstein}, Steven L. and {Bagley}, Micaela and {Dickinson}, Mark and {Ferguson}, Henry C. and {Giavalisco}, Mauro and {Graziani}, Luca and {Grogin}, Norman A. and {Hathi}, Nimish and {Hutchison}, Taylor A. and {Jung}, Intae and {Koekemoer}, Anton M. and {Larson}, Rebecca L. and {Papovich}, Casey and {Pirzkal}, Norbert and {Rojas-Ruiz}, Sof{\'\i}a and {Song}, Mimi and {Schneider}, Raffaella and {Somerville}, Rachel S. and {Wilkins}, Stephen M. and {Yung}, L.~Y. Aaron},
        title = "{On the Stellar Populations of Galaxies at z = 9-11: The Growth of Metals and Stellar Mass at Early Times}",
      journal = {\apj},
         year = 2022,
        month = mar,
       volume = {927},
       number = {2},
          eid = {170},
        pages = {170},
          doi = {10.3847/1538-4357/ac4cad},
archivePrefix = {arXiv},
       eprint = {2111.05351},
 primaryClass = {astro-ph.GA},
       adsurl = {https://ui.adsabs.harvard.edu/abs/2022ApJ...927..170T}
}

@ARTICLE{Taylor22,
       author = {{Taylor}, A.~J. and {Barger}, A.~J. and {Cowie}, L.~L.},
        title = "{Metallicities of Five z > 5 Emission-line Galaxies in SMACS 0723 Revealed by JWST}",
      journal = {\apjl},
         year = 2022,
        month = nov,
       volume = {939},
       number = {1},
          eid = {L3},
        pages = {L3},
          doi = {10.3847/2041-8213/ac959d},
archivePrefix = {arXiv},
       eprint = {2208.06418},
 primaryClass = {astro-ph.GA},
       adsurl = {https://ui.adsabs.harvard.edu/abs/2022ApJ...939L...3T}
}

@ARTICLE{Trump23,
       author = {{Trump}, Jonathan R. and {Arrabal Haro}, Pablo and {Simons}, Raymond C. and {Backhaus}, Bren E. and {Amor{\'\i}n}, Ricardo O. and {Dickinson}, Mark and {Fern{\'a}ndez}, Vital and {Papovich}, Casey and {Nicholls}, David C. and {Kewley}, Lisa J. and {Brunker}, Samantha W. and {Salzer}, John J. and {Wilkins}, Stephen M. and {Almaini}, Omar and {Bagley}, Micaela B. and {Berg}, Danielle A. and {Bhatawdekar}, Rachana and {Bisigello}, Laura and {Buat}, V{\'e}ronique and {Burgarella}, Denis and {Calabr{\`o}}, Antonello and {Casey}, Caitlin M. and {Ciesla}, Laure and {Cleri}, Nikko J. and {Cole}, Justin W. and {Cooper}, M.~C. and {Cooray}, Asantha R. and {Costantin}, Luca and {Croton}, Darren and {Ferguson}, Henry C. and {Finkelstein}, Steven L. and {Fujimoto}, Seiji and {Gardner}, Jonathan P. and {Gawiser}, Eric and {Giavalisco}, Mauro and {Grazian}, Andrea and {Grogin}, Norman A. and {Hathi}, Nimish P. and {Hirschmann}, Michaela and {Holwerda}, Benne W. and {Huertas-Company}, Marc and {Hutchison}, Taylor A. and {Jogee}, Shardha and {Juneau}, St{\'e}phanie and {Jung}, Intae and {Kartaltepe}, Jeyhan S. and {Kirkpatrick}, Allison and {Kocevski}, Dale D. and {Koekemoer}, Anton M. and {Lotz}, Jennifer M. and {Lucas}, Ray A. and {Magnelli}, Benjamin and {Matharu}, Jasleen and {P{\'e}rez-Gonz{\'a}lez}, Pablo G. and {Pirzkal}, Nor and {Rafelski}, Marc and {Rose}, Caitlin and {Seill{\'e}}, Lise-Marie and {Somerville}, Rachel S. and {Straughn}, Amber N. and {Tacchella}, Sandro and {Vanderhoof}, Brittany N. and {Weiner}, Benjamin J. and {Wuyts}, Stijn and {Yung}, L.~Y. Aaron and {Zavala}, Jorge A.},
        title = "{The Physical Conditions of Emission-line Galaxies at Cosmic Dawn from JWST/NIRSpec Spectroscopy in the SMACS 0723 Early Release Observations}",
      journal = {\apj},
         year = 2023,
        month = mar,
       volume = {945},
       number = {1},
          eid = {35},
        pages = {35},
          doi = {10.3847/1538-4357/acba8a},
archivePrefix = {arXiv},
       eprint = {2207.12388},
 primaryClass = {astro-ph.GA},
       adsurl = {https://ui.adsabs.harvard.edu/abs/2023ApJ...945...35T}
}

@ARTICLE{Vallini24,
       author = {{Vallini}, Livia and {Witstok}, Joris and {Sommovigo}, Laura and {Pallottini}, Andrea and {Ferrara}, Andrea and {Carniani}, Stefano and {Kohandel}, Mahsa and {Smit}, Renske and {Gallerani}, Simona and {Gruppioni}, Carlotta},
        title = "{Spatially resolved Kennicutt-Schmidt relation at z {\ensuremath{\approx}} 7 and its connection with the interstellar medium properties}",
      journal = {\mnras},
         year = 2024,
        month = jan,
       volume = {527},
       number = {1},
        pages = {10-22},
          doi = {10.1093/mnras/stad3150},
archivePrefix = {arXiv},
       eprint = {2309.07957},
 primaryClass = {astro-ph.GA},
       adsurl = {https://ui.adsabs.harvard.edu/abs/2024MNRAS.527...10V}
}

@ARTICLE{Venturi24,
       author = {{Venturi}, G. and {Carniani}, S. and {Parlanti}, E. and {Kohandel}, M. and {Curti}, M. and {Pallottini}, A. and {Vallini}, L. and {Arribas}, S. and {Bunker}, A.~J. and {Cameron}, A.~J. and {Castellano}, M. and {Ferrara}, A. and {Fontana}, A. and {Gallerani}, S. and {Gelli}, V. and {Maiolino}, R. and {Ntormousi}, E. and {Pacifici}, C. and {Pentericci}, L. and {Salvadori}, S. and {Vanzella}, E.},
        title = "{Gas-phase metallicity gradients in galaxies at z {\ensuremath{\sim}} 6{\textendash}8}",
      journal = {\aap},
         year = 2024,
        month = nov,
       volume = {691},
          eid = {A19},
        pages = {A19},
          doi = {10.1051/0004-6361/202449855},
archivePrefix = {arXiv},
       eprint = {2403.03977},
 primaryClass = {astro-ph.GA},
       adsurl = {https://ui.adsabs.harvard.edu/abs/2024A&A...691A..19V}
}

@ARTICLE{Ferland17,
       author = {{Ferland}, G.~J. and {Chatzikos}, M. and {Guzm{\'a}n}, F. and {Lykins}, M.~L. and {van Hoof}, P.~A.~M. and {Williams}, R.~J.~R. and {Abel}, N.~P. and {Badnell}, N.~R. and {Keenan}, F.~P. and {Porter}, R.~L. and {Stancil}, P.~C.},
        title = "{The 2017 Release Cloudy}",
      journal = {\rmxaa},
         year = 2017,
        month = oct,
       volume = {53},
        pages = {385-438},
          doi = {10.48550/arXiv.1705.10877},
archivePrefix = {arXiv},
       eprint = {1705.10877},
 primaryClass = {astro-ph.GA},
       adsurl = {https://ui.adsabs.harvard.edu/abs/2017RMxAA..53..385F}
}

@ARTICLE{Binette85,
       author = {{Binette}, L. and {Dopita}, M.~A. and {Tuohy}, I.~R.},
        title = "{Radiative shock-wave theory. II. High-velocity shocks and thermal instabilities.}",
      journal = {\apj},
         year = 1985,
        month = oct,
       volume = {297},
        pages = {476-491},
          doi = {10.1086/163544},
       adsurl = {https://ui.adsabs.harvard.edu/abs/1985ApJ...297..476B}
}

@ARTICLE{Sutherland93,
       author = {{Sutherland}, Ralph S. and {Dopita}, M.~A.},
        title = "{Cooling Functions for Low-Density Astrophysical Plasmas}",
      journal = {\apjs},
         year = 1993,
        month = sep,
       volume = {88},
        pages = {253},
          doi = {10.1086/191823},
       adsurl = {https://ui.adsabs.harvard.edu/abs/1993ApJS...88..253S}
}

@software{Sutherland18,
       author = {{Sutherland}, Ralph and {Dopita}, Mike and {Binette}, Luc and {Groves}, Brent},
        title = "{MAPPINGS V: Astrophysical plasma modeling code}",
 howpublished = {Astrophysics Source Code Library, record ascl:1807.005},
         year = 2018,
        month = jul,
          eid = {ascl:1807.005},
       adsurl = {https://ui.adsabs.harvard.edu/abs/2018ascl.soft07005S}
}

@ARTICLE{Smith22,
       author = {{Smith}, A. and {Kannan}, R. and {Garaldi}, E. and {Vogelsberger}, M. and {Pakmor}, R. and {Springel}, V. and {Hernquist}, L.},
        title = "{The THESAN project: Lyman-{\ensuremath{\alpha}} emission and transmission during the Epoch of Reionization}",
      journal = {\mnras},
         year = 2022,
        month = may,
       volume = {512},
       number = {3},
        pages = {3243-3265},
          doi = {10.1093/mnras/stac713},
archivePrefix = {arXiv},
       eprint = {2110.02966},
 primaryClass = {astro-ph.CO},
       adsurl = {https://ui.adsabs.harvard.edu/abs/2022MNRAS.512.3243S}
}

@ARTICLE{Kannan22,
       author = {{Kannan}, R. and {Garaldi}, E. and {Smith}, A. and {Pakmor}, R. and {Springel}, V. and {Vogelsberger}, M. and {Hernquist}, L.},
        title = "{Introducing the THESAN project: radiation-magnetohydrodynamic simulations of the epoch of reionization}",
      journal = {\mnras},
         year = 2022,
        month = apr,
       volume = {511},
       number = {3},
        pages = {4005-4030},
          doi = {10.1093/mnras/stab3710},
archivePrefix = {arXiv},
       eprint = {2110.00584},
 primaryClass = {astro-ph.GA},
       adsurl = {https://ui.adsabs.harvard.edu/abs/2022MNRAS.511.4005K}
}

@ARTICLE{Bhagwat24b,
       author = {{Bhagwat}, Aniket and {Napolitano}, Lorenzo and {Pentericci}, Laura and {Ciardi}, Benedetta and {Costa}, Tiago},
        title = "{Ly$\alpha$ with SPICE: Interpreting Ly$\alpha$ emission at $z>5$}",
      journal = {arXiv e-prints},
         year = 2024,
        month = aug,
          eid = {arXiv:2408.16063},
        pages = {arXiv:2408.16063},
          doi = {10.48550/arXiv.2408.16063},
archivePrefix = {arXiv},
       eprint = {2408.16063},
 primaryClass = {astro-ph.GA},
       adsurl = {https://ui.adsabs.harvard.edu/abs/2024arXiv240816063B}
}

@ARTICLE{Bhagwat24a,
       author = {{Bhagwat}, Aniket and {Costa}, Tiago and {Ciardi}, Benedetta and {Pakmor}, R{\"u}diger and {Garaldi}, Enrico},
        title = "{SPICE: the connection between cosmic reionization and stellar feedback in the first galaxies}",
      journal = {\mnras},
         year = 2024,
        month = jul,
       volume = {531},
       number = {3},
        pages = {3406-3430},
          doi = {10.1093/mnras/stae1125},
archivePrefix = {arXiv},
       eprint = {2310.16895},
 primaryClass = {astro-ph.GA},
       adsurl = {https://ui.adsabs.harvard.edu/abs/2024MNRAS.531.3406B}
}

@ARTICLE{Basu25,
       author = {{Basu}, Arghyadeep and {Bhagwat}, Aniket and {Ciardi}, Benedetta and {Costa}, Tiago},
        title = "{Variability of the UV luminosity function with SPICE}",
      journal = {arXiv e-prints},
         year = 2025,
        month = jan,
          eid = {arXiv:2501.18559},
        pages = {arXiv:2501.18559},
          doi = {10.48550/arXiv.2501.18559},
archivePrefix = {arXiv},
       eprint = {2501.18559},
 primaryClass = {astro-ph.GA},
       adsurl = {https://ui.adsabs.harvard.edu/abs/2025arXiv250118559B}
}

@ARTICLE{Forbes16,
       author = {{Forbes}, John C. and {Krumholz}, Mark R. and {Goldbaum}, Nathan J. and {Dekel}, Avishai},
        title = "{Suppression of star formation in dwarf galaxies by photoelectric grain heating feedback}",
      journal = {\nat},
         year = 2016,
        month = jul,
       volume = {535},
       number = {7613},
        pages = {523-525},
          doi = {10.1038/nature18292},
archivePrefix = {arXiv},
       eprint = {1605.00650},
 primaryClass = {astro-ph.GA},
       adsurl = {https://ui.adsabs.harvard.edu/abs/2016Natur.535..523F}
}

@ARTICLE{Hu17,
       author = {{Hu}, Chia-Yu and {Naab}, Thorsten and {Glover}, Simon C.~O. and {Walch}, Stefanie and {Clark}, Paul C.},
        title = "{Variable interstellar radiation fields in simulated dwarf galaxies: supernovae versus photoelectric heating}",
      journal = {\mnras},
         year = 2017,
        month = oct,
       volume = {471},
       number = {2},
        pages = {2151-2173},
          doi = {10.1093/mnras/stx1773},
archivePrefix = {arXiv},
       eprint = {1701.08779},
 primaryClass = {astro-ph.GA},
       adsurl = {https://ui.adsabs.harvard.edu/abs/2017MNRAS.471.2151H}
}

@ARTICLE{Emerick18,
       author = {{Emerick}, Andrew and {Bryan}, Greg L. and {Mac Low}, Mordecai-Mark},
        title = "{Stellar Radiation Is Critical for Regulating Star Formation and Driving Outflows in Low-mass Dwarf Galaxies}",
      journal = {\apjl},
         year = 2018,
        month = oct,
       volume = {865},
       number = {2},
          eid = {L22},
        pages = {L22},
          doi = {10.3847/2041-8213/aae315},
archivePrefix = {arXiv},
       eprint = {1808.00468},
 primaryClass = {astro-ph.GA},
       adsurl = {https://ui.adsabs.harvard.edu/abs/2018ApJ...865L..22E}
}

@ARTICLE{Lahen19,
       author = {{Lah{\'e}n}, Natalia and {Naab}, Thorsten and {Johansson}, Peter H. and {Elmegreen}, Bruce and {Hu}, Chia-Yu and {Walch}, Stefanie},
        title = "{The Formation of Low-metallicity Globular Clusters in Dwarf Galaxy Mergers}",
      journal = {\apjl},
         year = 2019,
        month = jul,
       volume = {879},
       number = {2},
          eid = {L18},
        pages = {L18},
          doi = {10.3847/2041-8213/ab2a13},
archivePrefix = {arXiv},
       eprint = {1905.09840},
 primaryClass = {astro-ph.GA},
       adsurl = {https://ui.adsabs.harvard.edu/abs/2019ApJ...879L..18L}
}

@ARTICLE{Agertz20,
       author = {{Agertz}, Oscar and {Pontzen}, Andrew and {Read}, Justin I. and {Rey}, Martin P. and {Orkney}, Matthew and {Rosdahl}, Joakim and {Teyssier}, Romain and {Verbeke}, Robbert and {Kretschmer}, Michael and {Nickerson}, Sarah},
        title = "{EDGE: the mass-metallicity relation as a critical test of galaxy formation physics}",
      journal = {\mnras},
         year = 2020,
        month = jan,
       volume = {491},
       number = {2},
        pages = {1656-1672},
          doi = {10.1093/mnras/stz3053},
archivePrefix = {arXiv},
       eprint = {1904.02723},
 primaryClass = {astro-ph.GA},
       adsurl = {https://ui.adsabs.harvard.edu/abs/2020MNRAS.491.1656A}
}

@ARTICLE{Tress20,
       author = {{Tress}, Robin G. and {Smith}, Rowan J. and {Sormani}, Mattia C. and {Glover}, Simon C.~O. and {Klessen}, Ralf S. and {Mac Low}, Mordecai-Mark and {Clark}, Paul C.},
        title = "{Simulations of the star-forming molecular gas in an interacting M51-like galaxy}",
      journal = {\mnras},
         year = 2020,
        month = feb,
       volume = {492},
       number = {2},
        pages = {2973-2995},
          doi = {10.1093/mnras/stz3600},
archivePrefix = {arXiv},
       eprint = {1909.10520},
 primaryClass = {astro-ph.GA},
       adsurl = {https://ui.adsabs.harvard.edu/abs/2020MNRAS.492.2973T}
}

@ARTICLE{Andersson24,
       author = {{Andersson}, Eric P. and {Mac Low}, Mordecai-Mark and {Agertz}, Oscar and {Renaud}, Florent and {Li}, Hui},
        title = "{Pre-supernova feedback sets the star cluster mass function to a power law and reduces the cluster formation efficiency}",
      journal = {\aap},
         year = 2024,
        month = jan,
       volume = {681},
          eid = {A28},
        pages = {A28},
          doi = {10.1051/0004-6361/202347792},
archivePrefix = {arXiv},
       eprint = {2308.12363},
 primaryClass = {astro-ph.GA},
       adsurl = {https://ui.adsabs.harvard.edu/abs/2024A&A...681A..28A}
}

@ARTICLE{Fotopoulou24,
       author = {{Fotopoulou}, Constantina M. and {Naab}, Thorsten and {Lah{\'e}n}, Natalia and {Cernetic}, Miha and {Rathjen}, Tim-Eric and {Steinwandel}, Ulrich P. and {Hislop}, Jessica M. and {Walch}, Stefanie and {Johansson}, Peter H.},
        title = "{The masses, structure, and lifetimes of cold clouds in a high-resolution simulation of a low-metallicity starburst}",
      journal = {\mnras},
         year = 2024,
        month = oct,
       volume = {534},
       number = {1},
        pages = {215-230},
          doi = {10.1093/mnras/stae2072},
archivePrefix = {arXiv},
       eprint = {2408.16887},
 primaryClass = {astro-ph.GA},
       adsurl = {https://ui.adsabs.harvard.edu/abs/2024MNRAS.534..215F}
}

@ARTICLE{Rathjen24,
       author = {{Rathjen}, Tim-Eric and {Walch}, Stefanie and {Naab}, Thorsten and {N{\"u}rnberger}, Pierre and {W{\"u}nsch}, Richard and {Seifried}, Daniel and {Glover}, Simon C.~O.},
        title = "{SILCC -- VIII: The impact of far-ultraviolet radiation on star formation and the interstellar medium}",
      journal = {arXiv e-prints},
         year = 2024,
        month = sep,
          eid = {arXiv:2410.00124},
        pages = {arXiv:2410.00124},
          doi = {10.48550/arXiv.2410.00124},
archivePrefix = {arXiv},
       eprint = {2410.00124},
 primaryClass = {astro-ph.GA},
       adsurl = {https://ui.adsabs.harvard.edu/abs/2024arXiv241000124R}
}

@ARTICLE{Pallottini19,
       author = {{Pallottini}, A. and {Ferrara}, A. and {Decataldo}, D. and {Gallerani}, S. and {Vallini}, L. and {Carniani}, S. and {Behrens}, C. and {Kohandel}, M. and {Salvadori}, S.},
        title = "{Deep into the structure of the first galaxies: SERRA views}",
      journal = {\mnras},
         year = 2019,
        month = aug,
       volume = {487},
       number = {2},
        pages = {1689-1708},
          doi = {10.1093/mnras/stz1383},
archivePrefix = {arXiv},
       eprint = {1905.08254},
 primaryClass = {astro-ph.GA},
       adsurl = {https://ui.adsabs.harvard.edu/abs/2019MNRAS.487.1689P}
}

@ARTICLE{Vallini21,
       author = {{Vallini}, L. and {Ferrara}, A. and {Pallottini}, A. and {Carniani}, S. and {Gallerani}, S.},
        title = "{High [O III]/[C II] surface brightness ratios trace early starburst galaxies}",
      journal = {\mnras},
         year = 2021,
        month = aug,
       volume = {505},
       number = {4},
        pages = {5543-5553},
          doi = {10.1093/mnras/stab1674},
archivePrefix = {arXiv},
       eprint = {2106.05279},
 primaryClass = {astro-ph.GA},
       adsurl = {https://ui.adsabs.harvard.edu/abs/2021MNRAS.505.5543V}
}

@ARTICLE{Schimek24b,
       author = {{Schimek}, A. and {Cicone}, C. and {Shen}, S. and {Decataldo}, D. and {Klaassen}, P. and {Mayer}, L.},
        title = "{Constraining the physical properties of gas in high-z galaxies with far-infrared and submillimetre line ratios}",
      journal = {\aap},
         year = 2024,
        month = jul,
       volume = {687},
          eid = {L10},
        pages = {L10},
          doi = {10.1051/0004-6361/202449903},
archivePrefix = {arXiv},
       eprint = {2406.11559},
 primaryClass = {astro-ph.GA},
       adsurl = {https://ui.adsabs.harvard.edu/abs/2024A&A...687L..10S}
}

@ARTICLE{Schimek24a,
       author = {{Schimek}, A. and {Decataldo}, D. and {Shen}, S. and {Cicone}, C. and {Baumschlager}, B. and {van Kampen}, E. and {Klaassen}, P. and {Madau}, P. and {Di Mascolo}, L. and {Mayer}, L. and {Montoya Arroyave}, I. and {Mroczkowski}, T. and {Warraich}, J.},
        title = "{High resolution modelling of [CII], [CI], [OIII], and CO line emission from the interstellar medium and circumgalactic medium of a star-forming galaxy at z {\ensuremath{\sim}} 6.5}",
      journal = {\aap},
         year = 2024,
        month = feb,
       volume = {682},
          eid = {A98},
        pages = {A98},
          doi = {10.1051/0004-6361/202346945},
archivePrefix = {arXiv},
       eprint = {2306.00583},
 primaryClass = {astro-ph.GA},
       adsurl = {https://ui.adsabs.harvard.edu/abs/2024A&A...682A..98S}
}

@ARTICLE{Garaldi24,
       author = {{Garaldi}, Enrico and {Kannan}, Rahul and {Smith}, Aaron and {Borrow}, Josh and {Vogelsberger}, Mark and {Pakmor}, R{\"u}diger and {Springel}, Volker and {Hernquist}, Lars and {Gal{\'a}rraga-Espinosa}, Daniela and {Yeh}, Jessica Y. -C. and {Shen}, Xuejian and {Xu}, Clara and {Neyer}, Meredith and {Spina}, Benedetta and {Almualla}, Mouza and {Zhao}, Yu},
        title = "{The THESAN project: public data release of radiation-hydrodynamic simulations matching reionization-era JWST observations}",
      journal = {\mnras},
         year = 2024,
        month = jun,
       volume = {530},
       number = {4},
        pages = {3765-3786},
          doi = {10.1093/mnras/stae839},
archivePrefix = {arXiv},
       eprint = {2309.06475},
 primaryClass = {astro-ph.CO},
       adsurl = {https://ui.adsabs.harvard.edu/abs/2024MNRAS.530.3765G}
}

@ARTICLE{Kannan22b,
       author = {{Kannan}, Rahul and {Smith}, Aaron and {Garaldi}, Enrico and {Shen}, Xuejian and {Vogelsberger}, Mark and {Pakmor}, R{\"u}diger and {Springel}, Volker and {Hernquist}, Lars},
        title = "{The THESAN project: predictions for multitracer line intensity mapping in the epoch of reionization}",
      journal = {\mnras},
         year = 2022,
        month = aug,
       volume = {514},
       number = {3},
        pages = {3857-3878},
          doi = {10.1093/mnras/stac1557},
archivePrefix = {arXiv},
       eprint = {2111.02411},
 primaryClass = {astro-ph.CO},
       adsurl = {https://ui.adsabs.harvard.edu/abs/2022MNRAS.514.3857K}
}

@ARTICLE{Ocvirk16,
       author = {{Ocvirk}, Pierre and {Gillet}, Nicolas and {Shapiro}, Paul R. and {Aubert}, Dominique and {Iliev}, Ilian T. and {Teyssier}, Romain and {Yepes}, Gustavo and {Choi}, Jun-Hwan and {Sullivan}, David and {Knebe}, Alexander and {Gottl{\"o}ber}, Stefan and {D'Aloisio}, Anson and {Park}, Hyunbae and {Hoffman}, Yehuda and {Stranex}, Timothy},
        title = "{Cosmic Dawn (CoDa): the First Radiation-Hydrodynamics Simulation of Reionization and Galaxy Formation in the Local Universe}",
      journal = {\mnras},
         year = 2016,
        month = dec,
       volume = {463},
       number = {2},
        pages = {1462-1485},
          doi = {10.1093/mnras/stw2036},
archivePrefix = {arXiv},
       eprint = {1511.00011},
 primaryClass = {astro-ph.GA},
       adsurl = {https://ui.adsabs.harvard.edu/abs/2016MNRAS.463.1462O}
}

@ARTICLE{Ocvirk20,
       author = {{Ocvirk}, Pierre and {Aubert}, Dominique and {Sorce}, Jenny G. and {Shapiro}, Paul R. and {Deparis}, Nicolas and {Dawoodbhoy}, Taha and {Lewis}, Joseph and {Teyssier}, Romain and {Yepes}, Gustavo and {Gottl{\"o}ber}, Stefan and {Ahn}, Kyungjin and {Iliev}, Ilian T. and {Hoffman}, Yehuda},
        title = "{Cosmic Dawn II (CoDa II): a new radiation-hydrodynamics simulation of the self-consistent coupling of galaxy formation and reionization}",
      journal = {\mnras},
         year = 2020,
        month = aug,
       volume = {496},
       number = {4},
        pages = {4087-4107},
          doi = {10.1093/mnras/staa1266},
archivePrefix = {arXiv},
       eprint = {1811.11192},
 primaryClass = {astro-ph.GA},
       adsurl = {https://ui.adsabs.harvard.edu/abs/2020MNRAS.496.4087O}
}

@ARTICLE{Lewis22,
       author = {{Lewis}, Joseph S.~W. and {Ocvirk}, Pierre and {Sorce}, Jenny G. and {Dubois}, Yohan and {Aubert}, Dominique and {Conaboy}, Luke and {Shapiro}, Paul R. and {Dawoodbhoy}, Taha and {Teyssier}, Romain and {Yepes}, Gustavo and {Gottl{\"o}ber}, Stefan and {Rasera}, Yann and {Ahn}, Kyungjin and {Iliev}, Ilian T. and {Park}, Hyunbae and {Th{\'e}lie}, {\'E}milie},
        title = "{The short ionizing photon mean free path at z = 6 in Cosmic Dawn III, a new fully coupled radiation-hydrodynamical simulation of the Epoch of Reionization}",
      journal = {\mnras},
         year = 2022,
        month = nov,
       volume = {516},
       number = {3},
        pages = {3389-3397},
          doi = {10.1093/mnras/stac2383},
archivePrefix = {arXiv},
       eprint = {2202.05869},
 primaryClass = {astro-ph.CO},
       adsurl = {https://ui.adsabs.harvard.edu/abs/2022MNRAS.516.3389L}
}

@ARTICLE{Pawlik17,
       author = {{Pawlik}, Andreas H. and {Rahmati}, Alireza and {Schaye}, Joop and {Jeon}, Myoungwon and {Dalla Vecchia}, Claudio},
        title = "{The Aurora radiation-hydrodynamical simulations of reionization: calibration and first results}",
      journal = {\mnras},
         year = 2017,
        month = apr,
       volume = {466},
       number = {1},
        pages = {960-973},
          doi = {10.1093/mnras/stw2869},
archivePrefix = {arXiv},
       eprint = {1603.00034},
 primaryClass = {astro-ph.GA},
       adsurl = {https://ui.adsabs.harvard.edu/abs/2017MNRAS.466..960P}
}

@ARTICLE{Rosdahl18,
       author = {{Rosdahl}, Joakim and {Katz}, Harley and {Blaizot}, J{\'e}r{\'e}my and {Kimm}, Taysun and {Michel-Dansac}, L{\'e}o and {Garel}, Thibault and {Haehnelt}, Martin and {Ocvirk}, Pierre and {Teyssier}, Romain},
        title = "{The SPHINX cosmological simulations of the first billion years: the impact of binary stars on reionization}",
      journal = {\mnras},
         year = 2018,
        month = sep,
       volume = {479},
       number = {1},
        pages = {994-1016},
          doi = {10.1093/mnras/sty1655},
archivePrefix = {arXiv},
       eprint = {1801.07259},
 primaryClass = {astro-ph.GA},
       adsurl = {https://ui.adsabs.harvard.edu/abs/2018MNRAS.479..994R}
}

@ARTICLE{Rosdahl22,
       author = {{Rosdahl}, Joakim and {Blaizot}, J{\'e}r{\'e}my and {Katz}, Harley and {Kimm}, Taysun and {Garel}, Thibault and {Haehnelt}, Martin and {Keating}, Laura C. and {Martin-Alvarez}, Sergio and {Michel-Dansac}, L{\'e}o and {Ocvirk}, Pierre},
        title = "{LyC escape from SPHINX galaxies in the Epoch of Reionization}",
      journal = {\mnras},
         year = 2022,
        month = sep,
       volume = {515},
       number = {2},
        pages = {2386-2414},
          doi = {10.1093/mnras/stac1942},
archivePrefix = {arXiv},
       eprint = {2207.03232},
 primaryClass = {astro-ph.GA},
       adsurl = {https://ui.adsabs.harvard.edu/abs/2022MNRAS.515.2386R}
}

@ARTICLE{Chabrier03,
       author = {{Chabrier}, Gilles},
        title = "{Galactic Stellar and Substellar Initial Mass Function}",
      journal = {\pasp},
         year = 2003,
        month = jul,
       volume = {115},
       number = {809},
        pages = {763-795},
          doi = {10.1086/376392},
archivePrefix = {arXiv},
       eprint = {astro-ph/0304382},
 primaryClass = {astro-ph},
       adsurl = {https://ui.adsabs.harvard.edu/abs/2003PASP..115..763C}
}

@ARTICLE{Nakajima23,
       author = {{Nakajima}, Kimihiko and {Ouchi}, Masami and {Isobe}, Yuki and {Harikane}, Yuichi and {Zhang}, Yechi and {Ono}, Yoshiaki and {Umeda}, Hiroya and {Oguri}, Masamune},
        title = "{JWST Census for the Mass-Metallicity Star Formation Relations at z = 4-10 with Self-consistent Flux Calibration and Proper Metallicity Calibrators}",
      journal = {\apjs},
         year = 2023,
        month = dec,
       volume = {269},
       number = {2},
          eid = {33},
        pages = {33},
          doi = {10.3847/1538-4365/acd556},
archivePrefix = {arXiv},
       eprint = {2301.12825},
 primaryClass = {astro-ph.GA},
       adsurl = {https://ui.adsabs.harvard.edu/abs/2023ApJS..269...33N}
}

@ARTICLE{Chemerynska24,
       author = {{Chemerynska}, Iryna and {Atek}, Hakim and {Dayal}, Pratika and {Furtak}, Lukas J. and {Feldmann}, Robert and {Greene}, Jenny E. and {Maseda}, Michael V. and {Nanayakkara}, Themiya and {Oesch}, Pascal A. and {Fujimoto}, Seiji and {Labb{\'e}}, Ivo and {Bezanson}, Rachel and {Brammer}, Gabriel and {Cutler}, Sam E. and {Leja}, Joel and {Pan}, Richard and {Price}, Sedona H. and {Wang}, Bingjie and {Weaver}, John R. and {Whitaker}, Katherine E.},
        title = "{The Extreme Low-mass End of the Mass{\textendash}Metallicity Relation at z {\ensuremath{\sim}} 7}",
      journal = {\apjl},
         year = 2024,
        month = nov,
       volume = {976},
       number = {1},
          eid = {L15},
        pages = {L15},
          doi = {10.3847/2041-8213/ad8dc910.1134/S1063772908080040},
archivePrefix = {arXiv},
       eprint = {2407.17110},
 primaryClass = {astro-ph.GA},
       adsurl = {https://ui.adsabs.harvard.edu/abs/2024ApJ...976L..15C}
}

@ARTICLE{Matthee23,
       author = {{Matthee}, Jorryt and {Mackenzie}, Ruari and {Simcoe}, Robert A. and {Kashino}, Daichi and {Lilly}, Simon J. and {Bordoloi}, Rongmon and {Eilers}, Anna-Christina},
        title = "{EIGER. II. First Spectroscopic Characterization of the Young Stars and Ionized Gas Associated with Strong H{\ensuremath{\beta}} and [O III] Line Emission in Galaxies at z = 5-7 with JWST}",
      journal = {\apj},
         year = 2023,
        month = jun,
       volume = {950},
       number = {1},
          eid = {67},
        pages = {67},
          doi = {10.3847/1538-4357/acc846},
archivePrefix = {arXiv},
       eprint = {2211.08255},
 primaryClass = {astro-ph.GA},
       adsurl = {https://ui.adsabs.harvard.edu/abs/2023ApJ...950...67M}
}

@ARTICLE{Saxena23,
       author = {{Saxena}, Aayush and {Robertson}, Brant E. and {Bunker}, Andrew J. and {Endsley}, Ryan and {Cameron}, Alex J. and {Charlot}, Stephane and {Simmonds}, Charlotte and {Tacchella}, Sandro and {Witstok}, Joris and {Willott}, Chris and {Carniani}, Stefano and {Curtis-Lake}, Emma and {Ferruit}, Pierre and {Jakobsen}, Peter and {Arribas}, Santiago and {Chevallard}, Jacopo and {Curti}, Mirko and {D'Eugenio}, Francesco and {De Graaff}, Anna and {Jones}, Gareth C. and {Looser}, Tobias J. and {Maseda}, Michael V. and {Rawle}, Tim and {Rix}, Hans-Walter and {Del Pino}, Bruno Rodr{\'\i}guez and {Smit}, Renske and {{\"U}bler}, Hannah and {Eisenstein}, Daniel J. and {Hainline}, Kevin and {Hausen}, Ryan and {Johnson}, Benjamin D. and {Rieke}, Marcia and {Williams}, Christina C. and {Willmer}, Christopher N.~A. and {Baker}, William M. and {Bhatawdekar}, Rachana and {Bowler}, Rebecca and {Boyett}, Kristan and {Chen}, Zuyi and {Egami}, Eiichi and {Ji}, Zhiyuan and {Kumari}, Nimisha and {Nelson}, Erica and {Perna}, Michele and {Sandles}, Lester and {Scholtz}, Jan and {Shivaei}, Irene},
        title = "{JADES: Discovery of extremely high equivalent width Lyman-{\ensuremath{\alpha}} emission from a faint galaxy within an ionized bubble at z = 7.3}",
      journal = {\aap},
         year = 2023,
        month = oct,
       volume = {678},
          eid = {A68},
        pages = {A68},
          doi = {10.1051/0004-6361/202346245},
archivePrefix = {arXiv},
       eprint = {2302.12805},
 primaryClass = {astro-ph.GA},
       adsurl = {https://ui.adsabs.harvard.edu/abs/2023A&A...678A..68S}
}

@ARTICLE{Sun23,
       author = {{Sun}, Fengwu and {Egami}, Eiichi and {Pirzkal}, Nor and {Rieke}, Marcia and {Baum}, Stefi and {Boyer}, Martha and {Boyett}, Kristan and {Bunker}, Andrew J. and {Cameron}, Alex J. and {Curti}, Mirko and {Eisenstein}, Daniel J. and {Gennaro}, Mario and {Greene}, Thomas P. and {Jaffe}, Daniel and {Kelly}, Doug and {Koekemoer}, Anton M. and {Kumari}, Nimisha and {Maiolino}, Roberto and {Maseda}, Michael and {Perna}, Michele and {Rest}, Armin and {Robertson}, Brant E. and {Schlawin}, Everett and {Smit}, Renske and {Stansberry}, John and {Sunnquist}, Ben and {Tacchella}, Sandro and {Williams}, Christina C. and {Willmer}, Christopher N.~A.},
        title = "{First Sample of H{\ensuremath{\alpha}}+[O III]{\ensuremath{\lambda}}5007 Line Emitters at z > 6 Through JWST/NIRCam Slitless Spectroscopy: Physical Properties and Line-luminosity Functions}",
      journal = {\apj},
         year = 2023,
        month = aug,
       volume = {953},
       number = {1},
          eid = {53},
        pages = {53},
          doi = {10.3847/1538-4357/acd53c},
archivePrefix = {arXiv},
       eprint = {2209.03374},
 primaryClass = {astro-ph.GA},
       adsurl = {https://ui.adsabs.harvard.edu/abs/2023ApJ...953...53S}
}

@ARTICLE{Meyer24,
       author = {{Meyer}, R.~A. and {Oesch}, P.~A. and {Giovinazzo}, E. and {Weibel}, A. and {Brammer}, G. and {Matthee}, J. and {Naidu}, R.~P. and {Bouwens}, R.~J. and {Chisholm}, J. and {Covelo-Paz}, A. and {Fudamoto}, Y. and {Maseda}, M. and {Nelson}, E. and {Shivaei}, I. and {Xiao}, M. and {Herard-Demanche}, T. and {Illingworth}, G.~D. and {Kerutt}, J. and {Kramarenko}, I. and {Labbe}, I. and {Leonova}, E. and {Magee}, D. and {Matharu}, J. and {Prieto Lyon}, G. and {Reddy}, N. and {Schaerer}, D. and {Shapley}, A. and {Stefanon}, M. and {Wozniak}, M.~A. and {Wuyts}, S.},
        title = "{JWST FRESCO: a comprehensive census of H {\ensuremath{\beta}} + [O III] emitters at 6.8 < z < 9.0 in the GOODS fields}",
      journal = {\mnras},
         year = 2024,
        month = nov,
       volume = {535},
       number = {1},
        pages = {1067-1094},
          doi = {10.1093/mnras/stae2353},
archivePrefix = {arXiv},
       eprint = {2405.05111},
 primaryClass = {astro-ph.GA},
       adsurl = {https://ui.adsabs.harvard.edu/abs/2024MNRAS.535.1067M}
}

@ARTICLE{Wold25,
       author = {{Wold}, Isak G.~B. and {Malhotra}, Sangeeta and {Rhoads}, James E. and {Weaver}, John R. and {Wang}, Bingjie},
        title = "{UNCOVERing the Faint End of the z {\ensuremath{\sim}} 7 [O III] Luminosity Function with JWST's F410M Medium Bandpass Filter}",
      journal = {\apj},
         year = 2025,
        month = feb,
       volume = {980},
       number = {2},
          eid = {200},
        pages = {200},
          doi = {10.3847/1538-4357/ada8a6},
archivePrefix = {arXiv},
       eprint = {2407.19023},
 primaryClass = {astro-ph.GA},
       adsurl = {https://ui.adsabs.harvard.edu/abs/2025ApJ...980..200W}
}

@ARTICLE{Rey25,
       author = {{Rey}, Martin P. and {Taylor}, Ethan and {Gray}, Emily I. and {Kim}, Stacy Y. and {Andersson}, Eric P. and {Pontzen}, Andrew and {Agertz}, Oscar and {Read}, Justin I. and {Cadiou}, Corentin and {Yates}, Robert M. and {Orkney}, Matthew D.~A. and {Scholte}, Dirk and {Saintonge}, Am{\'e}lie and {Breneman}, Joseph and {McQuinn}, Kristen B.~W. and {Muni}, Claudia and {Das}, Payel},
        title = "{EDGE: The emergence of dwarf galaxy scaling relations from cosmological radiation-hydrodynamics simulations}",
      journal = {arXiv e-prints},
         year = 2025,
        month = mar,
          eid = {arXiv:2503.03813},
        pages = {arXiv:2503.03813},
          doi = {10.48550/arXiv.2503.03813},
archivePrefix = {arXiv},
       eprint = {2503.03813},
 primaryClass = {astro-ph.GA},
       adsurl = {https://ui.adsabs.harvard.edu/abs/2025arXiv250303813R}
}

@ARTICLE{Marinacci19,
       author = {{Marinacci}, Federico and {Sales}, Laura V. and {Vogelsberger}, Mark and {Torrey}, Paul and {Springel}, Volker},
        title = "{Simulating the interstellar medium and stellar feedback on a moving mesh: implementation and isolated galaxies}",
      journal = {\mnras},
         year = 2019,
        month = nov,
       volume = {489},
       number = {3},
        pages = {4233-4260},
          doi = {10.1093/mnras/stz2391},
archivePrefix = {arXiv},
       eprint = {1905.08806},
 primaryClass = {astro-ph.GA},
       adsurl = {https://ui.adsabs.harvard.edu/abs/2019MNRAS.489.4233M}
}

@ARTICLE{McQuinn16,
       author = {{McQuinn}, Matthew},
        title = "{The Evolution of the Intergalactic Medium}",
      journal = {\araa},
         year = 2016,
        month = sep,
       volume = {54},
        pages = {313-362},
          doi = {10.1146/annurev-astro-082214-122355},
archivePrefix = {arXiv},
       eprint = {1512.00086},
 primaryClass = {astro-ph.CO},
       adsurl = {https://ui.adsabs.harvard.edu/abs/2016ARA&A..54..313M}
}

@ARTICLE{Wilkins23,
       author = {{Wilkins}, Stephen M. and {Vijayan}, Aswin P. and {Lovell}, Christopher C. and {Roper}, William J. and {Irodotou}, Dimitrios and {Caruana}, Joseph and {Seeyave}, Louise T.~C. and {Kuusisto}, Jussi K. and {Thomas}, Peter A. and {Parris}, Shedeur A.~K.},
        title = "{First light and reionization epoch simulations (FLARES) V: the redshift frontier}",
      journal = {\mnras},
         year = 2023,
        month = feb,
       volume = {519},
       number = {2},
        pages = {3118-3128},
          doi = {10.1093/mnras/stac3280},
archivePrefix = {arXiv},
       eprint = {2204.09431},
 primaryClass = {astro-ph.GA},
       adsurl = {https://ui.adsabs.harvard.edu/abs/2023MNRAS.519.3118W}
}

@ARTICLE{Ormerod24,
       author = {{Ormerod}, K. and {Conselice}, C.~J. and {Adams}, N.~J. and {Harvey}, T. and {Austin}, D. and {Trussler}, J. and {Ferreira}, L. and {Caruana}, J. and {Lucatelli}, G. and {Li}, Q. and {Roper}, W.~J.},
        title = "{EPOCHS VI: the size and shape evolution of galaxies since z   8 with JWST Observations}",
      journal = {\mnras},
         year = 2024,
        month = jan,
       volume = {527},
       number = {3},
        pages = {6110-6125},
          doi = {10.1093/mnras/stad3597},
archivePrefix = {arXiv},
       eprint = {2309.04377},
 primaryClass = {astro-ph.GA},
       adsurl = {https://ui.adsabs.harvard.edu/abs/2024MNRAS.527.6110O}
}

@ARTICLE{Jia24,
       author = {{Jia}, Cheng and {Wang}, Enci and {Wang}, Huiyuan and {Li}, Hui and {Yao}, Yao and {Song}, Jie and {Zhang}, Hongxin and {Rong}, Yu and {Chen}, Yangyao and {Yu}, Haoran and {Chen}, Zeyu and {Li}, Haixin and {Ma}, Chengyu and {Kong}, Xu},
        title = "{Size Growth on Short Timescales of Star-forming Galaxies: Insights from Size Variation with Rest-frame Wavelength with JADES}",
      journal = {\apj},
         year = 2024,
        month = dec,
       volume = {977},
       number = {2},
          eid = {165},
        pages = {165},
          doi = {10.3847/1538-4357/ad919a},
archivePrefix = {arXiv},
       eprint = {2411.07458},
 primaryClass = {astro-ph.GA},
       adsurl = {https://ui.adsabs.harvard.edu/abs/2024ApJ...977..165J}
}

@ARTICLE{Buitrago08,
       author = {{Buitrago}, Fernando and {Trujillo}, Ignacio and {Conselice}, Christopher J. and {Bouwens}, Rychard J. and {Dickinson}, Mark and {Yan}, Haojing},
        title = "{Size Evolution of the Most Massive Galaxies at 1.7 < z < 3 from GOODS NICMOS Survey Imaging}",
      journal = {\apjl},
         year = 2008,
        month = nov,
       volume = {687},
       number = {2},
        pages = {L61},
          doi = {10.1086/592836},
archivePrefix = {arXiv},
       eprint = {0807.4141},
 primaryClass = {astro-ph},
       adsurl = {https://ui.adsabs.harvard.edu/abs/2008ApJ...687L..61B}
}

@ARTICLE{vanDokkum10,
       author = {{van Dokkum}, Pieter G. and {Whitaker}, Katherine E. and {Brammer}, Gabriel and {Franx}, Marijn and {Kriek}, Mariska and {Labb{\'e}}, Ivo and {Marchesini}, Danilo and {Quadri}, Ryan and {Bezanson}, Rachel and {Illingworth}, Garth D. and {Muzzin}, Adam and {Rudnick}, Gregory and {Tal}, Tomer and {Wake}, David},
        title = "{The Growth of Massive Galaxies Since z = 2}",
      journal = {\apj},
         year = 2010,
        month = feb,
       volume = {709},
       number = {2},
        pages = {1018-1041},
          doi = {10.1088/0004-637X/709/2/1018},
archivePrefix = {arXiv},
       eprint = {0912.0514},
 primaryClass = {astro-ph.CO},
       adsurl = {https://ui.adsabs.harvard.edu/abs/2010ApJ...709.1018V}
}

@ARTICLE{Behroozi19,
       author = {{Behroozi}, Peter and {Wechsler}, Risa H. and {Hearin}, Andrew P. and {Conroy}, Charlie},
        title = "{UNIVERSEMACHINE: The correlation between galaxy growth and dark matter halo assembly from z = 0-10}",
      journal = {\mnras},
         year = 2019,
        month = sep,
       volume = {488},
       number = {3},
        pages = {3143-3194},
          doi = {10.1093/mnras/stz1182},
archivePrefix = {arXiv},
       eprint = {1806.07893},
 primaryClass = {astro-ph.GA},
       adsurl = {https://ui.adsabs.harvard.edu/abs/2019MNRAS.488.3143B}
}

@ARTICLE{Costantin23,
       author = {{Costantin}, Luca and {P{\'e}rez-Gonz{\'a}lez}, Pablo G. and {Vega-Ferrero}, Jes{\'u}s and {Huertas-Company}, Marc and {Bisigello}, Laura and {Buitrago}, Fernando and {Bagley}, Micaela B. and {Cleri}, Nikko J. and {Cooper}, Michael C. and {Finkelstein}, Steven L. and {Holwerda}, Benne W. and {Kartaltepe}, Jeyhan S. and {Koekemoer}, Anton M. and {Nelson}, Dylan and {Papovich}, Casey and {Pillepich}, Annalisa and {Pirzkal}, Nor and {Tacchella}, Sandro and {Yung}, L.~Y. Aaron},
        title = "{Expectations of the Size Evolution of Massive Galaxies at 3 {\ensuremath{\leq}} z {\ensuremath{\leq}} 6 from the TNG50 Simulation: The CEERS/JWST View}",
      journal = {\apj},
         year = 2023,
        month = apr,
       volume = {946},
       number = {2},
          eid = {71},
        pages = {71},
          doi = {10.3847/1538-4357/acb926},
archivePrefix = {arXiv},
       eprint = {2208.00007},
 primaryClass = {astro-ph.GA},
       adsurl = {https://ui.adsabs.harvard.edu/abs/2023ApJ...946...71C}
}

@ARTICLE{Arribas24,
       author = {{Arribas}, Santiago and {Perna}, Michele and {Rodr{\'\i}guez Del Pino}, Bruno and {Lamperti}, Isabella and {D'Eugenio}, Francesco and {P{\'e}rez-Gonz{\'a}lez}, Pablo G. and {Jones}, Gareth C. and {Crespo G{\'o}mez}, Alejandro and {Curti}, Mirko and {Lim}, Seunghwan and {{\'A}lvarez-M{\'a}rquez}, Javier and {Bunker}, Andrew J. and {Carniani}, Stefano and {Charlot}, St{\'e}phane and {Jakobsen}, Peter and {Maiolino}, Roberto and {{\"U}bler}, Hannah and {Willott}, Chris J. and {B{\"o}ker}, Torsten and {Chevallard}, Jacopo and {Circosta}, Chiara and {Cresci}, Giovanni and {Kumari}, Nimisha and {Parlanti}, Eleonora and {Scholtz}, Jan and {Venturi}, Giacomo and {Witstok}, Joris},
        title = "{GA-NIFS: The core of an extremely massive protocluster at the epoch of reionisation probed with JWST/NIRSpec}",
      journal = {\aap},
         year = 2024,
        month = aug,
       volume = {688},
          eid = {A146},
        pages = {A146},
          doi = {10.1051/0004-6361/202348824},
archivePrefix = {arXiv},
       eprint = {2312.00899},
 primaryClass = {astro-ph.GA},
       adsurl = {https://ui.adsabs.harvard.edu/abs/2024A&A...688A.146A}
}

@ARTICLE{Danhaive25,
       author = {{Danhaive}, A. Lola and {Tacchella}, Sandro and {\textbackslash''Ubler}, Hannah and {de Graaff}, Anna and {Egami}, Eiichi and {Johnson}, Benjamin D. and {Sun}, Fengwu and {Arribas}, Santiago and {Bunker}, Andrew J. and {Carniani}, Stefano and {Jones}, Gareth C. and {Maiolino}, Roberto and {McClymont}, William and {Parlanti}, Eleonora and {Simmonds}, Charlotte and {Villanueva}, Natalia C. and {Baker}, William M. and {Jaffe}, Daniel T. and {Eisenstein}, Daniel and {Hainline}, Kevin and {Helton}, Jakob M. and {Ji}, Zhiyuan and {Lin}, Xiaojing and {Pusk\textbackslash'as}, D\textbackslash'avid and {Rieke}, Marcia and {Rinaldi}, Pierluigi and {Robertson}, Brant and {Scholz}, Jan and {Williams}, Christina C. and {Willmer}, Christopher N.~A.},
        title = "{The dawn of disks: unveiling the turbulent ionised gas kinematics of the galaxy population at $z\sim4-6$ with JWST/NIRCam grism spectroscopy}",
      journal = {arXiv e-prints},
         year = 2025,
        month = mar,
          eid = {arXiv:2503.21863},
        pages = {arXiv:2503.21863},
          doi = {10.48550/arXiv.2503.21863},
archivePrefix = {arXiv},
       eprint = {2503.21863},
 primaryClass = {astro-ph.GA},
       adsurl = {https://ui.adsabs.harvard.edu/abs/2025arXiv250321863D}
}

@ARTICLE{DeGraaff24,
       author = {{de Graaff}, Anna and {Rix}, Hans-Walter and {Carniani}, Stefano and {Suess}, Katherine A. and {Charlot}, St{\'e}phane and {Curtis-Lake}, Emma and {Arribas}, Santiago and {Baker}, William M. and {Boyett}, Kristan and {Bunker}, Andrew J. and {Cameron}, Alex J. and {Chevallard}, Jacopo and {Curti}, Mirko and {Eisenstein}, Daniel J. and {Franx}, Marijn and {Hainline}, Kevin and {Hausen}, Ryan and {Ji}, Zhiyuan and {Johnson}, Benjamin D. and {Jones}, Gareth C. and {Maiolino}, Roberto and {Maseda}, Michael V. and {Nelson}, Erica and {Parlanti}, Eleonora and {Rawle}, Tim and {Robertson}, Brant and {Tacchella}, Sandro and {{\"U}bler}, Hannah and {Williams}, Christina C. and {Willmer}, Christopher N.~A. and {Willott}, Chris},
        title = "{Ionised gas kinematics and dynamical masses of z {\ensuremath{\gtrsim}} 6 galaxies from JADES/NIRSpec high-resolution spectroscopy}",
      journal = {\aap},
         year = 2024,
        month = apr,
       volume = {684},
          eid = {A87},
        pages = {A87},
          doi = {10.1051/0004-6361/202347755},
archivePrefix = {arXiv},
       eprint = {2308.09742},
 primaryClass = {astro-ph.GA},
       adsurl = {https://ui.adsabs.harvard.edu/abs/2024A&A...684A..87D}
}




\appendix

\section{\LOIII\ vs. SFR}
\label{appendixA}

\begin{figure}
    \centering
    \includegraphics[width=\linewidth]{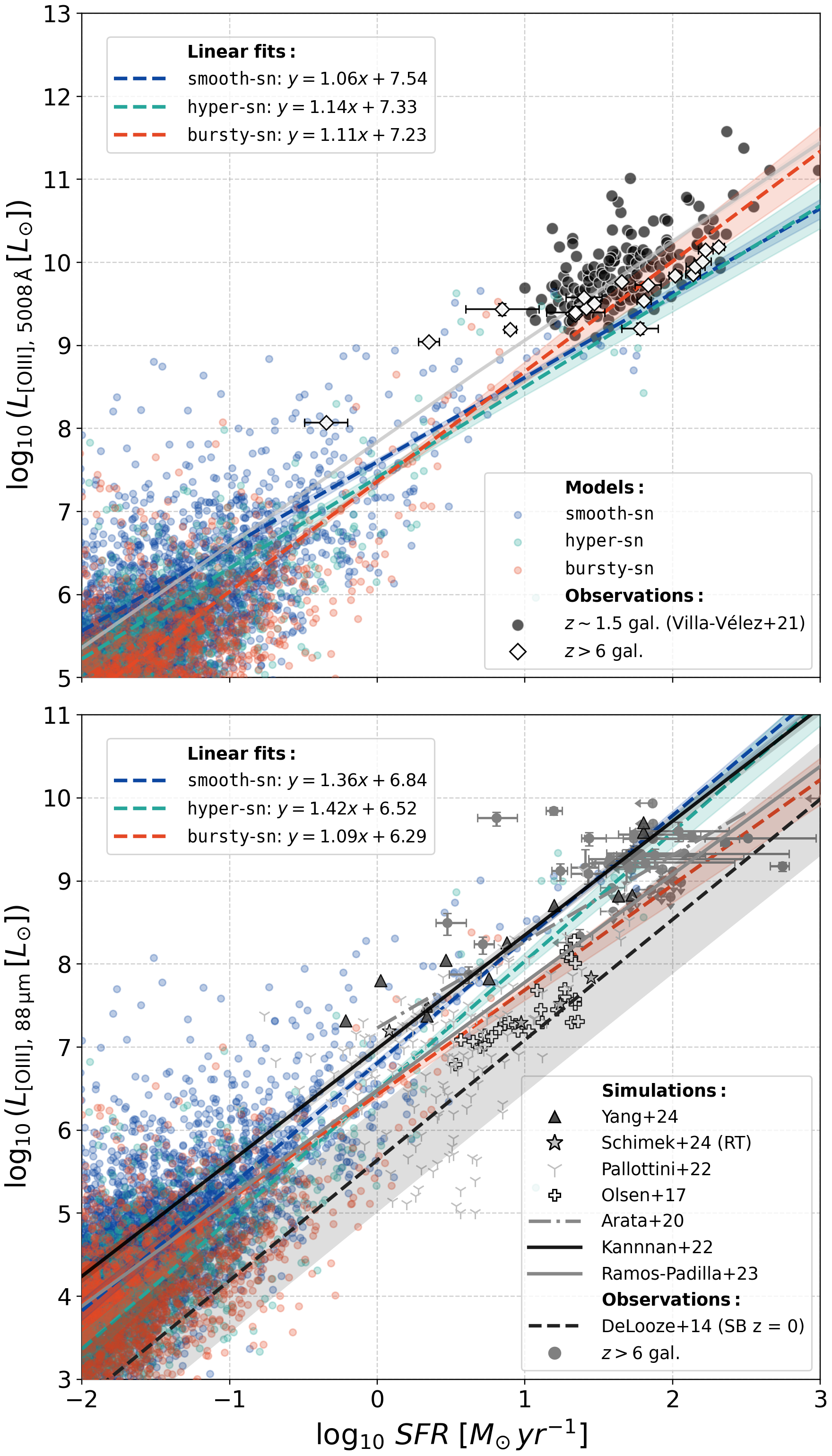}
    \caption{Dust-corrected \LOIII\ as a function of SFR for the emission at 5008 \AA\ (top) and 88 $\mu$m (bottom). Points represent galaxies at $z = 10, 7,$ and $5$ for the \texttt{bursty-sn} (red), \texttt{hyper-sn} (green), and \texttt{smooth-sn} (blue) models. Dashed colored lines indicate the stellar-mass-weighted linear fits. 
    We compare our results for the optical relation with observations of galaxies at $z \sim 1.5$ by \protect\cite{Villa-Velez21} (black points) and a compilation of $z > 6$ galaxies detected with JWST (white diamonds). We also compare the relation for the emission at 88 $\mu$m with $z > 6$ galaxies detected with ALMA (grey points) and local starburst galaxies from \protect\cite{DeLooze14} (black dashed line). Results from the cosmological simulations THESAN \protect\citep{Kannan22} and EAGLE \protect\citep{RamosPadilla23} 
    are shown as black and grey solid lines, respectively. 
    We also report results from individual zoom-in galaxies at $z = 6$ as grey symbols \citep{Olsen17, Arata20, Pallottini22, Schimek24a, Yang24}. Our linear trends are in good agreement with both observations and numerical studies.
    }
    \label{fig:LOIIIvsSFR}
\end{figure}

As a sanity check, in Figure \ref{fig:LOIIIvsSFR} we verified how our \OIII\ emission model reproduces the observed \LOIII–SFR relation, both for the optical 5008 \AA\ and the far-infrared 88 $\mu$m transitions. In both cases, we find a nearly linear trend, confirming that \OIII\ is a reliable tracer of star formation activity \citep{DeLooze14}.

For the 88 $\mu$m line, \texttt{SPICE} galaxies and the extrapolation of the linear fits at higher SFR are in agreement with recent ALMA observations of star-forming galaxies at $z>6$ displayed in Figure \ref{fig:LOIIIvsSFR} \citep[][]{Inoue16, Carniani17, Hashimoto18, Marrone18, Walter18, Hashimoto19, Tamura19, Harikane20, Akins22, Tadaki22, Witstok22, Wong22, Algera24, Bakx24, Fujimoto24, Zavala24, Algera25, Carniani25, Schouws25, vanLeeuwen25}, as well as with the post-processed \texttt{EAGLE} predictions for $z=6$ galaxies \citep{RamosPadilla23}. The \texttt{SPICE} galaxies also overlap with predictions from recent high-resolution zoom-in simulations \citep[e.g.,][]{Olsen17, Arata20, Pallottini22, Schimek24a, Yang24}, further reinforcing the consistency of the results across numerical approaches. Moreover, the \texttt{SPICE} galaxies reproduce the observed scatter, and their \LOIII–SFR relation closely follows the empirical fit derived for local starburst galaxies by \cite{DeLooze14}, with the best match found for the \texttt{bursty-sn} model. This indicates no clear redshift evolution in the normalization of the \LOIII–SFR relation, suggesting that \OIII\ emission traces similar ISM conditions across cosmic time.

The optical relation remains poorly explored observationally at high redshift, but we find that all three \texttt{SPICE} feedback models are consistent with the measurements from \cite{Villa-Velez21}, obtained for galaxies at $z \sim 1.5$ using the Fiber Multi-Object Spectrograph (FMOS) and for galaxies observed at $z > 6$ with JWST \citep{Curti23, Saxena23, Sun23, Rowland26}. As in the infrared case, this implies that the \LOIII–SFR relation exhibits little to no evolution with redshift, in agreement with results from the \texttt{THESAN} simulations \citep{Kannan22}.


\bsp	
\label{lastpage}
\end{document}